\documentclass[aps,prl,reprint,amsmath,amssymb,notitlepage,citeautoscript,superscriptaddress]{revtex4-2}

\usepackage{bm,mathrsfs,dcolumn,graphicx,color}
\usepackage{amsmath}
\usepackage{graphicx}
\usepackage[T1]{fontenc}
\usepackage{hyphenat}
\usepackage{setspace}
\usepackage[dvipsnames]{xcolor}
\usepackage{siunitx} 

\usepackage[normalem]{ulem}
\definecolor{darkerblue}{rgb}{0,0,0.75}
\definecolor{darkerred}{rgb}{0.8,0,0}
\usepackage[colorlinks=true,citecolor=darkerblue,linkcolor=darkerred]{hyperref}

\definecolor{ablue}{rgb}{0,0,0}
\definecolor{ired}{rgb}{0,0,0}

\begin{document}







\title{Non-linear and negative effective diffusivity of optical excitations in moiré-free heterobilayers}


\author{Edith Wietek}
\affiliation{Institute of Applied Physics and W\"urzburg-Dresden Cluster of Excellence ct.qmat, Technische Universit\"at Dresden, 01062 Dresden, Germany}
\author{Matthias Florian}
\affiliation{University of Michigan, Department of Electrical Engineering and Computer Science, Ann Arbor, Michigan 48109, USA}
\author{Jonas M. Göser}
\affiliation{Fakult\"at f\"ur Physik, Munich Quantum Center, and Center for NanoScience (CeNS), Ludwig-Maximilians-Universität M\"unchen, 80539 M\"unchen, Germany}
\author{Takashi Taniguchi}
\affiliation{Research Center for Materials Nanoarchitectonics, National Institute for Materials Science,  1-1 Namiki, Tsukuba 305-0044, Japan}
\author{Kenji Watanabe}
\affiliation{Research Center for Electronic and Optical Materials, National Institute for Materials Science, 1-1 Namiki, Tsukuba 305-0044, Japan}
\author{Alexander Högele}
\affiliation{Fakult\"at f\"ur Physik, Munich Quantum Center, and Center for NanoScience (CeNS), Ludwig-Maximilians-Universität M\"unchen, 80539 M\"unchen, Germany}
\affiliation{Munich Center for Quantum Science and Technology (MCQST), 80799 M\"unchen, Germany}
\author{Mikhail M. Glazov}
\affiliation{Ioffe Institute, 194021 Saint Petersburg, Russian Federation}
\author{Alexander Steinhoff}
\affiliation{Institut für Theoretische Physik, Universit\"at Bremen, 28334 Bremen, Germany}
\affiliation{Bremen Center for Computational Materials Science, Universit\"at Bremen, 28334 Bremen, Germany}
\author{Alexey Chernikov}
\email{alexey.chernikov@tu-dresden.de}
\affiliation{Institute of Applied Physics and W\"urzburg-Dresden Cluster of Excellence ct.qmat, Technische Universit\"at Dresden, 01062 Dresden, Germany}

\begin{abstract}
{Interlayer exciton diffusion is studied in atomically-reconstructed MoSe$_2$/WSe$_2$ heterobilayers with suppressed disorder.
Local atomic registry is confirmed by characteristic optical absorption, circularly-polarized photoluminescence, and \textit{g}-factor measurements.
Using transient microscopy we observe propagation properties of interlayer excitons that are independent from trapping at moiré- or disorder-induced local potentials.
Confirmed by characteristic temperature dependence for free particles, linear diffusion coefficients of interlayer excitons at liquid helium temperature and low excitation densities are almost 1000 times higher than in previous observations.
We further show that exciton-exciton repulsion and annihilation contribute nearly equally to non-linear propagation by disentangling the two processes in the experiment and simulations. 
Finally, we demonstrate effective shrinking of the light-emission over time across several 100's of picoseconds at the transition from exciton- to the plasma-dominated regimes.
Supported by microscopic calculations for bandgap renormalization to identify Mott threshold, this indicates transient crossing between rapidly expanding, short-lived electron-hole plasma and slower, long-lived exciton populations. 
}
\end{abstract}
\maketitle

Within the rich family of van der Waals heterostructures\,\cite{Geim2013, Novoselov2016}, artificially stacked bilayers of semiconducting transition metal dichalcogenides (TMDCs) emerged as a highly interesting platform for condensed matter research\,\cite{Jin2018,Regan2022}. 
They offer novel pathways to design electronic states by tuning local atomic registries through angular and lattice mismatch\,\cite{Yu2017,Wu2017,Zhang2018,Jin2019,Seyler2019,Alexeev2019,Tran2019}, revealing a plethora of many-body phenomena\,\cite{Xu2020,Regan2020,Tang2020}. 
Importantly, electron-hole excitations in these systems are governed by strong Coulomb interaction, forming tightly-bound excitons\,\cite{Jin2018,Merkl2019}.
They often involve spatially-separated electron and hole constituents, conceptually similar to the excitons in coupled quantum wells\,\cite{Butov1994,Rapaport2004}.
These interlayer excitons (IXs) can be long-lived\,\cite{Rivera2015}, manipulated and guided by external fields\,\cite{Unuchek2018}, exhibit long valley-polarization lifetimes\,\cite{Rivera2016}, and demonstrate correlated phenomena\,\cite{Wang2019,Tang2020}.
They emerged as the main carriers of energy and quantum information in van der Waals heterostructures.
Naturally, the question of how the excitons propagate attracted an increasing amount of attention, bridging the realms of optics and transport\,\cite{Rivera2016,Calman2018,Jauregui2019,Yuan2020,Choi2020,Li2021,Wang2021} with promising pathways towards excitonic devices\,\cite{Unuchek2018,Sun2022}.

Despite recent progress, however, it remains very challenging to disentangle intrinsic and extrinsic phenomena associated with van der Waals heterostructures that determine exciton transport.
Primarily, these include trapping due to disorder from strain and dielectric effects\,\cite{Raja2019} or within the moiré-induced potentials\,\cite{Yu2017,Wu2017,Zhang2018,Jin2019,Seyler2019,Alexeev2019,Tran2019}.
Both induce localization of excitons, evidenced by vanishing diffusion coefficients, observed at low temperatures\,\cite{Li2021,Wang2021}.
Exciton-exciton interactions further lead to strongly non-linear effects, typically considered to stem from dipolar repulsion in analogy to coupled quantum wells\,\cite{Rapaport2004,Voros2005,Stern2008,Yuan2020,Wang2021,Sun2022}.
In contrast, exciton-exciton annihilation often determines effective diffusion in the monolayer constituents\,\cite{Kulig2018,Goodman2020}.
Finally, the Mott transition from excitons to plasma already occurs at excitation densities of a few 10$^{12}$\,cm$^{-2}$\,\cite{Wang2019}, leading to observations of rapid effective diffusion\,\cite{Wang2021, Choi2023}.
To understand exciton propagation in van der Waals heterobilayers, the following questions thus need to be resolved:
What are the pristine properties of propagating interlayer excitons \textit{in the absence} of disorder and moiré potentials?
Which non-linear effects are important and how can they be \textit{distinguished}? 
Is there any \emph{qualitative} impact of excitons and plasma on the propagation dynamics at the Mott transition? 
\begin{figure*}[ht]
	\centering
			\includegraphics[width=15 cm]{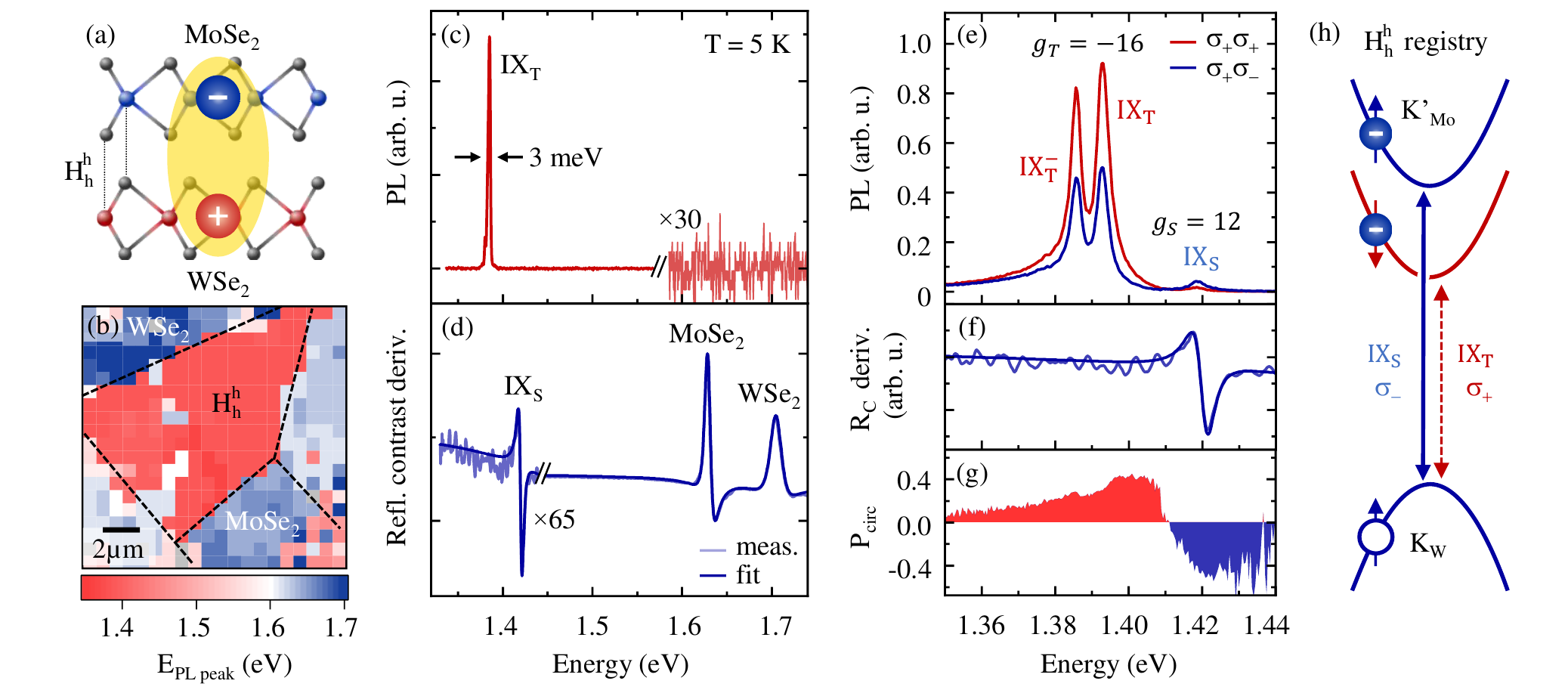}
		\caption{(a) Schematic illustration of an interlayer exciton in a H$^h_h$ reconstructed MoSe$_2$/WSe$_2$ heterobilayer. 
		(b) PL peak energy map at T = 5 K. 
		(c) Representative PL spectrum under low-density pumping. 
		(d) Reflectance contrast derivative with fit from transfer matrix analysis.
		(e) Circularly polarized PL after $\sigma_+$-polarized, pulsed excitation with $g$-factors from magneto-optical measurements.
		(f) Reflectance contrast derivative, R$_C$, and (g) degree of circular polarization, P$_{circ}$ = (PL$_{+}-$PL$_{-}$)/(PL$_{+}$+PL$_{-}$). 
		(h) Optical selection rules for H$^h_h$-stacked MoSe$_2$/WSe$_2$ heterobilayer (adapted from Ref.\,\cite{Forg2019}).
		}
	\label{fig1}
\end{figure*} 

Here, we address these questions by focusing on interlayer excitons in a prototypical heterobilayer system, MoSe$_2$/WSe$_2$, schematically illustrated in Fig\,\ref{fig1}.
Taking advantage of the recently demonstrated conditions for atomic reconstruction\,\cite{Weston2020,Rosenberger2020,Zhao2023}, we realize the scenario of sufficiently large, moiré-free domains with fixed interlayer registry.
The encapsulation in high-quality hBN suppresses extrinsic sources of disorder.
Employing time-resolved optical microscopy, we find that interlayer excitons are highly mobile even at low densities and cryogenic temperatures.
The measured diffusion coefficients are almost 1000 times higher than previously reported at comparable conditions in moiré systems\,\cite{Wang2021, Li2021}.
The absence of trapping is strongly supported by a characteristic temperature dependence for free propagation.
Further, we disentangle two main sources of non-linear diffusion, demonstrating nearly equal contributions from exciton-exciton repulsion and annihilation.
Finally, we show that propagation dynamics of optically induced excitations undergo a substantial change above the Mott transition.
Most strikingly, it leads to the appearance of \textit{negative} effective diffusion, indicating transient crossing between plasma- and exciton-dominated regimes.

The hBN-encapsulated MoSe$_2$/WSe$_2$ heterostructures were fabricated by mechanical exfoliation and dry visco-elastic transfer\,\cite{Castellanos-Gomez2014a}.
Matching angular alignment was confirmed by second-harmonic generation measurements, indicating H-type stacking; see Supplemental Material (SM).
The samples were placed in a liquid-helium microscopy-cryostat and closed-cycle magneto-cryostat for transient diffusion and $g$-factor measurements, respectively.
For photoluminescence (PL) studies we employed both continuous wave excitation (2.33\,eV photon energy) and Ti:Sapphire laser (80\,MHz repetition rate, 140\,fs pulses), tuned to the photon energy of the MoSe$_2$ A-exciton of 1.63\,eV.
Reflectance measurements were performed using a tungsten halogen lamp.
The signals were resolved spectrally and spatially by a grating and a mirror, respectively.
They were detected either by a charged-coupled device or by a streak camera.
Detailed description of the setup is outlined in Refs.\,\cite{Kulig2018,Wagner2020} and SM.

As illustrated in Fig.\,\ref{fig1}\,(b), the PL of the heterostructure region appears at photon energies around 1.4\,eV at T\,=\,5\,K, as expected for IXs in MoSe$_2$/WSe$_2$\,\cite{Rivera2015}.
A typical spectrum at low-density, continuous-wave excitation is dominated by a single peak with 3\,meV linewidth (Fig.\,\ref{fig1}\,(c)).
No emission from the intra-layer excitons is detected.
The IX PL is accompanied by a sharp resonance in the reflectance contrast derivative spectrum, presented in Fig.\,\ref{fig1}\,(c).
Its oscillator strength is determined to be about 2\% of the intra-layer excitons, consistent with Refs.\,\cite{Forg2019,Barre2022}.
This implies good interlayer contact for sufficient electron-hole wavefunction overlap and the presence of atomic reconstruction.
Most importantly, the reflectance contrast spectrum is a superposition of the individual MoSe$_2$ and WSe$_2$ monolayer spectra, slightly shifted and broadened.
No resonance splitting and redistribution of the oscillator strengths are observed, demonstrating the absence of moiré potentials\,\cite{Jin2019}. 
Circularly-polarized PL indicates emission from spin-singlets IX$_S$ and spin-triplets IX$_T$ with high and low oscillator strengths, respectively, presented in Figs.\,\ref{fig1}\,(d)-(f).
They correspond to the K$_{W}$-K$'_{Mo}$ electronic transitions, consistent with the optical selection rules of H$^h_h$ registry, illustrated in Fig.\,\ref{fig1}\,(g)\,\cite{Forg2019}.
The lowest energy peak stems from IX$_T^-$ trions \cite{Jauregui2019,Brotons-Gisbert2021, Liu2021, Wang2021a} due to weak non-intentional doping below 10$^{11}$\,cm$^{-2}$, as observed in gate-dependent measurements (see SM).
Finally, both the absolute peak energies and the $g$-factors obtained from magneto-PL of $g_S = 12$ and $g_T = -16$ strongly support the K-K$'$ transitions and H$^h_h$ registry assignment\,\cite{Zhao2023, FariaJunior2023}.
\begin{figure}[t]
	\centering
			\includegraphics[width=8.5 cm]{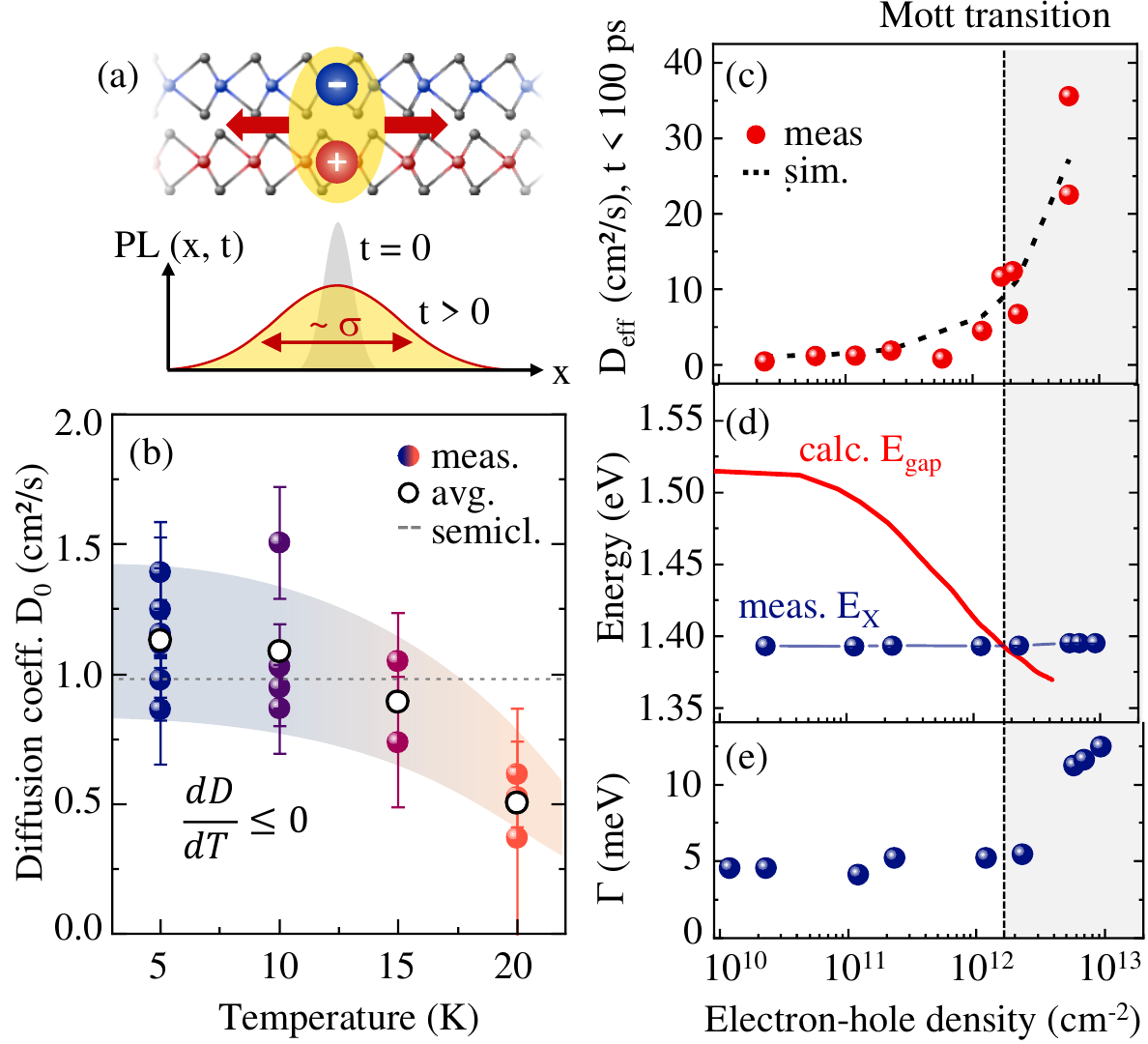}
		\caption{(a) Schematic illustration of transient microscopy.
		(b) Temperature-dependent, linear diffusion coefficient $D_0$ at injection density of 1.2$\times10^{11}$\,cm$^{-2}$ electron-hole pairs per pulse (energy density: 0.5\,$\mu$J\,cm$^{-2}$).
		(c) Effective diffusion coefficients extracted during the first 100\,ps after the excitation. 
		(d) Calculated free particle bandgap, renormalized due to the presence of free charge carriers, and measured exciton peak energy. 
		(e) Lineshape broadening at the Mott transition.
		}
	\label{fig2}
\end{figure} 

To demonstrate that IXs are free to move in the absence of moiré localization and disorder across a spatially-extended, reconstructed domain we perform time- and spatially-resolved exciton diffusion experiments, Fig.\,\ref{fig2}\,(a), at low pump power and as a function of temperature.
Measured, linear diffusion coefficients $D_0$, extracted within the IX lifetime below 1\,ns (see SM), are presented in Fig.\,\ref{fig2}\,(b).
Notably, the average value of $1.1\pm0.2$\,cm$^2$/s at T\,=\,5\,K is about 1000 times higher than previously found for moiré potentials at low densities\,\cite{Wang2021, Li2021}.
The diffusion matches the semi-classical expectation of $D_0 = 0.98$\,cm$^2$/s using the scattering time obtained from the spectral broadening (see SM) and it decreases with temperature.
The physics of IX propagation in the heterostructure thus closely resemble those of WSe$_2$ monolayers\,\cite{Wagner2021}, with the 5\,K diffusion being determined by the exciton-phonon interaction. 
Rapid diffusion of free IXs further rationalizes comparatively short lifetimes due to associated increase of the non-radiative capture probability\,\cite{Zipfel2020}.

Non-linear propagation is presented in Fig.\,\ref{fig2}\,(c) as a function of optically injected electron-hole pair density $n_{eh}$.
The effective diffusion coefficient, evaluated during the first 100\,ps after the excitation, increases above 30\,cm$^2$/s at $n_{eh}=6\times10^{12}$\,cm$^{-2}$.
These densities cover the Mott transition from excitons to electron-hole plasma, typically found in MoSe$_2$/WSe$_2$ heterostructures in this range\,\cite{Wang2019}.
Using the theoretical approach from Refs.\,\cite{Steinhoff2017,Wang2019}, we determine the Mott threshold by calculating the renormalization of the quasiparticle bandgap as a function of electron-hole density (see SM) for the studied H$^h_h$ registry.
The assumptions of the model based on free carriers should be applicable close to the Mott transition.
The absolute energy of the bandgap is fixed for the lowest densities to the sum of the exciton peak energy and the exciton binding energy. 
The latter was set to 130\,meV in heterobilayers according to both calculated value and literature\,\cite{Merkl2019}.
The Mott threshold is estimated from the crossing of the calculated bandgap and measured exciton peak energies, see Fig.\,\ref{fig2}\,(d).
The obtained value of $2\times10^{12}$\,cm$^{-2}$ is consistent with literature\,\cite{Wang2019} and further supported by the onset of spectral broadening (Fig.\,\ref{fig2}\,(e)).

\begin{figure*}[ht]
	\centering
			\includegraphics[width=15 cm]{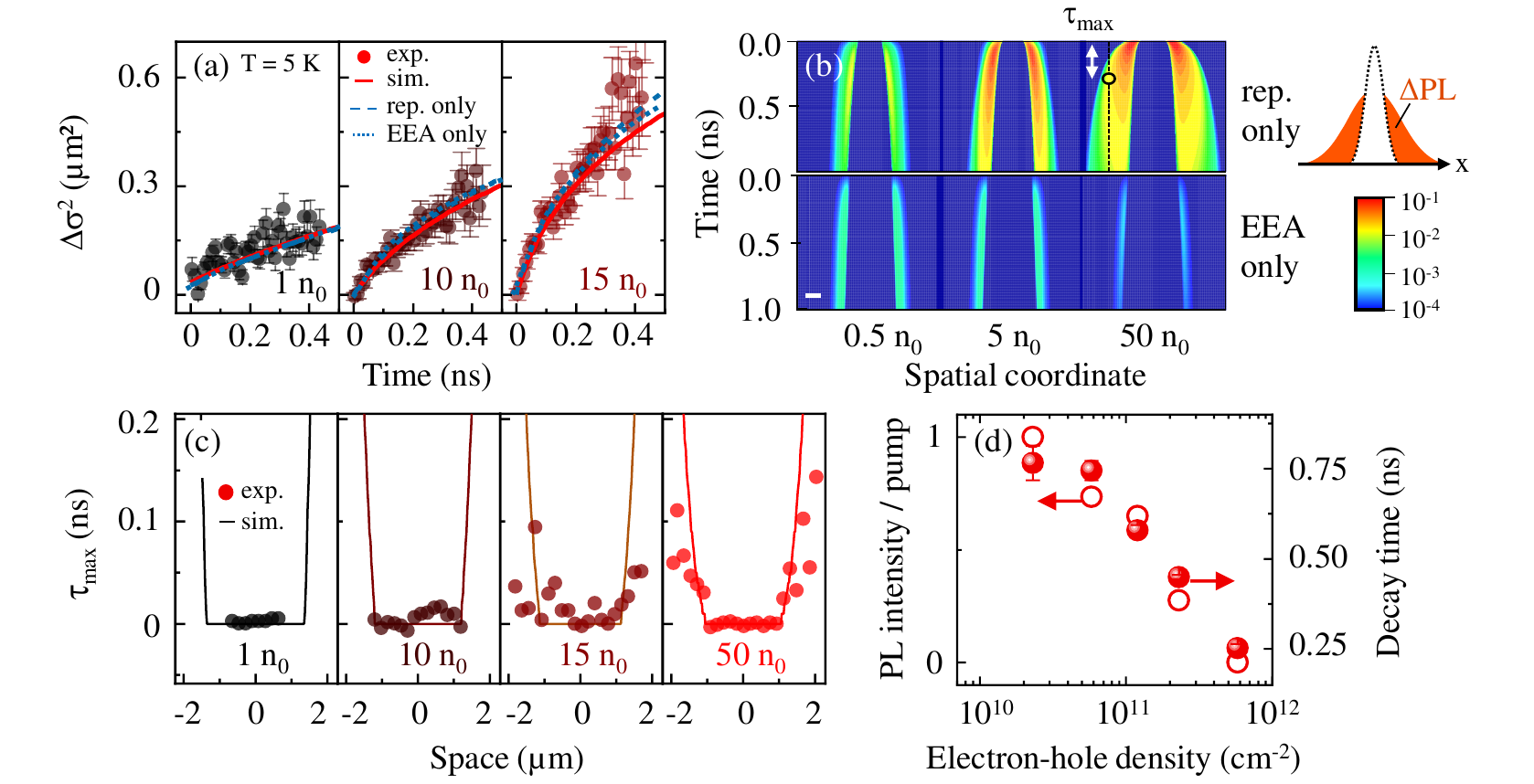}
		\caption{(a) Mean squared displacement of IX for selected densities in units of $n_0=1.2\times10^{11}$\,cm$^{-2}$.
	 Lines represent simulation results using Eq.\,\eqref{diffusion} due to annihilation only (dotted), repulsion only (dashed) and their combination (solid).
	(b) Simulated relative increase of $\Delta$\,PL$ = [$PL$(t)-$PL$(0)]/$PL$_{\rm max}$; scale bar is 1\,$\mu$m. 
	(c) Spatially-dependent time $\tau_{\rm max}$ for the PL to reach its maximum, extracted from measurements and simulation using best fit $U_0 = 1.7\times10^{-15}$\,eV\,cm$^{2}$. 
	(d) Measured density-dependence of the PL decay time and relative yield, corresponding to the exciton-exciton annihilation coefficient $R_A=5\times10^{-3}$\,cm$^{2}$/s.
		}
	\label{fig3}
\end{figure*} 
The origin of the non-linear increase of the effective diffusion coefficient is commonly attributed to exciton-exciton \textit{annihilation} (EEA) in monolayers\,\cite{Kulig2018,Goodman2020} and exciton-exciton \textit{repulsion} in heterostructures\,\cite{Yuan2020,Wang2021} and coupled quantum wells\,\cite{Rapaport2006}.
These two processes enter the diffusion equation as the last two terms in:
\begin{equation}
\label{diffusion}
\frac{\partial n}{\partial t} = D_0\Delta n - \frac{n}{\tau} - R_A n^2 + \frac{U_0 D_0}{k_B T}\nabla\cdot(n\nabla n),
\end{equation}
where $D_0$ and $\tau$ are the linear diffusion coefficient and exciton lifetime at low-densities, respectively, $R_A$ the annihilation rate coefficient, $U_0$ the interaction constant ($U_0n$ corresponds to the blue shift of the exciton peak), and $k_B$ is the Boltzmann constant.
We solve Eq.\,\eqref{diffusion} numerically by using the initial shape of the excitation spot as a boundary condition, fixing $D_0$ and $\tau$ to the measured values, and varying $R_A$ and $U_0$ parameters (see SM). 
The resulting changes of the mean-squared-displacement, $\Delta\sigma^2$, corresponding to the increase of the PL emission area over time\,\cite{Ginsberg2020}, are presented in Fig.\,\ref{fig3}\,(a) together with the measured data at selected densities.
Interestingly, setting either $R_A$ or $U_0$ strictly to zero demonstrates that neither the effective diffusivity $D_{\rm eff}\propto\partial\sigma^2/\partial t$ nor the $\Delta\sigma^2(t)$ traces identify the origin of the non-linear propagation.

To distinguish annihilation and repulsion, we consider relative increase of the spatially- and time-dependent PL instead, presented in Fig.\,\ref{fig3}\,(b) as $\Delta$\,PL$ =[$PL$(t)-$PL$(0)]/$PL$_{\rm max}$ from simulations.
In the case of repulsion, the increase of the PL counts on the flanks of the excitation spot stem from the drift of the excitons towards outer regions. 
For annihilation only, the increase of $\Delta$\,PL stems exclusively from linear diffusion and is very small.
Experimentally, changes in $\Delta$\,PL can be detected by extracting either differential spatial profiles at different times or PL transients at different positions. 
For example, the delay time $\tau_{\rm max}$ of the maximum PL intensity rises with the increasing distance from the excitation spot in case of repulsion, as illustrated in Fig.\,\ref{fig3}\,(b).

The results of this analysis are presented in Fig.\,\ref{fig3}\,(c), demonstrating a characteristic feature of repulsion and determining the interaction constant $U_0 = 1.7\times10^{-15}$\,eV\,cm$^{2}$ (see SM for detailed discussion).
This interaction constant accounts for about one half of the observed increase of the effective diffusivity with excitation density.
It is accompanied by a decrease of the PL lifetime and relative yield (Fig.\,\ref{fig3}\,(c)), that are characteristic features of an efficient annihilation process with $R_A=5\times10^{-3}$\,cm$^{2}$/s.
Consequently, both annihilation and repulsion are found to contribute to the non-linear propagation of IXs in roughly equal measures; the corresponding simulated result is shown in Figs.\,\ref{fig2}\,(b) and \ref{fig3}\,(a).

\begin{figure*}[ht]
	\centering
			\includegraphics[width=15 cm]{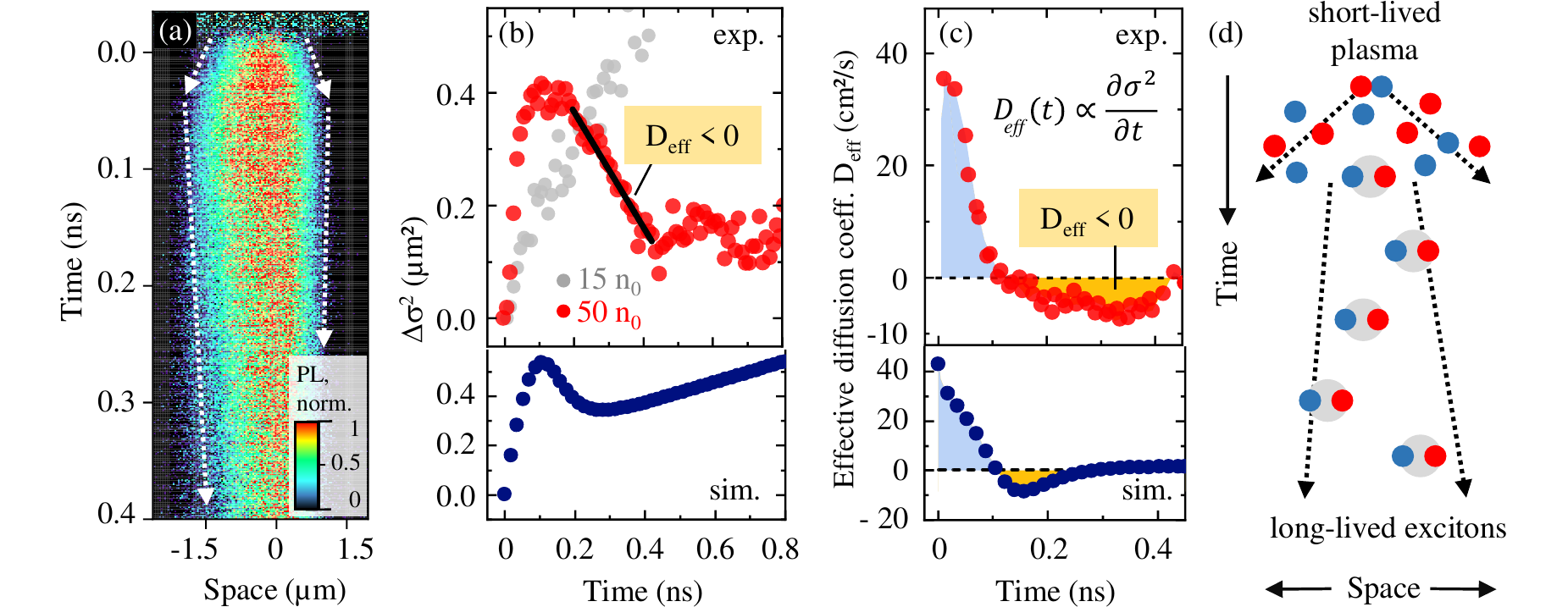}
		\caption{(a) Streak camera image of the spatially- and time-resolved PL at T\,=\,5\,K in the high-density regime of about $6\times10^{12}$\,cm$^{-2}$ (50\,$n_0$, pump energy density: 25\,$\mu$J\,cm$^{-2}$) above the Mott transition; the PL intensity is normalized at each time step. 
		(b) (upper panel) Time-resolved mean squared displacement in experiment comparing high (50\,$n_0$) and intermediate (15\,$n_0$) densities. 
			(lower panel) Simulation results showing the implications of a generic two-component model: rapidly expanding, short-lived plasma and longer-lived, slower excitons. 
			(c) Corresponding time-dependent effective diffusion coefficients extracted from $D_{\rm eff}\propto\partial\sigma^2/\partial t$. 
			(d) Schematic illustration of the two-component model.
		}
	\label{fig4}
\end{figure*} 
In general, regardless of the origin of the density dependent process, one typically finds sub-diffusive behavior of the mean-squared-displacement\,\cite{Kulig2018,Goodman2020,Ginsberg2020,Yuan2020,Wang2021,Choi2023}.
It involves rapid initial diffusion that decreases towards smaller values at later times.
Interestingly, it is not the case for the studied high-density conditions in the vicinity of the Mott transition.
This is demonstrated by the normalized space- and time-resolved PL image in Fig.\,\ref{fig4}\,(a) and the extracted $\Delta\sigma^2$ values in Fig.\,\ref{fig4}\,(b) (upper panel).
The data shows a rapid initial expansion during the first 50\,ps followed by a contraction over several 100's of ps.
This corresponds to an effective diffusion coefficient starting from 35\,cm$^2$/s and then decreasing to a \textit{negative} value of $-5$\,cm$^2$/s, see upper panel of Fig.\,\ref{fig4}\,(c).
At later times, the contraction slows down and the effective diffusion coefficient increases again. 

The effectively negative diffusivity of IXs is a consistent observation in the studied heterobilayer (see SM), albeit a highly unusual one.
This implies a substantial change in the dynamics of the optically exited electron-hole pairs associated with the Mott transition.
While a contraction of a quasiparticle cloud can be related to density-induced many-particle states such as droplet formation (see SM), interplay of multiple types of propagating, transient excitations is often a more likely cause\,\cite{Rosati2020, Ziegler2020,Beret2023}.
At the studied density conditions close to the Mott threshold, a possible scenario would involve rapid propagation of short-lived plasma, followed by a slower diffusion of a small fraction of long-lived excitons.
As schematically shown in Fig.\,\ref{fig4}\,(d), this would effectively lead to a substantial broadening of the electron-hole distribution and a subsequent shrinking after the plasma recombines.

To illustrate the consequence of such a scenario, we use a simplified model with two components that both diffuse according to Eq.\,\eqref{diffusion} (see SM for details).
The resulting, simulated $\Delta\sigma^2$ and extracted, time-dependent diffusivity are presented in the lower panels of Fig.\,\ref{fig4}\,(b) and (c), respectively.
This shows that a two-component diffusion involving plasma and excitons could indeed lead to the observed behavior.
We note, however, that an appropriate microscopic description of these intriguing findings involving a dense many-particle system above Mott threshold would be highly desirable.

In summary, we have shown that interlayer excitons in van der Waals heterostructures diffuse rapidly in the absence of moiré- and disorder-induced localization.
Corresponding low-density diffusion coefficients are about 1000 times higher compared to those found in moiré superlattices, with temperature dependence confirming free diffusion. 
At elevated excitation densities we have demonstrated how the effects of exciton-exciton annihilation and repulsion can be distinguished with both providing substantial contributions to the non-linear propagation. 
Density range of the Mott transition revealed a highly unusual behavior involving rapid initial expansion followed by a long-lived effectively negative diffusion.
Altogether, these findings should allow for disentanglement of complexities associated with dipolar excitons propagation in van der Waals heterostructures in both linear and non-linear regimes, demonstrating a promising platform for controllable exciton transport.
The intriguing dynamics of electron-hole propagation at the exciton-plasma crossover should be of particular interest for a broad community studying physics of interacting electron-hole quasiparticles and correlated many-body states.



%

\section{Acknowledgments}
We thank Sivan Refaely-Abramson, Ronen Rapaport, Andreas Beer, Jonas Zipfel, and Archana Raja for helpful discussions, as well as Imke Gronwald, Christian B\"auml, and Nicola Paradiso for their assistance with pre-patterned substrate preparation. 
Financial support by the DFG via SPP2244 (Project-ID: 443405595), Emmy Noether Initiative (CH 1672/1, Project-ID: 287022282), SFB 1277 (project B05), the Würzburg-Dresden Cluster of Excellence on Complexity and Topology in Quantum Matter (ct.qmat) (EXC 2147, Project-ID 390858490) and the Munich Center for Quantum Science and Technology (MCQST) (EXC 2111, Project-ID 390814868) is gratefully acknowledged. 
A. H. acknowledges funding by the European Research Council (ERC) (Grant Agreement No. 772195).
K.W. and T.T. acknowledge support from the JSPS KAKENHI (Grant Numbers 20H00354, 21H05233 and 23H02052) and World Premier International Research Center Initiative (WPI), MEXT, Japan.
A. S. and M. F. would like to acknowledge resources for computational time at the HLRN (Göttingen/Berlin). M. F. acknowledges support by the Alexander von Humboldt Foundation.

\end{document}








\title{Supplementary material: \\
	Non-linear and negative diffusion of optical excitations in Moiré-free heterobilayers}

\author{Edith Wietek}
\affiliation{Institute of Applied Physics and Würzburg-Dresden Cluster of Excellence ct.qmat, Technische Universität Dresden, 01062 Dresden, Germany}
\author{Matthias Florian}
\affiliation{University of Michigan, Department of Electrical Engineering and Computer Science, Ann Arbor, Michigan 48109, USA}
\author{Jonas M. Göser}
\affiliation{Fakultät für Physik, Munich Quantum Center, and Center for NanoScience (CeNS), Ludwig-Maximilians-Universität München, 80539 München, Germany}
\author{Takashi Taniguchi}
\affiliation{Research Center for Materials Nanoarchitectonics, National Institute for Materials Science,  1-1 Namiki, Tsukuba 305-0044, Japan}
\author{Kenji Watanabe}
\affiliation{Research Center for Electronic and Optical Materials, National Institute for Materials Science, 1-1 Namiki, Tsukuba 305-0044, Japan}
\author{Alexander Högele}
\affiliation{Fakultät für Physik, Munich Quantum Center, and Center for NanoScience (CeNS), Ludwig-Maximilians-Universität München, 80539 München, Germany}
\affiliation{Munich Center for Quantum Science and Technology (MCQST), 80799 München, Germany}
\author{Mikhail M. Glazov}
\affiliation{Ioffe Institute, 194021 Saint Petersburg, Russian Federation}
\author{Alexander Steinhoff}
\affiliation{Institut für Theoretische Physik, Universität Bremen, 28334 Bremen, Germany}
\affiliation{Bremen Center for Computational Materials Science, Universität Bremen, 28334 Bremen, Germany}
\author{Alexey Chernikov}
\email{alexey.chernikov@tu-dresden.de}
\affiliation{Institute of Applied Physics and Würzburg-Dresden Cluster of Excellence ct.qmat, Technische Universität Dresden, 01062 Dresden, Germany}

\maketitle
\newpage
\tableofcontents

\newpage
\section{Sample characterization and peak assignment}

\subsection{Sample fabrication}
The studied MoSe$_2$/WSe$_2$ heterostructures, encapsulated in high-quality, hexagonal boron-nitride (hBN), were fabricated by mechanical exfoliation from bulk crystals and subsequent, all-dry viscoelastic stamping. 
Standard SiO$_2$/Si substrates (oxide layer thickness, 285\,nm) were cleaned using an ultrasonic bath in acetone and isopropanol for two minutes and subsequent blow drying using N$_2$. 
To further remove any adsorbates the substrates underwent an O$_2$ cleaning for 2\,min at 200\,W. \\
\begin{figure}[ht]
	\centering
	\includegraphics[width=\textwidth]{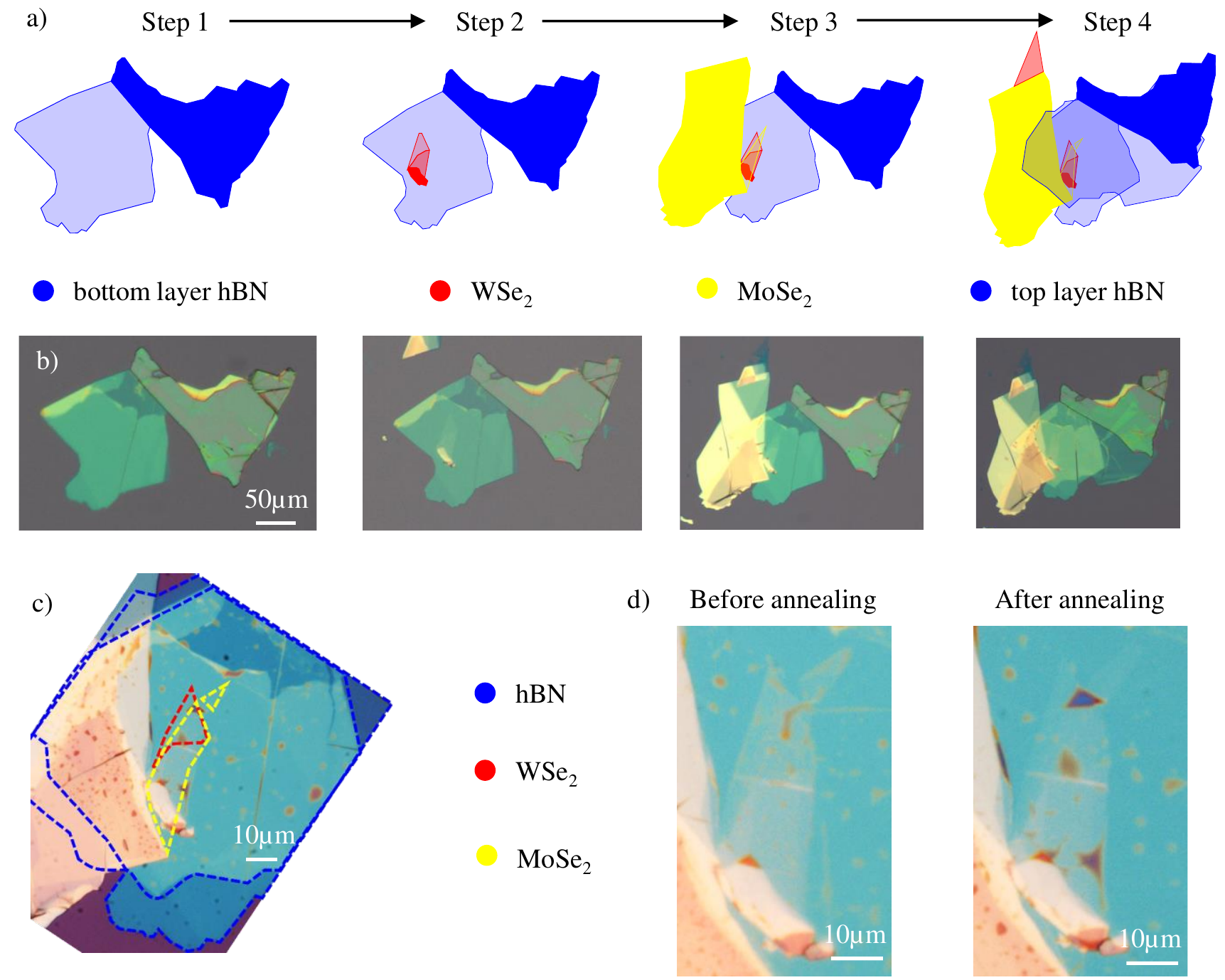}
	\caption{Fabrication process of a MoSe$_2$/WSe$_2$ heterostructure going from left to right. a) An outline model of the transferred hBN, WSe$_2$, MoSe$_2$ flakes is color coded in blue, red and yellow respectively. 
	The opacity of the sketch indicates the thickness of the flakes. 
	The lower panels of b) show the respective micrographs. 
	c) Magnified micrograph of the final heterostructure with outlined flakes. 
	d) Micrographs before and after annealing. The adsorbates become mobile during the annealing process and accumulate in pockets, leaving behind clean areas.}
	\label{fig:samplefabrication}
\end{figure} 

The heterostructures were fabricated as follows, illustrated in \fig{fig:samplefabrication} a) - c): hBN crystals, supplied by our colleagues from the National Institute for Material Science (NIMS, Tsukuba) were cleaved and thinned down using Scotch Magic Tape. 
A tape with sufficiently thin hBN crystals was then placed on polydimethylsiloxane (PDMS; WF-20-X4, Gel-Pak) and suitable, approximately 10\,nm thin hBN flakes selected by their optical contrast. 
These flakes were stamped onto the substrates, preheated to 70\,$^\circ$C at ambient conditions. 
Single layers of both MoSe$_2$ and WSe$_2$ (crystals acquired from HQgraphene) were fabricated in a similar method using blue tape (polyvinyl chloride tape 223PR, Nitto). 
Single layers with clean edges were chosen in order to be aligned along the crystallographic axis. 
First the WSe$_2$ flake was stamped on top of the previously deposited hBN, followed by the MoSe$_2$ monolayer, rotated to align the two crystals along one of the axis.
An hBN layer was then placed on top.
The stack was annealed after every stamping process at 10\,mbar and 150\,$^\circ$C for 3 to 4 hours. 
This annealing process typically ensures good interlayer contact by making the adsorbates mobile and being able to accumulate in pockets leaving behind clean areas, see \fig{fig:samplefabrication} d) \cite{Jain2018}.

\section{Stacking nomenclature}

\begin{figure}[ht]
	\centering
	\includegraphics[width=0.65\textwidth]{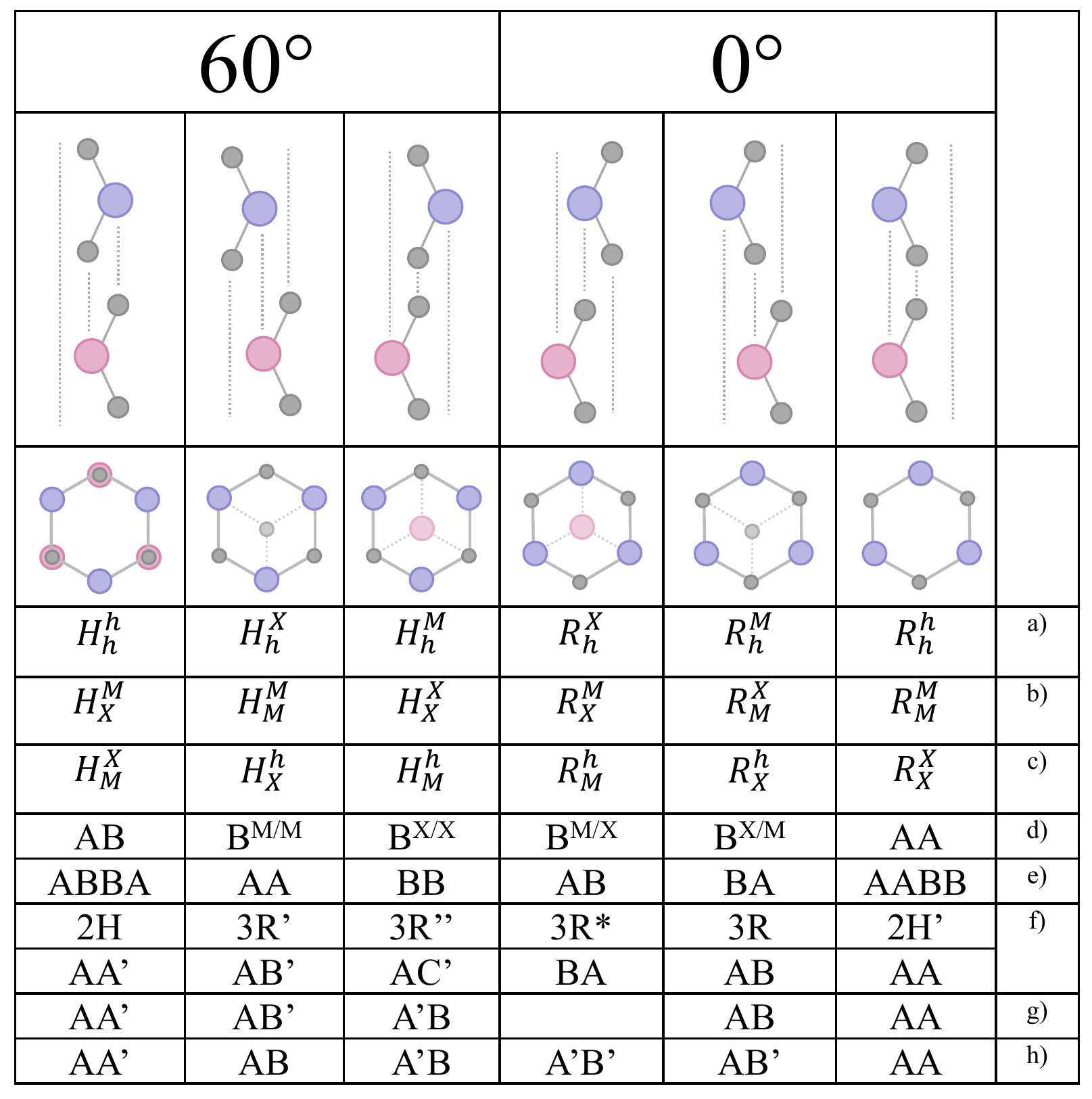}
	\caption{Overview of common nomenclatures for high symmetry points in TMDC heterobilayers used in the recent literature: a) \cite{Zhao2023, Yu2017, Tran2019a}, b) \cite{Tong2017}, c) variation of a) following the same principle, d) \cite{Naik2019}, e) \cite{phillips2019, Rosenberger2020}, f) \cite{Zhao2018}, g) \cite{Hu2016}, and h) \cite{Foerg2019}.}
	\label{fig:Nomenclature}
\end{figure}

There are different types of nomenclature currently used for high points of symmetry in TMDC heterobilayers.  
Here, we present an overview of the most common descriptions of different stacking orders in both 60$^\circ$ and 0$^\circ$, as summarized in the table of \fig{fig:Nomenclature}. Schematic illustrations of the crystal orientation from the side view and top view are shown in row one and two respectively.
Different transition metal atomic species are indicated by either red or pink color, the chalcogens are shaded gray.

In this work we use the nomenclature illustrated in row a). 
Hereby the 60$^\circ$ twist angle between the layers is indicated by the \textit{H} and 0$^\circ$ stacking by \textit{R}. 
This nomenclature is based on the hexagonal stacking phase \textit{2H} abundant in natural TMDC bilayers, where the layers are twisted 180$^\circ$ with respect to each other (multiple of 60$^\circ$) and is therefore centrosymmetric. 
The 0$^\circ$ stacking corresponds to the \textit{3R} natural stacking pattern on TMDCs, where no centrosymmetric structure is present in bilayer form \cite{Suzuki2014}.
Both are stable configurations possible for TMDCs. 
The notations in rows a) - c) all follow the sample principle, by only varying the initial position in the unit cell from which the other points are described. 
These notations describe visually which element of the upper TMDC layer is placed above which element of the lower layer. 
The superscript character gives reference to the upper layer and the subscript to the lower layer, respectively. 
In the case of TMDC bilayers, $M$ stands for \textit{transition metal}, $X$ for \textit{chalcogen} and $h$ for \textit{center of hexagon} (or ``\textit{hole}''). 
Using the stacking configuration studied in this work as an example, the structure is described as follows: $H_h^h$, stacking of 60$^\circ$ from $H$, and the center of hexagon from the top layer $\bhexagon^h$ is above the center of hexagon of the bottom layer $\bhexagon_h$, ($H + \bhexagon^h + \bhexagon_h = H_h^h$). 
In a similar manner, this type of stacking can be described as a 60$^\circ$ where the transition metal above the chalcogen ($H^M_X$, \fig{fig:Nomenclature} b)) or the chalcogen is above the transition metal ($H^X_M$, \fig{fig:Nomenclature} c)), respectively.

There are also alternative notations to describe stacking domains.
Row d) of \fig{fig:Nomenclature} follows the nomenclature of bilayer graphene, where \textit{B} refers to the Bernal stacking (\,=\,half of the atoms from one layer are located over the center of the hexagon from the other layer). 
Analogous to rows a-c) the superscript notation of $B^{\bhexagon}$ describes the stacking nature of the remaining elements.
For example, $B^{M/M}$ is a Bernal stacked heterobilayer with transition metal of the top layer lies above the transition metal of the bottom layer. 
In the absence of Bernal stacking, for 0$^\circ$ twist angle the bilayers are denoted as $AA$ stacking and $AB$ stacked for 60$^\circ$.
In the notation presented in row e) $A$ and $B$ are used to indicate transition metal and chalcogen respectively.
Each couple of letters describes again which element of the top layer is placed above which element from the bottom layer when read from left to right. 
Following this scheme, the $H_h^h$ registry can also be described as transition metal (A) top laying over chalcogen (B) of bottom  and chalcogen (B) of top laying over transition metal (A) of the bottom layer ($AB+BA=ABBA$).
Variations of the natural stacking pattern are used in the notation in the top row of f). 
\textit{2H} and \textit{3R} indicate the stacking phases as found in nature, while the following superscripts indicate what modification was applied to obtain the remaining registries: \textit{'} indicates the upper layer being flipped horizontally, \textit{''} the bottom layer being flipped horizontally, and \textit{*} a flip in the vertical direction. 
Bottom row of f), g) and h) follow this superscript notation, but refer to a horizontal shift of a layer as changing it from notation A to B (A and B are different as in there is no mirror-reflection axis).

\subsection{Twist angle determination through second harmonic generation (SHG)}
The twist angle of the heterobilayer was determined using polarization-resolved second-harmonic-generation (SHG) microscopy.
Linearly polarized laser light is focused to a spot diameter of 1\,$\mu$m and passes a superachromatic $\lambda/2$\,--\,waveplate (wavelength range of 310\,nm -- 1100\,nm with less than 1.3\,\% deviation of retardation) mounted on a continuous stepper motor rotation stage. 
The laser is a 80\,MHz Ti:sapphire, tuned to a wavelength of 800\,nm and a power of 5\,mW,  giving rise to a SHG signal at 400\,nm. The back scattered second harmonic signal passes through the analyzing linear polarizer, which is parallel to the polarization of the initial incident laser beam.
High linear polarization was achieved with a Glan-Taylor calcite polarizer, acting both as input and analyzing polarizer.  \\

TMDCs belong to the $D_{3h}$ point group, showing three-fold rotational symmetry. Therefore the parallel and perpendicular polarization components of the corresponding second-harmonic response follow a six-fold symmetry. 
The intensity of the SHG signal shows an intensity maximum when the direction of the polarization axis is parallel (perpendicular) with the armchair (zigzag) direction of the crystal flake.
Rotating the waveplate and therefore rotating the angle of the incident linear polarization leads to a variation in the intensity of the parallel signal component $I_{\parallel} $ which follows the relation $I_{\parallel} = I_0 \cos^2(3\theta)$. Here, $I_0$ is the intensity and $\theta$ the polarization angle (relative to the initial setting of the waveplate). 
The difference in angle for the maximum intensities of two flakes then corresponds to the twist angle $\theta_t$ \cite{Hsu2014, Paradisanos2022, Li2013, Kim2020}.

\begin{figure}[ht]
	\centering
	\includegraphics[width=\textwidth]{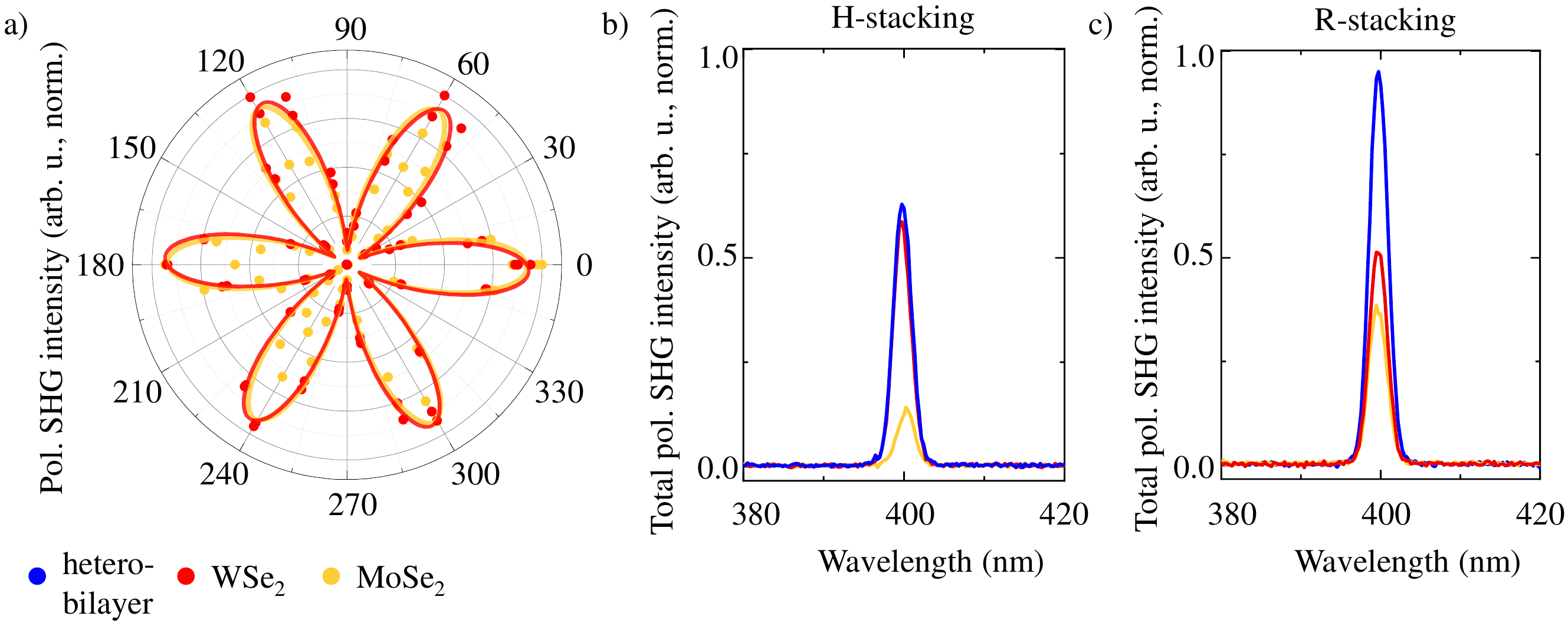}
	\caption{a) SHG intensity as a function of rotational angle. 
	The yellow and red curves correspond to the monolayer flakes of MoSe$_2$ and WSe$_2$, respectively. 
	The twist angle between the flakes is $\theta_t = 1.3^{\circ} \pm 0.6^{\circ}$. 
	b) SHG spectra for MoSe$_2$ and WSe$_2$ monolayers together with that of the stacked heterobilayer region for H-stacking 
	c) Same plot for R-stacking, showing the difference in decrease and enhancement of SHG intensity for the R-stacked bilayer region.}
	\label{fig:SHG}
\end{figure} 

The resulting SHG intensity is plotted in \fig{fig:SHG}\,a) as a function of rotational angle between sample axis and linear polarization axis of the incident laser.
The data is plotted for MoSe$_2$ and WSe$_2$ monolayers in yellow and red curves, respectively. 
The resulting twist angle between the two flakes is $\theta_t = 1.3^{\circ} \pm 0.6^{\circ} $.
To assign the angle to either R- or H-type stacking, we consider the intensity changes of the SHG signal in the heterobilayer region with respect to its monolayer constituents. 
In one of the heterostructures, the SHG signal corresponds to that of the highest monolayer component, see \fig{fig:SHG} b).
In the second heterostructure, it is substantially enhanced, see \fig{fig:SHG} c), allowing us to assign it to R-stacking (0$^{\circ}$ twisted) due to the lack of inversion symmetry in the positions of the atoms in contrast to H-stacking, cf.~\cite{VanderZande2014}. 
For the sample studied in the diffusion measurements (\fig{fig:SHG} b) this already indicates H-type stacking (0$^{\circ}$ twisted) due to the lack of SHG enhancement.

\subsection{Gate-dependent photoluminescence (PL)}\label{sec:GatePL}
In order to assign the interlayer exciton species found in PL spectroscopy with respect to neutral excitons and those dressed by free charges, a gate-tunable heterostructure was fabricated. Reconstructed areas of H$^h_h$ domains were verified  by characteristic features in PL, as elaborated in the main text and \cite{Zhao2023}.
The experiments were performed at a nominal temperature of 5\,K, under 532\,nm cw-excitation with a laser spot of 1\,$\mu$m in diameter and power of 10\,$\mu$W. 
\begin{figure}[ht]
	\centering
	\includegraphics[width=\textwidth]{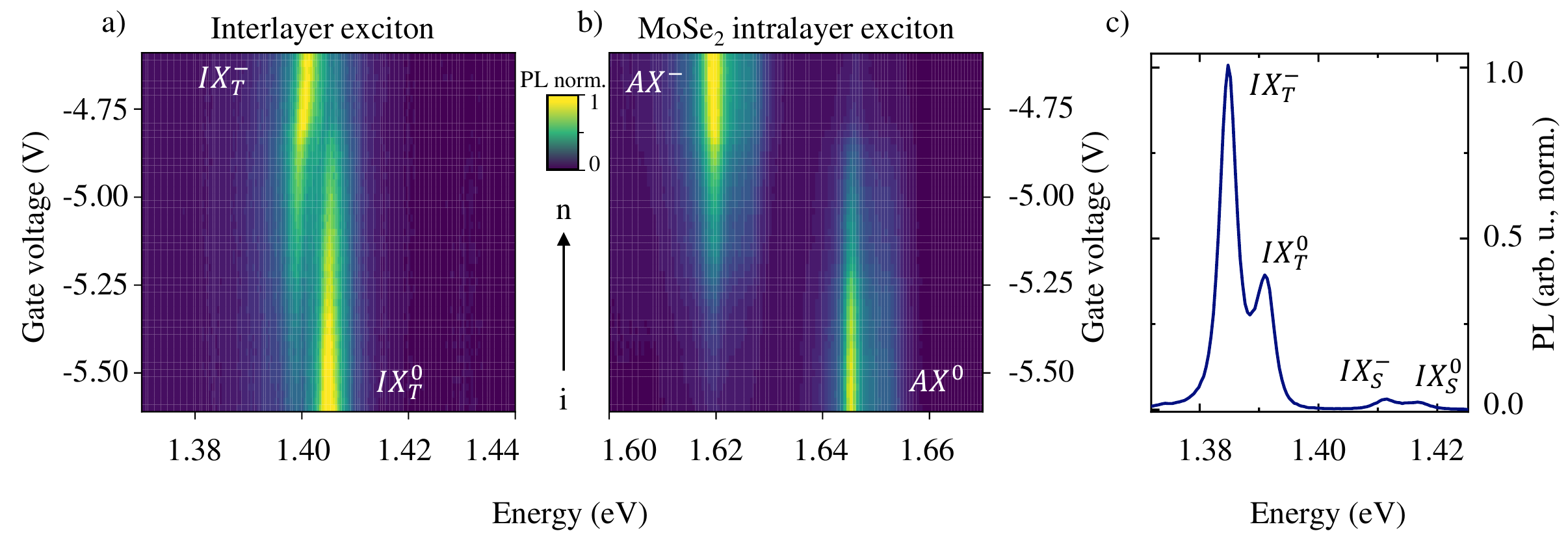}
	\caption{Gate dependent PL of heterostructure. Injection of free charge carriers through voltage. Simultaneous measurement of charge carrier influence from both bilayer and monolayer region. Corresponding behavior of both systems by entering n-doped regime. Emergence of the negatively charged exciton (trion) and diminishing of the neutral exciton. Significant difference in binding energy. 6\,meV for interlayer and 25\,meV for intralayer. c) Time integrated spectra at moderate doping showing trion of both interlayer triplet and singlet state. Both emerging with 6\,meV binding energy. }
	\label{fig:GatePL}
\end{figure} 
Figures \ref{fig:GatePL} a) and b) show PL spectra as function of photon energy and gate-voltage for the interlayer exciton in MoSe$_2$/WSe$_2$ heterobilayer and intralayer exciton in MoSe$_2$ monolayer, respectively.
As the voltage is tuned from charge neutrality to negatively doped regime, the PL from neutral excitons decreases and that from the trions (attractive Fermi polarons) emerges.
The trion binding energy is extracted from the energy separation of the two peaks at vanishing carrier densities, corresponding to the lowest gate voltages in Figs.\ref{fig:GatePL}\,a) and b).
It is determined as 6\,meV for the MoSe$_2$/WSe$_2$ heterobilayer and 25\,meV for the MoSe$_2$ monolayer; the latter matches previously measured values in hBN-encapsulated MoSe$_2$ monolayers\,\cite{Zipfel2022} and literature results \cite{Singh2014, Ross2013}.
The slight blueshift of the interlayer trion as seen in \fig{fig:GatePL} a) is partially attributed to the Stark effect, as we used only one gate in this sample. 
Interlayer trions are also observed for both triplet and singlet states, as shown in the exemplary spectrum at higher doping \fig{fig:GatePL} c). \\

\begin{figure}[ht]
	\centering
	\includegraphics[width=\textwidth]{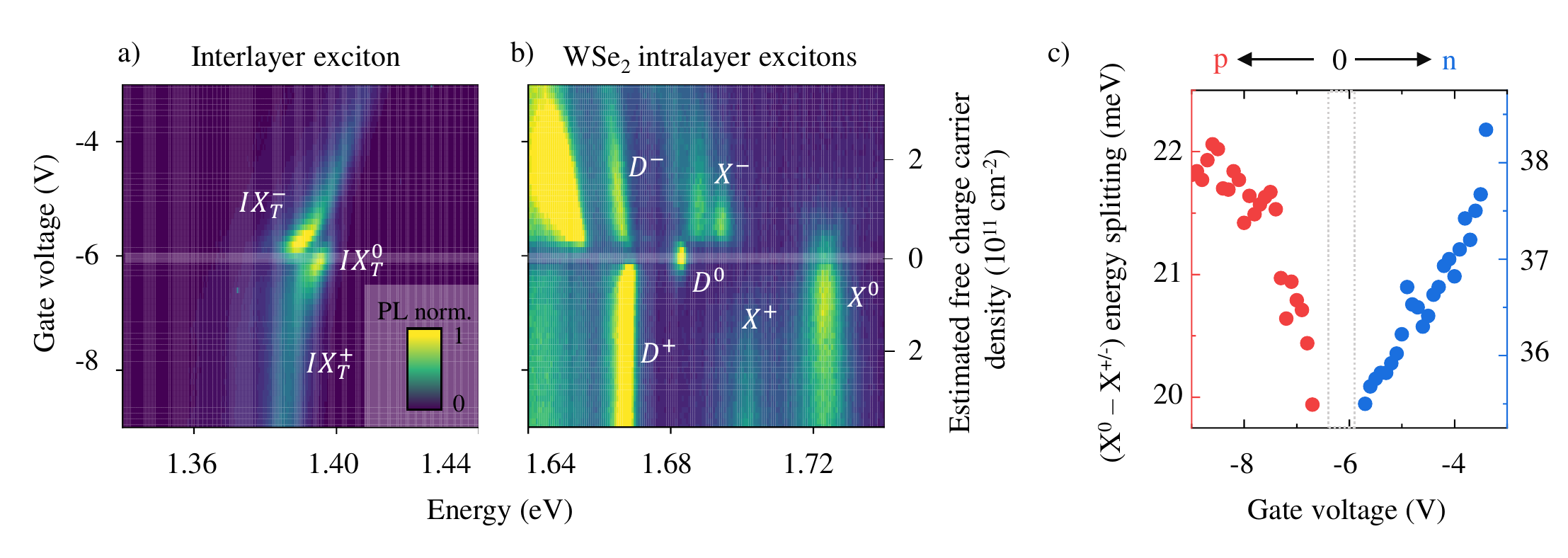}
	\caption{a) PL of interlayer excitons at T = 5\,K as a function of photon energy and gate voltage for MoSe$_2$/WSe$2$ heterostructure and b) simultaneously measured  WSe$_2$ monolayer region. Transparently shaded area indicates the charge neutrality regime. 1\,V corresponds to the change of free charge carriers density of $1.1 \times 10^{11}$\, cm$^{-2}$ for the n-doped and $2.6 \times 10^{11}$\, cm$^{-2}$  (for $-6.7$\,V to $-7.5$\,V) and $6.2 \times 10^{10}$\, cm$^{-2}$ (starting from $-7.5$\,V)  the for the p-doped regime. c) Energy separation of neutral exciton ($X^0$) and charged exciton ($X^+$ and low energy $X^-$) of WSe$_2$ which was used for the estimation of the free charge carrier doping densities.}
	\label{fig:GatePLDensity}
\end{figure} 

For further identification the same kind of gate-dependent measurement on the heterostructure was performed, while simultaneously detecting also PL from the WSe$_2$ monolayer region where the distinctly different PL signal for negatively and positively doped region also allows for clear assignment of negatively and positively charged exciton respectively, see \fig{fig:GatePLDensity} a) and b). Analogously to \fig{fig:GatePL} a) by tuning the sample into the n-doped regime, the negatively charged interlayer exciton ($IX^-_T$) emerges, while now also doping into the p-doped regime gives rise to the positively charged interlayer exciton ($IX^+_T$). Simultaneously on the WSe$_2$ monolayer tuning into the n- and p-doped regime the characteristic excitonic states of neutral exciton ($X^0$), positive trion ($X^+$), intravalley and intervalley trion ($X^-$), spin forbidden neutral dark exciton $D^0$, positive dark trion ($D^+$) and negative dark trion  ($D^-$) as marked in \fig{fig:GatePLDensity} b) \cite{Jones2013, Liu2019b, Zhang2015a, He2020}. The charge neutrality regime is determined by the appearance of the neutral spin dark exciton $D^0$ and absence of charged excitonic states. The zero-doping regime is denoted by transparent shading in \fig{fig:GatePLDensity} b) and extended to \fig{fig:GatePLDensity} a), indicating the neutrality regime of the heterostructure with the neutral interlayer exciton $IX_T^0$. As the applied gate voltage covers a much wider range than in \fig{fig:GatePL} the induced Stark effect in the heterostructure is much more pronounced, shifting the charged interlayer exciton energies additionally into both blue and red for n- and p-doped regime respectively.  For the WSe$_2$ monolayer we extract a trion binding energy $E_{B, tr}$ of 19.4\,meV for $X^+$, and 35.2\,meV $/$ 29.2\,meV for the $X^-$ doublet, as given from the energy splitting of neutral and charged excitons $(X^0 - X^{+/-}$), see \fig{fig:GatePLDensity} c). The free charge carrier density is estimated by the relative energy separation $\Delta_{XT} = (X^0 - X^{+/-}) - E_{B, tr}$ between exciton and trion which is set equal to the Fermi energy \cite{Mak2012}. Deviations in actual charge carrier density of up to 50\,\% may arise, depending on the trion model ($\Delta_{XT}\approx 2/3 E_F$) \cite{Esser2001} or Fermi polaron model ($\Delta_{XT}\approx 3/2 E_F$) \cite{Efimkin2021}. We thus estimate the charge carrier density as
\begin{equation}
	n_{e,h} = \Delta_{XT} \times \dfrac{m_{e,h}}{\pi \hbar^2}
\end{equation}
using $m_e=0.36m_0$ and $m_h=0.4m_0$ \cite{Kormanyos2014} and therefore obtain a scaling of $1.1 \times 10^{11}$\, cm$^{-2}$/V for the n-doped and $2.6 \times 10^{11}$\, cm$^{-2}$/V  (for $-6.7$\,V to $-7.5$\,V) and $6.2 \times 10^{10}$\, cm$^{-2}$/V (starting from $-7.5$\,V) for the p-doped regime. Assuming the injected charge carrier density is mostly equal in the heterobilayer region we estimate an upper limit of the residual doping in the sample used for diffusion measurements of $10^{11}$\,cm$^{-2}$.

\subsection{Lande $g$-factors from magneto-PL}
\label{gfactor}

To support the assignment of the optical resonances to specific electronic transitions, $g$-factors were measured using magneto-optical photoluminescence spectroscopy. 
For these measurements, we utilized a closed-cycle magneto-cryostat (attoDRY1000) at a sample base temperature of 3.8\,K under pi-polarized confocal laser excitation at 2.33\,eV with 300\,$\mu$W excitation power and external magnetic fields up to 8\,T. 
To obtain the data presented in the main text, the measurements were performed on the same position (or in close proximity thereof) within the area of 2H stacking, see Fig. 1b of the main text. 
Due to the multiple cool-down and sample-preparation cycles a change of internal doping resulted in a stronger PL signal from the interlayer trions, allowing also for determination of their $g$-factors.\\

\begin{figure}[ht]
	\centering
	\includegraphics[width=\textwidth]{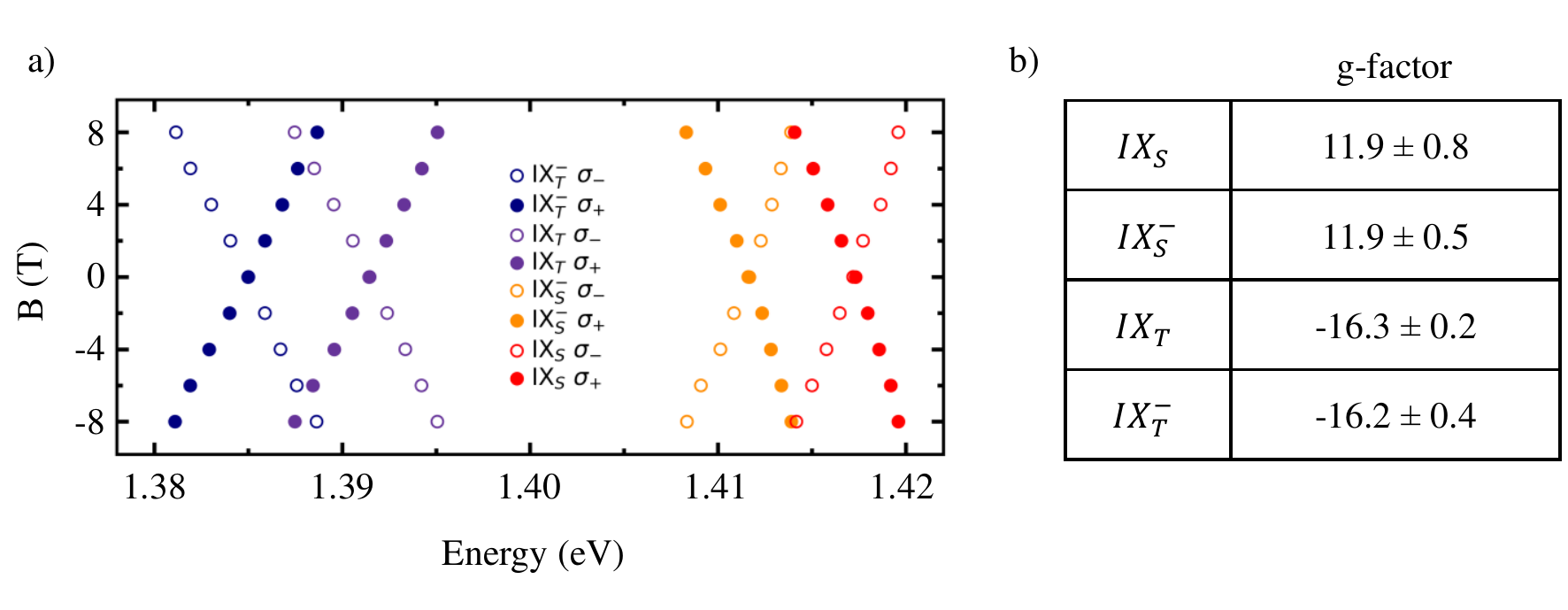}
	\caption{a) Plot of the fitted peak positions for different magnetic fields under $\sigma_{+}$- and $\sigma_{-}$ polarized detection. 
	b) Table of the extracted $g$-factors. 
	All measurements were performed at 3.8\,K with excitation wavelength corresponding to 2.33\,eV and a pump power of 300\,$\mu$W.}
\label{fig:Gfactor}
\end{figure} 

In the $g$-factor measurement, \fig{fig:Gfactor}, a magnetic field perpendicular to the sample surface was varied from -8 to 8\,T in steps of 1\,T.
At each field value a combination of $\lambda$/2 waveplate and linear polarizer were used to discriminate between $\sigma_+$ and $\sigma_{-}$ polarized light in the signal detection direction addressing K and K’ valley-polarizations respectively. 
The resulting Zeeman energy shift extracted from peaks in both polarizations is proportional to the characteristic g-factor according to $\Delta E=g\cdot \mu_B \cdot B$. 
The peak positions of each spectrum are determined using a Lorentzian fit function and are plotted in \fig{fig:Gfactor} a). 
From the data we determine $g$-factor values of $11.9\pm 0.8$ and $11.9\pm 0.5$ for $IX_S$ and $IX_S^-$ as well as $-16.3\pm 0.2$ and $-16.2\pm 0.4$ for $IX_T$ and $IX_T^-$, respectively, see \fig{fig:Gfactor} b).
Following Ref.\onlinecite{Zhao2023}, the $g$-factor of $IX_S$ matches the $K'-K$, spin-singlet of the $H^h_h$ registry of MoSe$_2$/WSe$_2$ heterobilayer, theoretically predicted to be 12.9 and experimentally measured as 11.9.
The $g$-factor of $IX_T$ is close to that of the corresponding $K'-K$ triplet, theoretically predicted to be -17.6 and experimentally measured as -15.8.
Further considering that the $H^h_h$ registry is the energetically lowest, preferred configuration of H-type MoSe$_2$/WSe$_2$ heterobilayer, this strongly supports the assignment of the optical resonances.
We also note, that a very weak emission of $Q-K$ valley-indirect excitons may be masked by the $K'-K$ transitions and thus cannot be excluded.
However, the $g$-factors of the lowest, spin-allowed $Q-K$ and $Q'-K$ excitons of 9.6 and 10.0, respectively, are not consistent with the measured $g$-factors of $IX_S$ and $IX_T$, providing a strong argument against such assignment.

\subsection{Interlayer exciton lifetime}\label{Lifetime}

\begin{figure}[ht]
	\centering
	\includegraphics[width=0.75\textwidth]{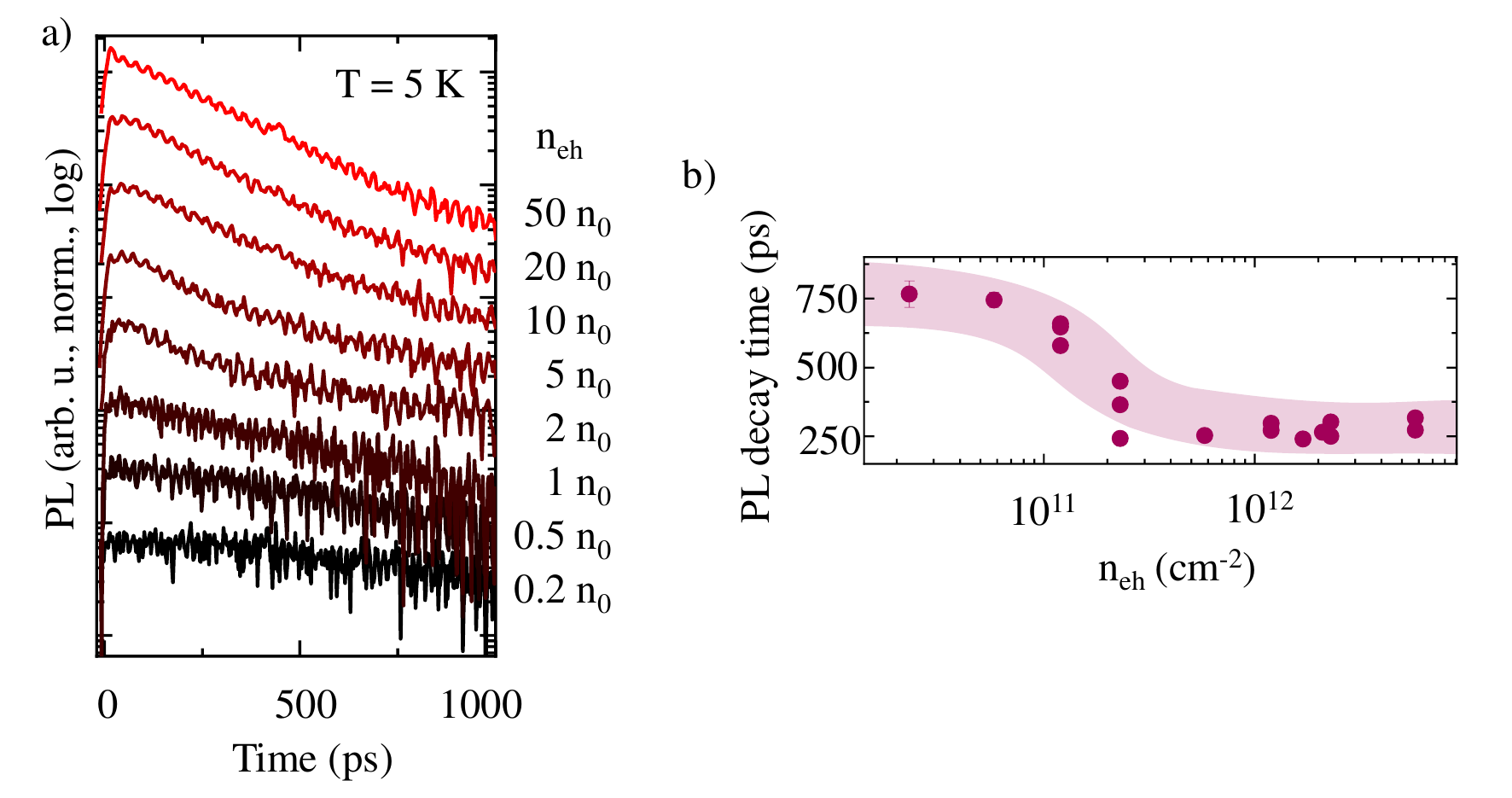}
	\caption{a) PL decay time transients at a temperature of T\,=\,5\,K, normalized and vertically offset for clarity. 
	The estimated injected electron-hole pair density ranges  from $1.2\times10^{10}$\,cm$^{-2}$ (0.1\,$n_0$) to $6\times10^{12}$\,cm$^{-2}$ (50\,$n_0$), with $n_0 = 1.2\times10^{11}$\,cm$^{-2}$. 
	b) Density-dependent PL decay times extracted from a) during the first 500\,ps. 
	Shaded area is a guide-to-the-eye.}
	\label{fig:Auger}
\end{figure}

\begin{figure}[ht]
	\centering
	\includegraphics[width=0.9\textwidth]{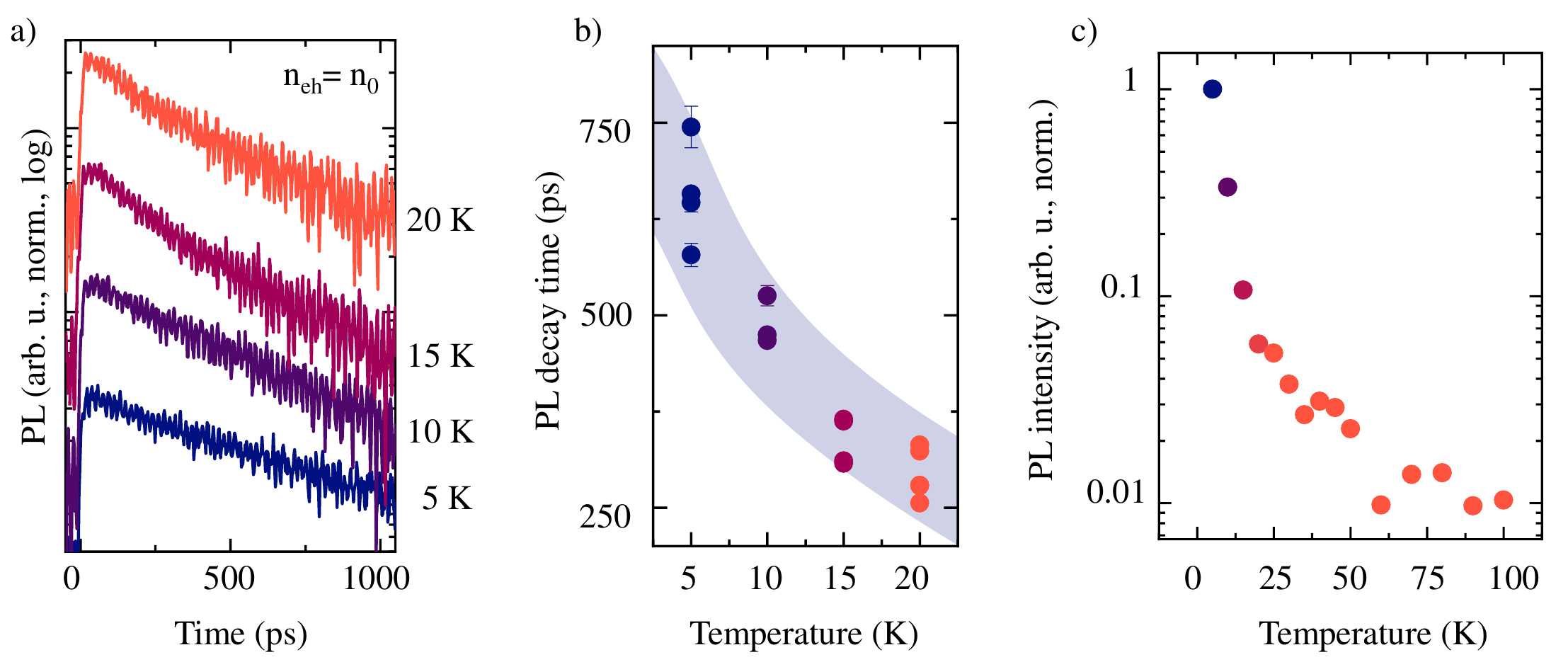}
	\caption{a)  Temperature dependent PL transients for temperatures ranging from T\,=\,5\,K to T\,=\,20\,K at a constant injected electron-hole pair density of $n_0$.  
		b) Temperature-dependent PL decay times as extracted from a) during the first 500\,ps. 
		Shaded area as guide-to-the-eye. c) Decrease of total PL intensity with temperatures from 5\,K to 100\,K.}
	\label{fig:Temperature}
\end{figure}

Here, we present the PL decay time of interlayer excitons as function of injected electron-hole pairs (\fig{fig:Auger}) at a nominal temperature of 5\,K and as function of temperature (\fig{fig:Temperature}), where the excitation density is fixed to the linear regime of $1.2\times10^{11}$ cm$^{-2}$ to exclude effects of exciton-exciton annihilation or repulsion.
The PL lifetimes were extracted from time- and spatially-resolved measurements presented in the main manuscript.\\

In the linear regime, at T\,=\,5\,K, we find the lifetime of the interlayer exciton PL of $690\pm 90$\,ps. As outlined in the supporting information of Ref.\cite{Ziegler2020, Wagner2021} diffusion influences the measured PL lifetime, due to excitons diffusing out of the detection area. Accounting for this, the corrected lifetime of the exciton population can be extracted as $780\pm 100$\,ps.
This PL lifetime is much shorter than frequently measured for interlayer excitons in TMDC heterobilayers and can often be as long as several hundreds of nanoseconds.
In the studied samples, the majority of the interlayer excitons (more than 80\%) decays during the first nanosecond and only a small fraction could be long-lived.
Moreover, we do not observed any residual PL intensity at negative times, prior to excitation. 
Since the pulse-to-pulse time is 12.5\,ns, it means that no substantial contributions from long-lived exciton fractions are detected.
Considering substantial oscillator strength of the spin-singlet $IX_S$ interlayer exciton on the order of 1 to 2\% of the intralayer transitions with picosecond lifetimes\,\cite{Robert2016a,Fang2019}, the corresponding radiative lifetime should be on the order of 100's of ps to a nanosecond by scaling argument. 
The majority of the interlayer excitons, however, reside in spin-dark $IX_T$ states with much longer lifetimes.
It means that the dominant recombination channel is non-radiative.
Since the excitons are mobile, they are also able to reach non-radiative traps rather fast\,\cite{Zipfel2020} in contrast to more typical scenarios studied in the literature of inter-layer excitons localized in disorder or moiré potentials.\\

As the density of injected electron-hole pairs increases, the PL signal decays faster. 
The initial PL decay time, extracted during the first 500\,ps decreases with increasing electron-hole pair density from its initial value of 750\,ps at low densities to 250\,ps at the highest studied density of $6\times10^{12}$\,cm$^{-2}$. 
As shown in the main manuscript, this decrease is accompanied by a saturation of the PL intensity relative to the pump power. 
The reduction in PL lifetime in combination with PL yield saturation is a strong evidence of non-radiative exciton-exciton annihilation \cite{Kumar2014, Mouri2014, Yuan2017, Sun2014, Hoshi2017}. 
In this scenario two exciton interact leading to the non-radiative recombination of one of them by transferring its energy to the second exciton and exciting it to a higher state.
In a basic bimolecular model, the probability of exciton-exciton annihilation increases linearly with increasing density leading to a decrease of the PL lifetime and relative intensity.\\

Furthermore, the interlayer exciton lifetime also decreases with increasing temperature, see \fig{fig:Temperature} a) and corresponding extracted PL lifetimes in \fig{fig:Temperature} b). 
The presented PL transients were acquired at a temperature ranging from 5\,K to 20\,K in the linear density regime of $n_0=1.2\times10^{11}$\,cm$^{-2}$.
In this temperature range, the lifetime decreases from 750 to 250\,ps.
The decrease of the exciton lifetime with temperature is accompanied by the reduced PL intensity, indicating faster non-radiative recombination, e.g. by capture into deep traps, as a likely origin.

\subsection{Transfer matrix analysis of reflectance contrast}
\begin{figure}[ht]
	\centering
	\includegraphics[width=0.9\textwidth]{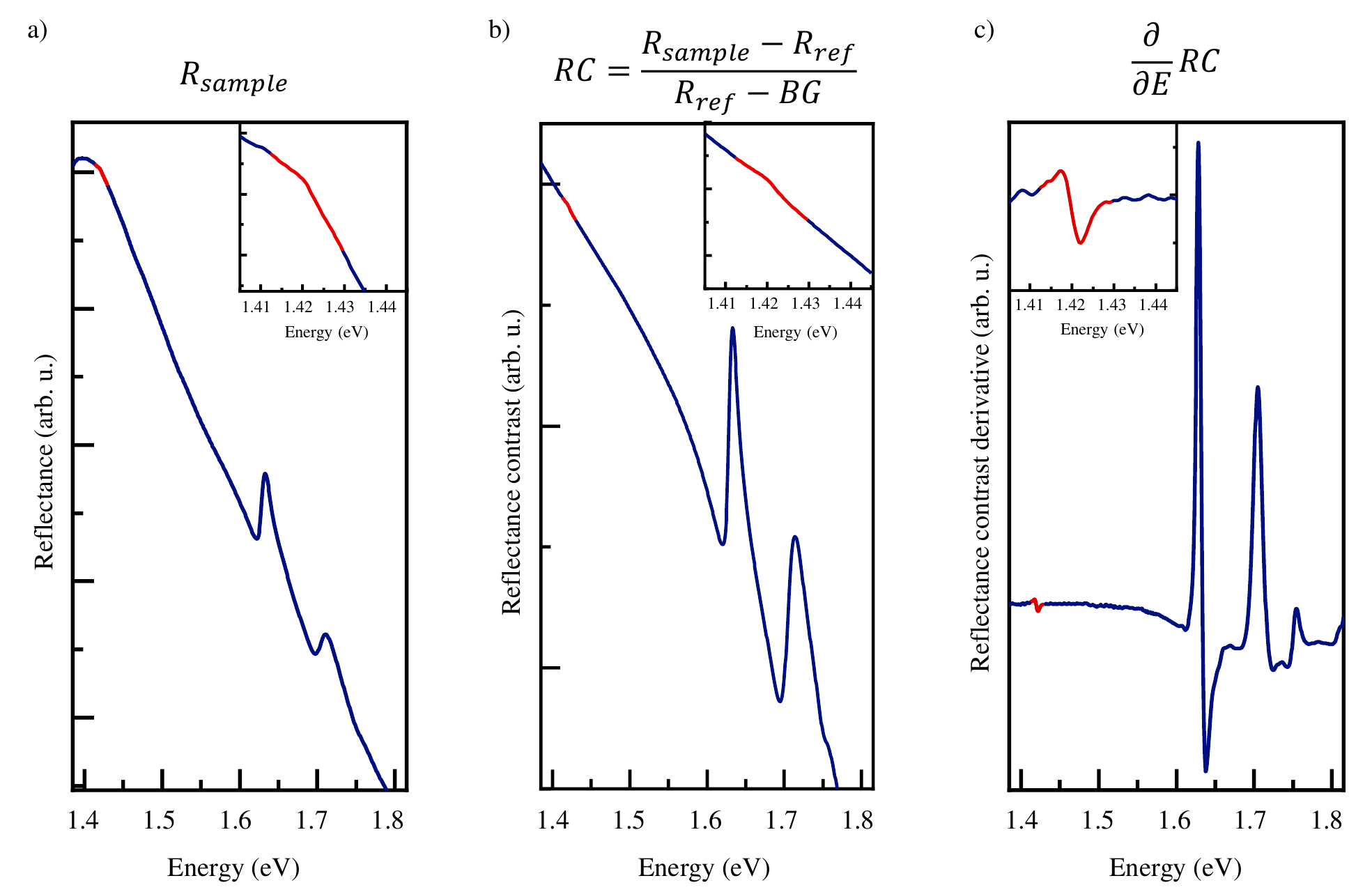}
	\caption{Feature of interlayer exciton, marked in red and enlarged in inset, in a) reflectance, b) calculated reflectance contrast and c) first derivation of reflectance contrast. All curves smoothed over 10 points. }
	\label{fig:IXreflectance}
\end{figure}

\begin{figure}[ht]
	\centering
	\includegraphics[width=\textwidth]{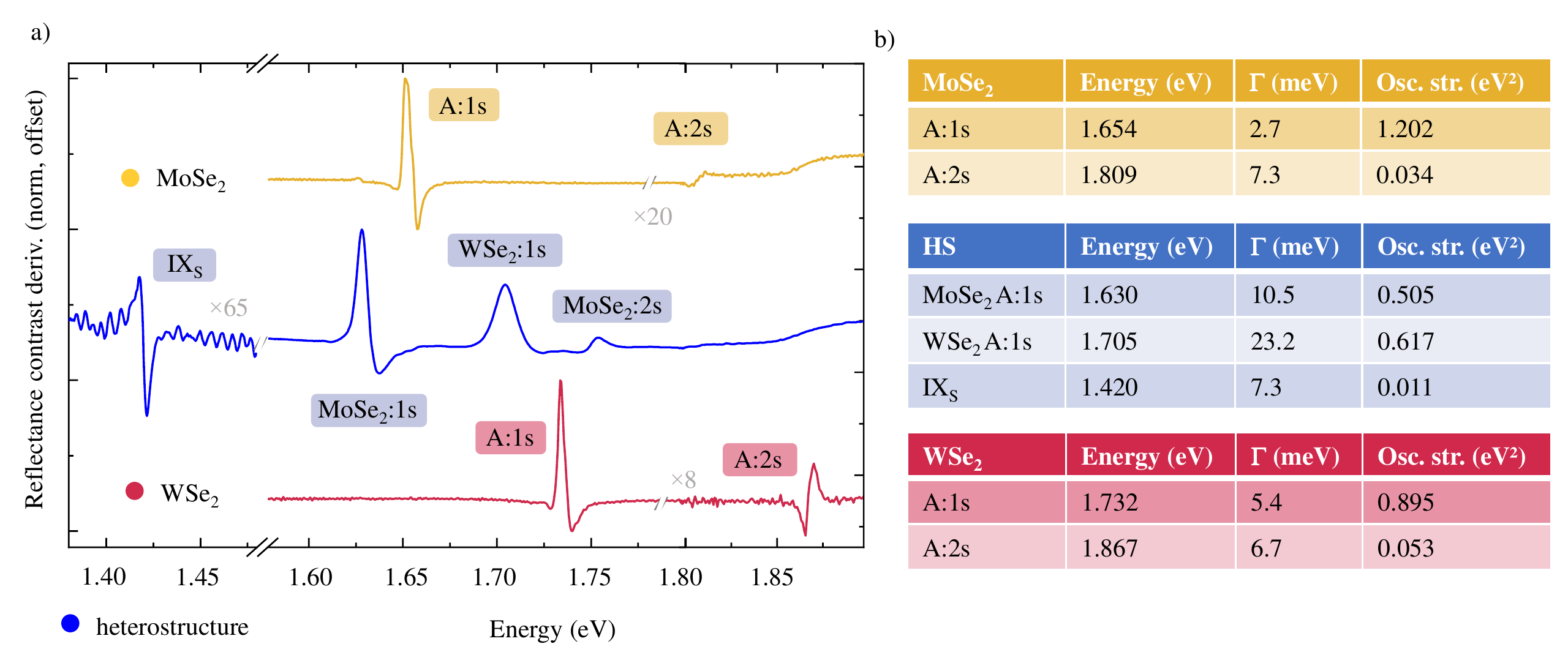}
	\caption{a) Reflectance contrast derivative spectra at corresponding spots on the heterostructure with free monolayer WSe$_2$, (bottom, red), complete stack of the heterostructure MoSe$_2$WSe$_2$ (middle, blue) and free monolayer part of MoSe$_2$ (top, yellow. 
	b) Parameters for resonance energy, linewidth and oscillator strength for each feature labeled in a), extracted from the transfer matrix analysis.}
	\label{fig:RC}
\end{figure}

As discussed in the main text, linear reflectance is a useful tool to probe the presence or absence of moiré potentials due to characteristic spectral features associated with them\,\cite{Jin2019, Alexeev2019}. We studied both the overlapping region of the MoSe$_2$/WSe$_2$ and also the individual MoSe$_2$ and WSe$_2$ monolayers.
The reflectance spectroscopy experiments in this work were performed at a temperature of 5\,K using a white-light halogen lamp focused to a spot with a diameter of less than 2\,$\mu$m. The light was cut with a 650 LP spectral filter to omit unnecessary reflectance from the silicon substrate. 
Integration time was set to 240\,ms in order to obtain maximum counts below saturation threshold of the camera. Due to the small oscillator strength of the interlayer exciton it is necessary to average over a reasonable amount of spectra for a sufficient signal to noise ratio. The minimum frames in this experiment to obtain detectable signal from the interlayer resonance was 500 frames. For the data shown in the main manuscript and \fig{fig:RC} the data was averaged over 5000 frames. The most suitable region for detecting the interlayer exciton resonance on a heterostructure is in the very middle of the sample and far from any sources of disorder including bubbles of insufficient interlayer contact and areas of consistent sample quality. An exemplary reflectance spectrum $R_{sample}$ at 5\,K is shown in \fig{fig:IXreflectance} a) with the interlayer exciton feature marked in red and enlarged in the inset.
A reference spectrum $R_{ref}$ was acquired on a region including both top and bottom layer of the encapsulating hBN adjacent to the MoSe$_2$/WSe$_2$ heterostructure. The reflectance contrast $RC$ is then calculated according to 
\begin{equation}
	RC = \dfrac{R_{sample} - R_{ref}}{R_{ref} - BG}
\end{equation}
where $BG$ is the background signal with the light-source blocked. The resulting reflectance contrast is shown in \fig{fig:IXreflectance} b) accordingly. First derivative with respect to energy is then taken on the resulting RC spectrum to better visualize the resonances, as seen in \fig{fig:IXreflectance} c).
\\
Further obtained RC derivative spectra are shown in \fig{fig:RC} a) for monolayer WSe$_2$ (bottom, red), heterostructure MoSe$_2$/WSe$_2$ (middle, blue) and monolayer MoSe$_2$ (top, yellow). 
Labeled are most prominent resonances of A1:s (ground) and A2:s (first excited) states stemming from the intralayer excitons of the individual monolayers and the spin-singlet interlayer exciton $IX_S$. \\

Due to interferences of the multi-stack structure of the sample a transfer-matrix approach is used to model the dielectric response \cite{Byrnes2016}.
Extracted values from transfer matrix analysis for labeled resonances in terms of resonance energy, linewidth and oscillator strength are summarized in the table of \fig{fig:RC} b). 
The monolayer parameters are typical for hBN-encapsulated structures in terms of central peak energies and spectral linewidths.
Moreover, we do not observe any substantial contributions from trions (attractive Fermi polarons), implying that the residual doping is below 10$^{11}$\,cm$^{-2}$\,\cite{Wagner2020}.

Intralayer exciton resonances also appear in the reflectance contrast spectrum of the heterobilayer area. 
In this region, the spectrum corresponds to the superposition of the two monolayer spectra, with slightly shifted and broadened A-exciton resonances.
Most importantly, no resonance splitting and redistribution of the oscillator strength into several peaks is observed, that is otherwise a characteristic feature of the moiré lattice \cite{Jin2019}.
It thus signifies large-area atomic reconstruction with one predominant lattice registry\,\cite{Zhao2023}.
An additional feature evidencing atomic reconstruction is the appearance of a spectrally narrow resonance matching the energy of the spin-singlet interlayer exciton PL of the H$^h_h$ registry (identified by the positive degree of circular polarization and $g$-factor of 11.9, see Fig.1 of the main manuscript and Sect.\ref{gfactor}).
The extracted oscillator strength is 2\,\% of that from the intralayer A-excitons in the same spectrum. 
It is consistent with the measurements of interlayer excitons using cavity-enhanced\,\cite{Forg2019} and electro-modulated spectroscopy\,\cite{Barre2022} as well as theoretical calculations from first-principles many-body theory\,\cite{Barre2022}.
The latter identifies the origin of the interlayer exciton absorption from the hybridization of the hole wavefunction of K/K' valleys that is located predominantly in WSe$_2$, but has also small contributions in MoSe$_2$.
This increases the overlap with the electron wavefunction located in the MoSe$_2$ layer and leads to a substantial oscillator strength of the interlayer exciton, as also observed in the experiments.

The shift of the intralayer resonances to lower energies is attributed to the change in dielectric screening and strain; the latter is discussed in more detail in the following Sect.\,\ref{strain}. 
The former effect stems from a large dielectric constant of an additional TMDC layer in contrast to hBN.
This is further evidenced by a smaller energy separation between 1s and 2s resonances of the MoSe$_2$ in the heterostructure, shown in \fig{fig:RC} a).

In addition, all intralayer exciton resonances are broadened in the heterostructure.
This effect is often observed in heterobilayers and can be attributed to the ultrafast charge transfer due to type-II alignment on sub-picosecond time-scales\,\cite{Rigosi2015}. 

\subsection{Strain dependence across sample}
\label{strain}
\begin{figure}[ht]
	\centering
	\includegraphics[width=\textwidth]{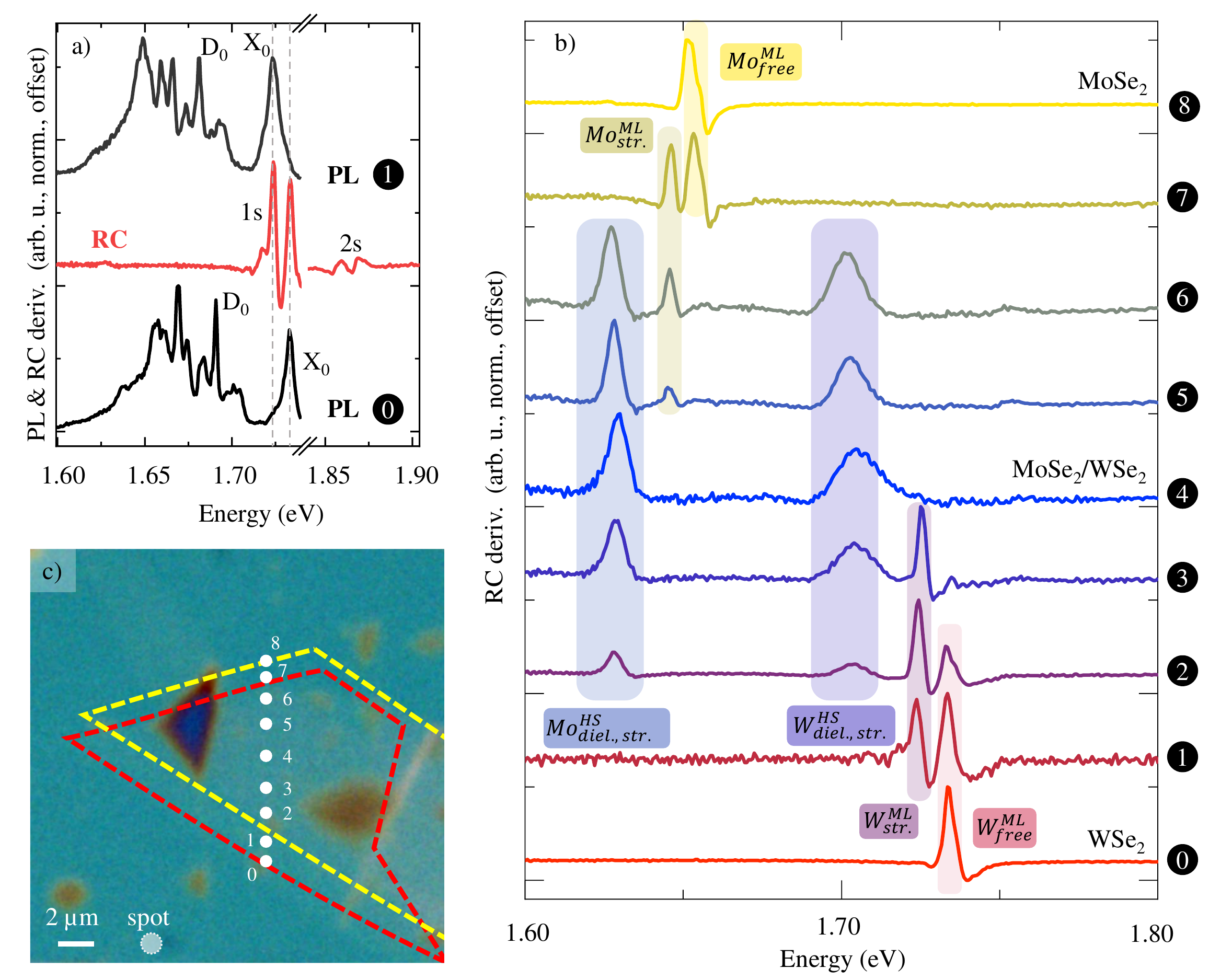}
	\caption{Signature of strain in reconstructed heterostructure. a) Consequences of strain as visualized in freestanding monolayer constituent WSe$_2$. Selected PL spectra at positions 0 and 1, as marked in micrograph c) and RC spectra acquired at a spot between position 0 and 1. b) RC deriv. spectra following a straight linecut across the heterostructure from monolayer, to overlap to monolayer. c) Micrograph of heterostructure with indicated path of shown spectra from a) and b) starting enumerated 0 to 8. Monolayer WSe$_2$ and MoSe$_2$ are indicated by dashed lines in red and yellow respectively.}
	\label{fig:Strain}
\end{figure}

In general one can expect additional strain, when layers are stacked on top of each other.
Moreover, atomic reconstruction implies additional local strain across the sample \cite{Li2021b, Rosenberger2020, Carr2018, Enaldiev2021, Rodriguez2023}.
To estimate strain we study PL and reflectance contrast in the WSe$_2$ monolayer in close proximity to the heterostructure region.
\fig{fig:Strain} a) shows PL and RC spectra in for several positions with respect to the edge of the heterostructure region. The PL spectra were acquired at position 0 (at the free edge of the WSe$_2$ monolayer) and position 1 (approximately distanced 1\,$\mu$m from position 0, close to the heterostructure overlap region), as marked in the micrograph \fig{fig:Strain} c). These PL spectra show position-dependent shift of the emission to lower energies of about 10\,meV, indicated by the dashed gray lines for the bright exciton ($X_0$).
The same shift is also observed in the reflectance contrast spectrum, acquired at the position between the two PL spectra. 
Due to the larger size of the whitelight spot used for reflectance (indicated by the sketch in \fig{fig:Strain}) the signal is thus collected from both regions. 
This leads to the appearance of a double peak feature of the A:1s resonance, with the same splitting of 10\,meV, as extracted in the PL spectra.
The same energy separation is also found for the A:2s state, demonstrating that it cannot originate from the changes of the dielectric environment.
In the latter case, the energy splitting would differ for 1s and 2s excitons, due to increased sensitivity of the excited states to the dielectric surroundings.
This is expected, since the measurements are close to the heterostructure, but still in the monolayer region. Consequently, the energy shift of 10\,meV in PL and reflectance is attributed to the impact of strain. In the case of a biaxial strain, this shift would correspond to strain of 0.1\,\%. 
Alternatively, the strain can also be estimated from the change in energy difference between bright ($X_0$) and neutral dark excitons ($D_0$). 
The reduction of $X_0-D_0$ difference on the scale of 500\,$\mu$eV also gives a strain on the order of 0.1\,\% \cite{Dirnberger2021}.\\

We also note that strain is not only present in the monolayer WSe$_2$ constituent, but also evolves gradually across the sample. 
\fig{fig:Strain} b) shows selected reflectance contrast derivative spectra along a line through the studied heterostructure: from unstrained to strained WSe$_2$, across the overlap region of MoSe$_2$/WSe$_2$, to strained and lastly unstrained MoSe$_2$ (as indicated by the color transition going from red to blue to yellow) and labelled according to their position as illustrated in the micrograph of \fig{fig:Strain} c).\\

The reflectance contrast follows a continuous evolution as one can assume for strain. 
The most bottom spectrum of \fig{fig:Strain} b) features a single peak ($W_{free}^{ML}$), as expected for WSe$_2$ monolayer. 
Closer to heterostructure, the impact of strain becomes noticeable as a double peak feature ($W_{str.}^{ML}$ and $W_{free}^{ML}$) appears, as discussed above.
Due to the spatial overlap from the size of the whitelight spot, it is observed also in the proximate region of the heterostructure. 
In the center of it, the WSe$_2$ peak shifts further to lower energies and broadens from the additional effects of charge transfer and change in dielectric environment ($W_{diel., str.}^{HS}$). 
Similar features are observed at the sample positions closer to the MoSe$_2$ monolayer.
The splitting of the MoSe$_2$ 1s resonance due to strain occurs already in the heterobilayer region ($Mo_{str.}^{ML}$ ).
Thus, even weak resonances in the heterostructure region in close proximity to its monolayer constituents can be attributed to the combined effect of strain and finite size of the measurement spot.
In the center of the heterostructure, however, only the intralayer resonances of MoSe$_2$ and WSe$_2$, $Mo_{diel., str.}^{HS}$ and $W_{diel., str.}^{HS}$ respectively, are observed, and no additional features related to moiré-like superlattice potentials are detected.

\newpage
\section{Experimental setup}
\subsection{Time-dependent measurements}
Time resolved experiments were carried out using a streak camera (Hamamatsu Photonics), type C10910-01 equipped with a S-20 photocathode. The sample was excited resonantly into the MoSe$_2$ A:1s exciton resonance with a  140\,fs Ti:sapphire laser with a repetition rate of 80\,MHz, focused onto the sample through a 60\,x objective. The emitted PL signal of interlayer exciton emission was subsequently resolved in space and its lateral cross section was guided  into the streak camera, where the evolution of the PL broadening is then resolved with respect to time (see \cite{Wagner2021, Zipfel2020, Kulig2018}). The data was acquired in single photon counting mode for optimal signal-to-noise ratio.
The challenge in measuring diffusion on TMDC heterostructures lies in the requirement of resolving a PL spot of good quality Gauss shape, that is symmetrical and can be fitted with $\propto \exp(-x^2)$. Especially for heterobilayers this can be challenging, as insufficient interlayer coupling can lead to variations in interlayer distance between the two TMDC layers and formation of bubbles causing a distortion in the laser spot focused onto it. Through means of thermal annealing interlayer distance can be decreased efficiently leaving behind smooth areas to perform transport measurements on \cite{Alexeev2017, Luong2017}, see \fig{fig:samplefabrication}.\\

\begin{figure}[ht]
	\centering
	\includegraphics[width=\textwidth]{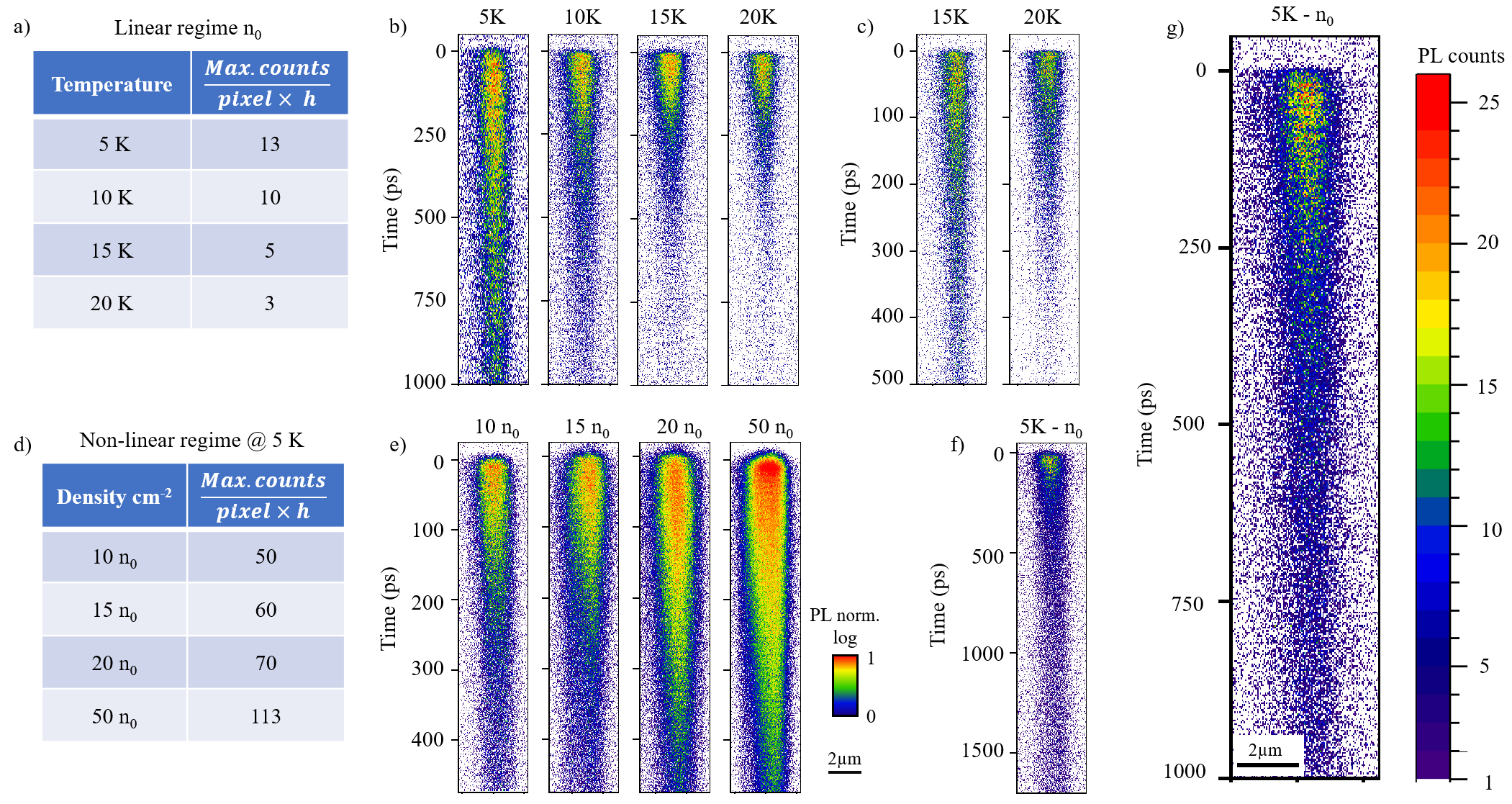}
	\caption{a) Maximum counts per pixel within one hour for the temperature dependent diffusion measurements in the linear regime. b) Selected streak images for the studied temperature range in a 1000\,ps time window. c) Streak images for 15\,K and 20\,K in a 500\,ps time window.
		d) Maximum counts per pixel within two hours for non-linear pump density dependent experiments at 5\,K. e) Selected streak images for densities of 10\,$n_0$, 15\,$n_0$, 20\,$n_0$ and 50\,$n_0$. b), c) and e) normalized to pixel with maximum counts and logarithmic color scale. f) Streak image of interlayer exciton diffusion at 5\,K in the linear regime for 1700\,ps time window and g) 1000\,ps in linear color scale. }
	\label{fig:Sensitivity}
\end{figure}

Depending on the kind of photocathode installed in the streak camera, integration times can vary due to a decrease of sensitivity of the spectral response in the near infrared \cite{StreakManual} compared to the visible range. Furthermore the drop in intensity of the interlayer exciton PL signal with temperature accompanied with a reduction of the lifetime, as illustrated in \fig{fig:Temperature}, results in longer integration times as well. In the case of the streak camera used in this study the maximum counts per pixel of the camera chip within one hour was 13 counts at 5\,K for the linear regime and decreased by 80\% to 3 counts for 20\,K, as listed in \fig{fig:Sensitivity} a). These values are corrected for dark counts, which contribute to 1 count per pixel per hour. Therefore for the temperature dependent diffusion experiments in the linear regime total integration times varied from 2 hours at 5\,K temperature to $8-10$ hours at 20\,K. Selected streak images, normalized to their respective maxima and shaded in logarithmic color scale, are presented for long (1000\,ps) and short (500\,ps) time ranges, \fig{fig:Sensitivity} b) and c) respectively. While a 1000\,ps time window is sufficient for lowest temperatures, the reduction in PL lifetime makes in necessary to record the PL broadening in a smaller time window of 500\,ps, in order to resolve the broadening accurately. \\

Pump density dependent measurements at 5\,K on the other hand were all performed with integration times ranging from 30 minutes to 1 hour. The maximum counts per pixel within one hour over the camera chip are listed in the table of \fig{fig:Sensitivity} d). The effects of exciton-exciton annihilation results in the non-linear increase of the counts with increasing pump density. Selected streak images for the nonlinear regime for densities of 10\,$n_0$, 15\,$n_0$, 20\,$n_0$ and 50\,$n_0$ are presented in \fig{fig:Sensitivity} e), where a short time window of 500\,ps was chosen in order to resolve the intricate nonlinear broadening and transport dynamics. \\

\subsection{Estimation of electron-hole pair density}
\begin{figure}[ht]
	\centering
	\includegraphics[width=0.6\textwidth]{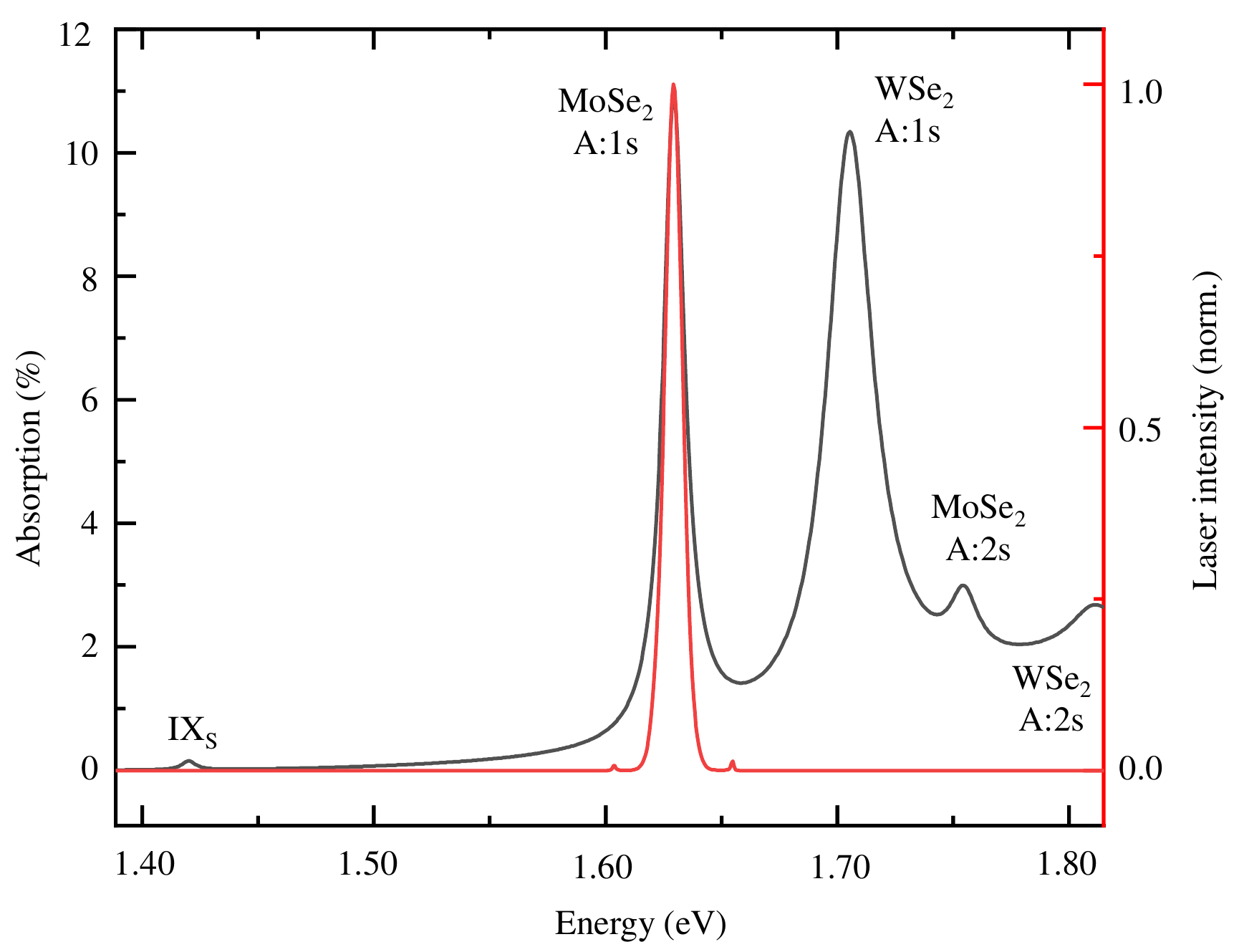}
	\caption{Absorption of MoSe$_2$/WSe$_2$ heterostructure as evaluated from transfer matrix analysis of reflectance measurements (black line). Labelled are resonances of interlayer exciton $X_S$, MoSe$_2$ A:1 and A:2s states as well as WSe$_2$ A:1s and A:2s states. The spectrum of the pump laser ($\lambda = 761$\,nm) in red, overlapping with the absorption peak of the MoSe$_2$ A:1s state.}
	\label{fig:Absorption}
\end{figure}

The injected electron-hole pair density was estimated using the following equation
\begin{equation*}
	n_{ex}=\frac{P \cdot \alpha}{f_{rep} \cdot \pi \cdot  r^2 \cdot E_{ph}}
\end{equation*}
with $P$ laser power (range of 100\,nW -- 25\,$\mu$W), $\alpha=0.12$ absorption, $f_{rep}$\,=\,80\,MHz, $E_{ph}=1.63$\,eV photon energy and $r=(0.7\pm 0.1)$\,$\mu$m is the radius of the focused laser spot. The radius is chosen so that the effective density is equal to the maximum density of the spot center and was evaluated from the FWHM of the laser profile as $r=\text{FWHM}/(2\sqrt{\ln 2})\approx \text{FWHM}/1.6652$. Therefore $n_{ex} \times \pi r^2 = N_0$ is close to the maximum exciton density at the center of the pump spot \cite{Wagner2023}. Due to efficient and ultra fast charge transfer it is likely that the majority of the excitons created in MoSe$_2$ by the pump pulse form interlayer excitons. 
The absorption value was estimated from transfer matrix analysis of reflectance measurements and its spectral overlap with the spectral function of the laser, see black and red lines of \fig{fig:Absorption} respectively.

\begin{figure}[ht]
	\centering
	\includegraphics[width=\textwidth]{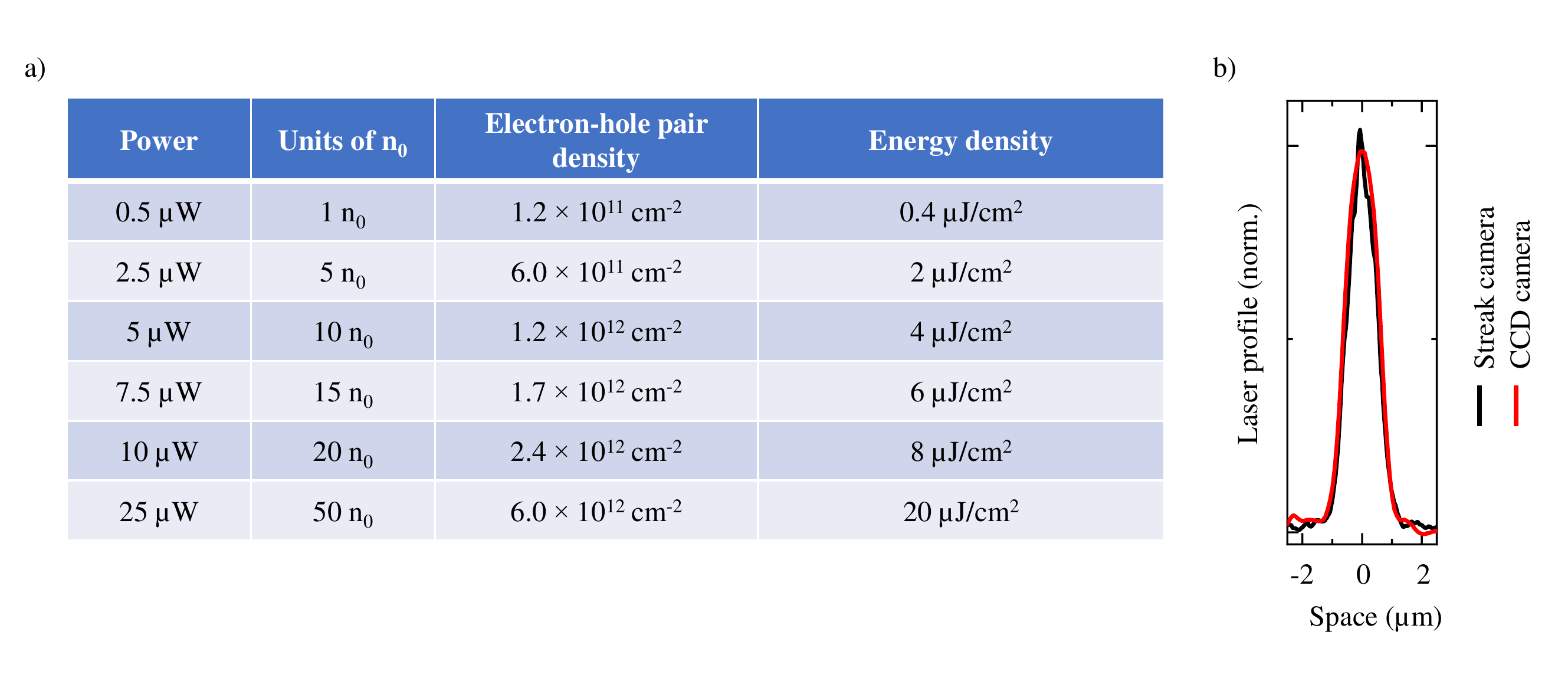}
	\caption{a) Summary of conversion from laser power to units of $n_0$, electron-hole pair density and respective energy density. Average error on the order of 20\% to 30\% due to sensitivity of calculation with regards to laser spot radius $r$ and present error of laser profile width extraction. b) Normalized profile of laser on CCD and streak camera in red and black lines respectively.}
	\label{fig:Density}
\end{figure}

A table of conversion from measured laser power to units of $n_0$, electron-hole pair density and respective energy density is given in \fig{fig:Density} a). Due to the sensitivity of the calculation on the inserted radius $r$, an approximate error of about 20\% to 30\% in the density is reasonable. This is partially due to slightly fluctuating laser profile width during different measurement sessions and alignment process. An exemplary laser profile, recorded on both CCD camera (red line) and streak camera (black line) is presented in \fig{fig:Density} b). Here the extracted width $FWHM \approx 1.2$\,$\mu$m corresponds to $r=0.73$\,$\mu$m.

\newpage
\section{Drift-diffusion model}
\subsection{Expectation from semi-classical approximation}
\begin{figure}[ht]
	\centering
	\includegraphics[width=\textwidth]{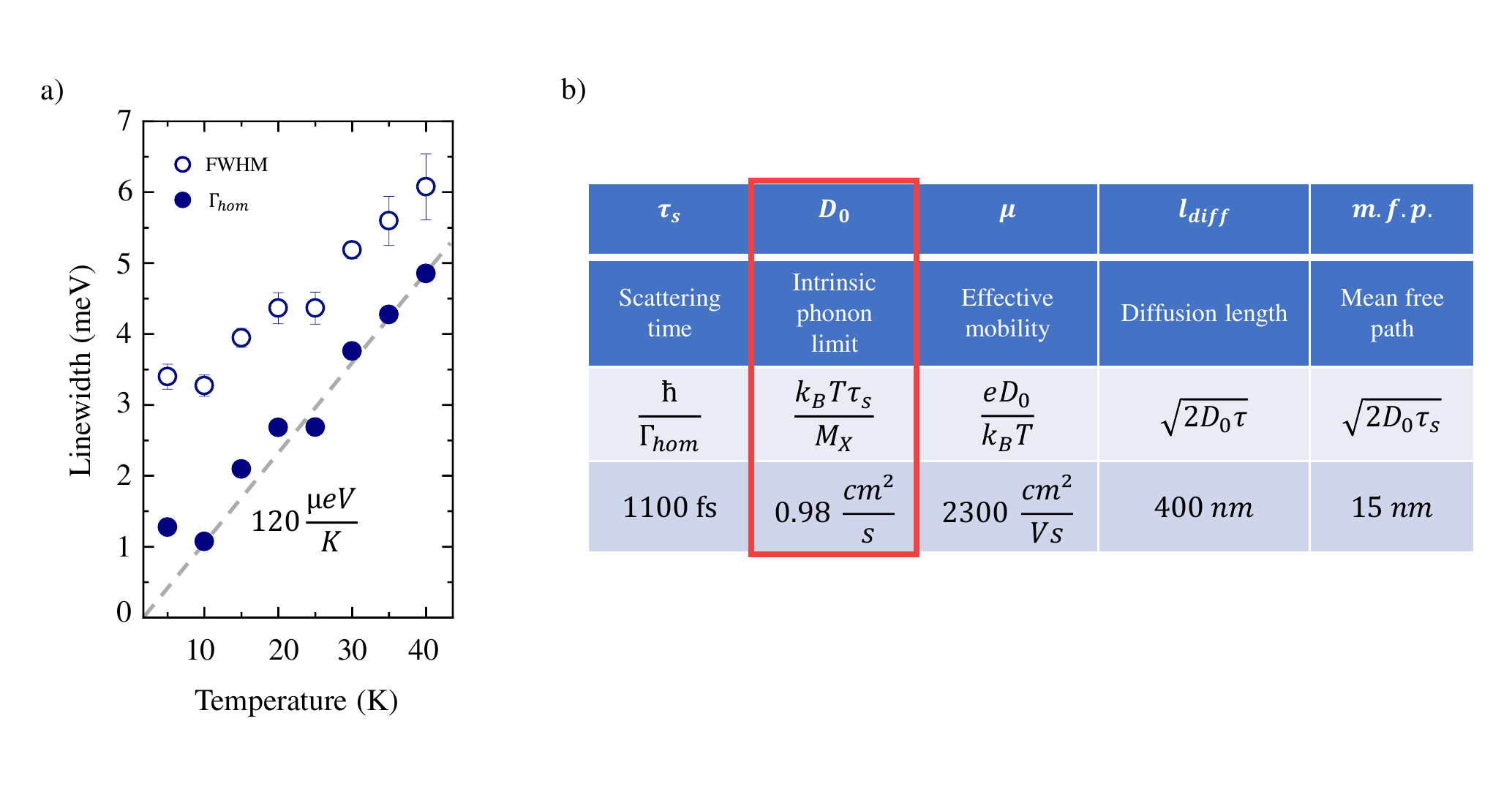}
	\caption{a) Total spectral linewidths as full-width-at-half-maximum of interlayer exciton triplets ($IX_T$) as a function of temperature extracted from PL spectra (empty datapoints), pulsed excitation at a density of $2\times10^{11}$\,cm$^{-2}$. Deconvoluted linewidth with inhomogeneous broadening of $\Gamma_{inhom} = 2.65$\,meV, (solid datatpoints). b) Corresponding values for scattering time $\tau_s$, diffusion coefficient $D_0$, effective mobility $\mu$, diffusion length $l_{diff}$ (with exciton lifetime $\tau$), and mean free path from the semiclassical model (Eq.\eqref{semiclassic}). }
	\label{fig:Expect}
\end{figure}

Here, we estimate the expected value for the measured diffusion coefficient of interlayer excitons at 5\,K from semi-classical diffusion model: 
\begin{equation}
	D_0 = \frac{k_BT \hbar}{M_X\Gamma},
	\label{semiclassic}
\end{equation}
where $T$ is the temperature, $M_X$ is the exciton translational mass, $\Gamma / \hbar$ is the momentum scattering rate and $k_B$ is the Boltzmann constant. 
For the case of isotropic scattering, $\Gamma$ corresponds to the homogeneous contribution to the spectral linewidth broadening $\Gamma_{hom}$.
The total linewidths of the interlayer exciton ($IX_T$) extracted from PL spectra at different temperatures are plotted in \fig{fig:Expect} a).
The are usually a convolution of homogeneous and inhomogeneous broadening from various sources of disorder.
Assuming that the latter is independent from temperature in the studied range, we use a deconvolution procedure to estimate the homogeneous component (see \cite{Wagner2021}).
The linear increase of the linewidth with temperature is typical for the exciton scattering to low-momentum acoustic phonons with the extracted value of 120\,$\mu$eV/K. 
This broadening rate is larger than that of the intralayer A-excitons in WSe$_2$ monolayer, but closer to the values for A-excitons in MoSe$_2$ monolayer\,\cite{Selig2016}.

Using the extracted broadening rate, we obtain the diffusion coefficient and related parameters, such as the scattering time $\tau_s$, effective mobility $\mu$, diffusion length $l_{diff}$, and mean free path $m.f.p.$, summarized in \fig{fig:Expect} b).
The value of 1\,cm$^2$/s for the diffusion coefficient matches the experimental observations presented in the main text.
It is thus likely that the propagation of interlayer excitons in absence of disorder- or moiré-induced potentials is limited by the scattering with acoustic phonons in analogy to the individual monolayers\,\cite{Wagner2021}.
Combined with the diffusion length of several hundreds of nanometers, it means that the domain boundaries between fixed registries in the reconstructed heterostructure do not contribute substantially to the diffusion.
For the studied case of the $H^h_h$ reconstruction, the domain boundaries have $H^X_h$ registry with a lower energy of interlayer exciton transitions\,\cite{Zhao2023}. 
Thus, the should act as potential traps and inhibit diffusion.
In contrast to that, measured diffusion coefficients are high and match the expectation from intrinsic scattering with phonons.
Consequently, we conclude that the size of the reconstructed domains is on the order of the diffusion length, corresponding to a recently demonstrated scenario of atomic reconstruction on mesoscopic length scales\,\cite{Zhao2023}.
Finally, a mean free path of 15\,nm at T\,=\,5\,K is on the order of typical sizes of moiré potentials.
It implies that localization in potential minima is expected for moiré periods larger than that, if the potentials are sufficiently steep. 
For the case of smaller moiré periods, one could expect tunneling and the formation of minibands to determine the transport instead.

\subsection{Semi-classical drift-diffusion equation in interacting exciton gas}

We consider exciton diffusion in a 2D system in the presence of exciton-exciton interactions. Corresponding diffusion equation can be derived combining the continuity equation for the exciton density $n(\bm r,t)$ with $\bm r$ being the in-plane coordinate and $t$ being time
\begin{equation}
	\label{cont}
	\frac{\partial n}{\partial t} + \divv{\bm j}  + \frac{n}{\tau} + R_A n^2 =0,
\end{equation}
where $\tau$ is the exciton lifetime and $R_A$ is the bimolecular recombination, i.e., exciton-exciton annihilation, coefficient, and the material equation for the exciton current density $\bm j$
\begin{equation}
	\label{j:eq}
	\bm j = - {D_0} \bm\nabla n + {\frac{\hbar}{M \Gamma}} \bm f n.
\end{equation}
Here {$D_0$} is the linear diffusion coefficient (related to the velocity-velocity autocorrelation function~\cite{Glazov:2022aa}), $\bm f$ is the force acting on the excitons~\cite{PhysRevLett.120.207401,Perea-Causin:2019aa,PhysRevB.100.045426,PhysRevMaterials.6.094006}, and $\Gamma/\hbar$ is the momentum scattering rate. It follows from Eqs.~\eqref{cont} and \eqref{j:eq} that
\begin{equation}
	\label{diffusion}
	\frac{\partial n}{\partial t} + \frac{n}{\tau} + R_A n^2 = {D_0} \Delta n - {\frac{\hbar}{M\Gamma}} \divv{(\bm f n)}.
\end{equation}

For excitons with parabolic dispersion $E_k = \hbar^2 k^2/2M$ ($M$ is the exciton effective mass) occupying single band, the exciton diffusion coefficient ${D_0} = k_B T {\hbar/(M\Gamma)}$ itself in Eq.~\eqref{j:eq} (with $T$ being the temperature and $k_B$ being the Boltzmann constant) is not affected by exciton-exciton collisions.
This follows from the Einstein relation between the diffusion coefficient and the mobility and the Kohn theorem~\cite{PhysRev.123.1242}. 
However, it is often practical to consider the observed \emph{effective} diffusion coefficient that describes the rate of exciton cloud expansion and depends on exciton-exciton interactions. 

We take the repulsion force $\bm F$ between two excitons in the simplest possible form
\begin{equation}
	\label{Force:2d}
	\bm F(\bm r) = -\frac{\partial V(\bm r)}{\partial \bm r},
\end{equation}
where $\bm r$ is the interexciton distance and $V(\bm r)$ is the potential energy of their interaction. The latter is typically understood as the consequence of the dipole-dipole repulsion between the particles~\cite{PhysRevB.80.195313,PhysRevB.78.045313,https://doi.org/10.48550/arxiv.2204.09760} and is sufficiently short-ranged so that the integral $U_0 = \int d\bm r V(\bm r)$ converges. Thus, we approximate 
\begin{equation}
	\label{Force:2d:approx}
	V(\bm r) = U_0 \delta(\bm r ), \quad U_0>0,
\end{equation}
and obtain the force $\bm f$ in the exciton ensemble
\begin{equation}
	\label{force:f}
	\bm f = \int \bm F(\bm r - \bm r') n(\bm r') d \bm r' = -\int V(\bm r - \bm r') \frac{\partial n(\bm r')}{\partial \bm r'} d\bm r' =-U_0 \frac{\partial n(\bm r)}{\partial \bm r}.
\end{equation} 
Hence, we arrive from Eq.~\eqref{diffusion} at the drift-diffusion equation in the form
\begin{equation}
	\label{diffusion:1}
	\frac{\partial n}{\partial t} + \frac{n}{\tau} + R_A n^2 = {D_0} \Delta n + \frac{U_0 {D_0}}{k_BT} \bm \nabla\cdot(n\bm \nabla n).
\end{equation}
We note that in this expression $U_0n$ corresponds to the shift of the exciton line to higher energies (blue shift) at a given density $n$. Equation~\eqref{diffusion:1} agrees with Eq.~(7) of Ref.~\cite{PhysRevMaterials.6.094006}.

\subsection{Analytical description and numerical solution}

\begin{figure}[ht]
	\includegraphics[width=\textwidth]{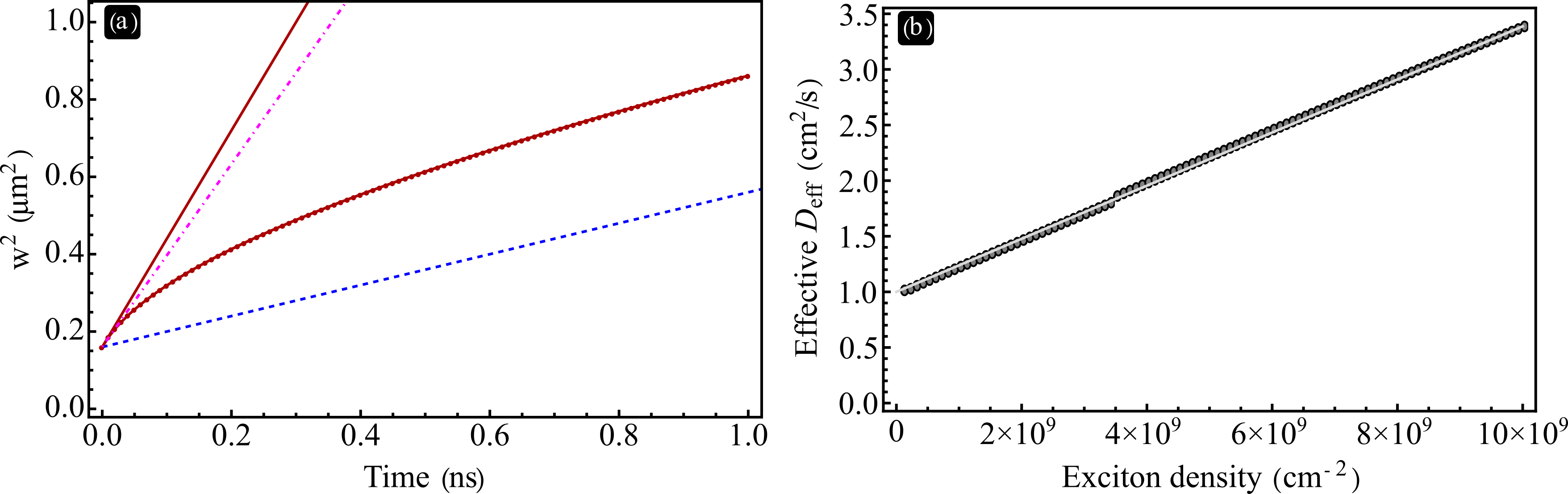}
	\caption{(a) Squared widths of the exciton cloud as a function of time calculated by numerical solution of Eq.~\eqref{diffusion:1} (dark red points) and approximations (lines): Blue dashed line shows the expansion with the linear diffusion coefficient, $w^2 = r_0^2 + 4Dt$, magenta dot-dashed line shows the result with $D_{\rm eff}$ evaluated after analytical formula~\eqref{D:eff} $w^2 = r_0^2 + 4D_{\rm eff} t$ (in the limit of $t \rightarrow 0$, corresponding to early times after the excitation), dark red solid curve shows the fit of initial points in $w^2(t)$ found numerically. The squared widths correspond to the mean squared displacement $\Delta \sigma^2$ according to $w^2 = 2\sigma^2$ with $\Delta \sigma^2 = \sigma^2-\sigma_0^2$.
		(b) Effective diffusion coefficient at $t \rightarrow 0$ as a function of exciton density: points are the results of numerical calculation and line is plotted after Eq.~\eqref{D:eff}. A small kink in the points is due to numerical errors in fitting the exciton cloud by a Gaussian. Parameters of the calculation are as follows: linear diffusion coefficient ${D_0}=1$~cm$^2$/s, Auger recombination rate $R_A=10^{-1}$~cm$^2$/s, effective repulsion constant $U_0 D_0/k_B T = 10^{-10}$~cm$^{4}$/s, $r_0=0.4$~$\mu$m. 
		In panel (a) the initial density is $5\times 10^{10}$~cm$^{-2}$.}\label{fig:SI:theory} 
\end{figure}

To describe the basic physics we provide an estimate for the effective diffusion coefficient. While this approach is simplified compared to the semi-analytical model developed in Refs.~\cite{PhysRevLett.120.207401,deQuilettes:2022tb}, it gives a very good agreement with numerical solution of Eq.~\eqref{diffusion:1}. To that end, we consider the short-time evolution of the initial distribution of the exciton density
\begin{equation}
	\label{initial}
	n(\bm r,0 ) = \frac{N_0}{\pi r_0^2} e^{-r^2/r_0^2},
\end{equation}
where $r_0$ determines the spot size (full-width-at-half-maximum is equal to $2\sqrt{\ln 2}\,r_0\approx 1.665\,r_0$) and $N_0$ is the number of excitons injected optically, $N_0 = \int N(\bm r,0) d\bm r$. Substituting Eq.~\eqref{initial} in Eq.~\eqref{diffusion:1} and taking $r=0$ we have
\begin{equation}
	\label{evolution}
	\left.\frac{\partial n}{\partial t}\right|_{\bm r=0,t=0} = -\left[ \frac{1}{\tau} \frac{N_0}{\pi r_0^2}  +   {D_0}\frac{4N_0}{\pi r_0^4} + R_A \left(\frac{N_0}{\pi r_0^2}\right)^2 + \frac{U_0 {D_0}}{k_B T} \frac{4N_0^2}{\pi^2 r_0^6}\right].
\end{equation}
Equation~\eqref{evolution} describes the decay of the peak density as a result of three effects: (i) recombination (mono- and bimolecular), (ii) diffusion, and (iii) repulsive interactions. The number of excitons obeys the following equation which can be obtained from Eq.~\eqref{diffusion:1} by integrating it over the sample surface~\cite{PhysRevLett.120.207401}
\begin{equation}
	\label{evolution:N0}
	\frac{dN_0}{dt} = - \frac{N_0}{\tau} - R_A \frac{N_0^2}{2\pi r_0^2}.
\end{equation}
To exclude from the decay of the peak value $n(\bm 0,t)$ the contribution due to the decay of the total exciton population we consider the quantity
\begin{multline}
	\label{evolution:1}
	\left.N_0\frac{\partial }{\partial t} \frac{n}{N_0}\right|_{\bm r=0,t=0} 
	= -\left[ \frac{1}{\tau} \frac{N_0}{\pi r_0^2}  + D \frac{4N_0}{\pi r_0^4}  + R_A \left(\frac{N_0}{\pi r_0^2}\right)^2  + \frac{U_0 {D_0}}{k_B T} \frac{4N_0^2}{\pi^2 r_0^6}\right] 
	+ \frac{1}{\pi r_0^2} \left( \frac{N_0}{\tau} + R_A \frac{N_0^2}{2\pi r_0^2}\right)\\
	= - {D_0}\frac{4N_0}{\pi r_0^4}   - R_A \frac{N_0^2}{2\pi^2 r_0^4}    - \frac{U_0 {D_0}}{k_B T} \frac{4N_0^2}{\pi^2 r_0^6}.
\end{multline}
If we associate the resulting evolution of the renormalized peak density with exciton diffusion then the effective diffusion coefficient (in the limit of $t \rightarrow 0$, corresponding to early times after the excitation) reads
\begin{equation}
	\label{D:eff}
	D_{\rm eff} = -  \frac{\pi r_0^4}{4N_0} \left[N_0\frac{\partial }{\partial t} \frac{n}{N_0}\right]_{\bm r=0,t=0}  = {D_0} + R_A \frac{N_0}{8\pi} + \frac{U_0 {D_0}}{k_B T} \frac{N_0}{\pi r_0^2}.
\end{equation}
The comparison of the analytical expresion~\eqref{D:eff} for the effective diffusion coefficient with numerical results is presented in Fig.~\ref{fig:SI:theory} where panel (a) shows the squared width of the exciton cloud $w^2$  found numerically (points) by fitting the results of solving Eq.~\eqref{diffusion:1} with a Gaussian $\propto \exp{[-r^2/w^2(t)]}$ compared with various approximations. The squared width $w^2$ is connected to the mean squared displacement $\Delta\sigma^2$ according to $w^2 = 2\sigma^2$ with  $\Delta \sigma^2 = \sigma^2-\sigma_0^2$. It is seen that the result of Eq.~\eqref{D:eff} (magenta dot-dashed line) provides reasonably accurate description of the initial expansion. Figure~\ref{fig:SI:theory}(b) shows the effective diffusion coefficient extracted from the initial expansion of the exciton cloud found numerically (points) and analytical expression~\eqref{D:eff} (line) as function of the initial exciton density. The two dependencies agree very well with each other.

\subsection{Discussion of superfluid scenario}\label{superfluid:theo}

The bosonic nature of excitons was predicted to result in their superfluidity at low temperatures and moderate densities in two-dimensions~\cite{1975JETPL..22..274L}, see Ref.~\cite{GlazovSuris_2021} for review. Let us analyze the expansion of the exciton liquid in the superfluid state. If any dissipative processes (exciton-exciton annihilation, lifetime) are disregarded, the continuity equation takes the form
\begin{equation}
	\label{superfluid:cont}
	\frac{\partial n}{\partial t} + \divv \bm j =0,
\end{equation}
where the flux density $\bm j = n\bm v$ ($\bm v$ is the velocity of excitons) is given by~\cite{ll6_eng}
\begin{equation}
	\label{superfluid:curr}
	\frac{\partial j_\alpha}{\partial t} = - \frac{1}{m} \frac{\partial \Pi_{\alpha\beta}}{\partial x_\beta},
\end{equation}
where $\Pi_{\alpha\beta}$ is the momentum flux density tensor, $\alpha$ and $\beta=x,y$ are Cartesian subscripts. 

Integrating Eq.~\eqref{superfluid:cont} over the sample area we see that the total number of excitons $N = \int n(\bm r) d\bm r$ is conserved. We are interested in the expansion of the exciton cloud, to that end we introduce the effective area occupied by the excitons as
\begin{equation}
	\label{R2:def}
	R^2(t)= \frac{1}{N} \int r^2 n(\bm r,t) d\bm r.
\end{equation}
Combining Eqs.~\eqref{superfluid:cont} and \eqref{superfluid:curr} we obtain
\begin{equation}
	\label{2nd:order}
	\frac{\partial^2 n}{\partial t^2} =  {\frac{1}{m}}\frac{\partial^2 \Pi_{\alpha\beta}}{\partial x_\alpha \partial x_\beta} \quad \Rightarrow \quad  N\frac{d^2 R^2(t)}{dt^2} = {\frac{1}{m}} \int r^2\frac{\partial^2 \Pi_{\alpha\beta}}{\partial x_\alpha \partial x_\beta} d\bm r = {\frac{2}{m}}\int \Pi_{\alpha\alpha} d\bm r.
\end{equation}
Taking into account that for the ideal fluid $\Pi_{\alpha\beta} = p\delta_{\alpha\beta} + n m v_\alpha v_\beta$. Since the pressure $p$ is minus derivative of the total (internal) energy of the fluid over the area (in two-dimensions) we have in agreement with Ref.~\cite{Kuznetsov:2020aa}
\begin{equation}
	\label{SF:velocity}
	\frac{d^2 R^2(t)}{dt^2} = 4 \frac{\mathcal E_{tot}}{m N} \quad \Rightarrow \quad R(t) = \sqrt{ \frac{ 2 \mathcal E_{tot}}{m N}} t \approx \sqrt{\frac{2U_0 n}{m}} t,
\end{equation}
where $U_0$ is the interaction constant, $\mathcal E_{tot} =  U_0 n N$. The effective expansion velocity $v_x = (2U_0 n/m)^{1/2}$ corresponds to the speed of sound in the exciton condensate. In that case $\sigma^2 \propto R^2 \propto t^2$ compared to the drift-diffusive scenario with $\sigma^2 \propto t$ described above.

\subsection{Experimental observables}
\subsubsection{Disentanglement of exciton-exciton annihilation and repulsion}

\begin{figure}[ht]
	\centering
	\includegraphics[width=0.8\textwidth]{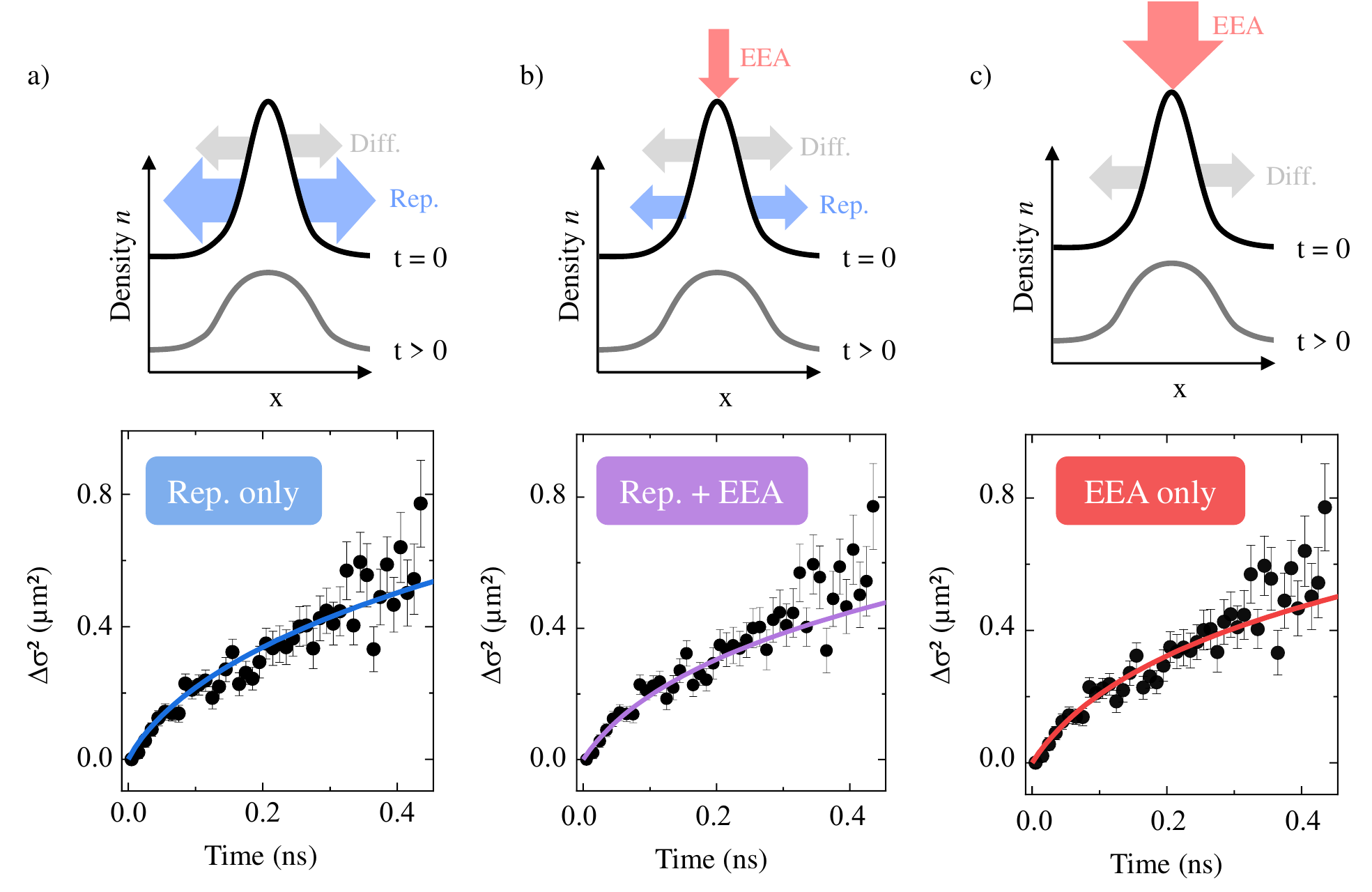}
	\caption{
		(Top row) Schematic illustration of the exciton density distribution starting as a Gaussian at time t=0 and acquiring a super-Gaussian lineshape at later times t>0 as a result of exciton-exciton repulsion (blue), exciton-exciton annihilation (red), or both processes. 
		The broadening due to diffusive propagation is indicated as light gray arrows. 
		(Bottom row) Measured mean squared displacement of interlayer exciton distribution at an initial density of 15\,$n_0$. 
		Fits to the experimental data using a non-linear drift-diffusion model from Eq.\,\eqref{diffusion:1} and taking into account (a) exciton exciton repulsion only (repulsion constant $U_0 D_0/k_B T = 1\times 10^{-11}$~cm$^{4}$/s), (b) equal contributions of repulsion and annihilation (repulsion constant $U_0 D_0/k_B T = 1\times 10^{-11}$~cm$^{4}$/s, Auger coefficient $R_A = 5\times 10^{-3}$~cm$^{2}$/s), and (c) exciton-exciton annihilation only (Auger coefficient $R_A = 10\times 10^{-3}$~cm$^{2}$/s). }
	\label{fig:SimSigma}
\end{figure}

\begin{figure}[ht]
	\centering
	\includegraphics[width=0.7\textwidth]{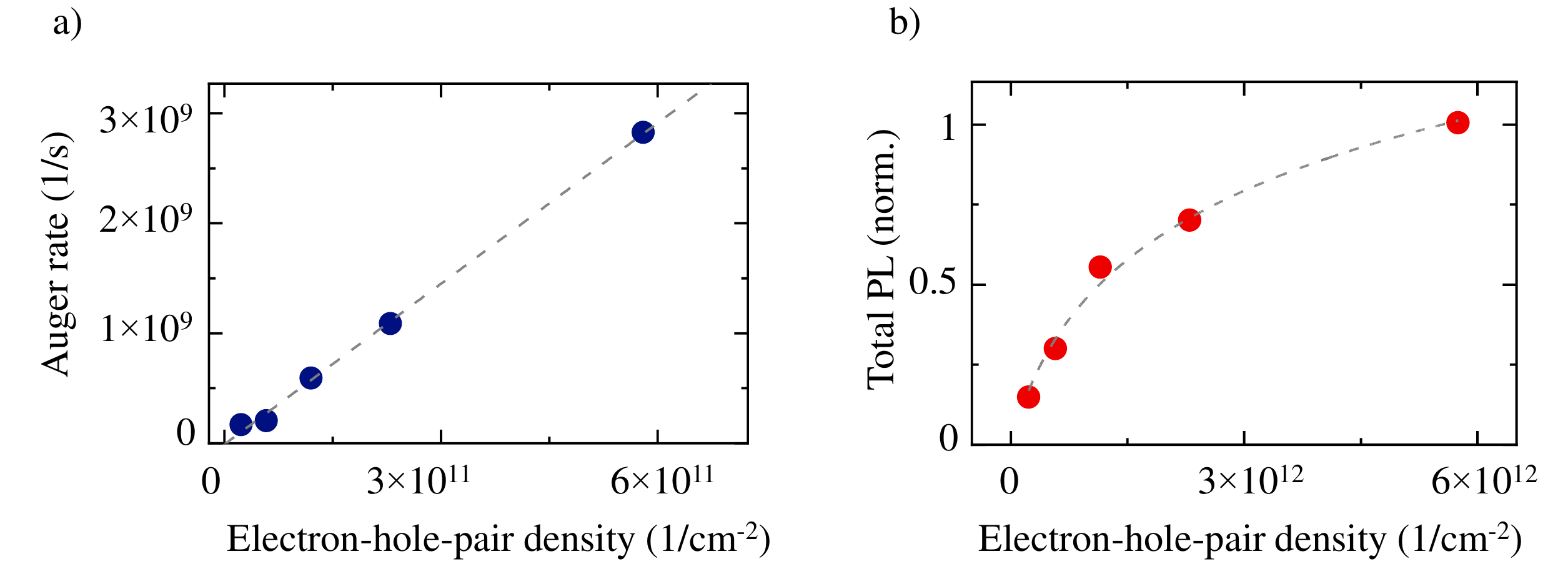}
	\caption{a) Linear increase of recombination rate as function of density, giving a Auger coefficient of $R_A = 5\times10^{-3}$\,cm$^2$/s b) Accompanied saturation of total PL, shown for larger density range, resulting in a Auger coefficient of $R_A = 3\times10^{-3}$\,cm$^2$/s. }
	\label{fig:augerrate}
\end{figure}

As presented in the main manuscript, one is not able to distinguish between the effects of exciton-exciton repulsion or annihilation based on the time-dependent mean squared displacement or effective diffusion coefficients. 
Schematic illustrations of the impact of these processes on the spatial distribution of excitons are presented in the top row of \fig{fig:SimSigma}. 
The black line indicates the initially density distribution at time $t=0$, assuming a Gaussian $\propto\exp(-x^2)$. 
For linear diffusion the distribution would remain Gaussian over time.
However, for both exciton-exciton annihilation and repulsion it evolves over time into a super-Gaussian $\propto\exp(-x^n)$ with $n>2$, i.e., a flat-top\,\cite{Kulig2018}. 
In terms of exciton-exciton\textit{ repulsion} this can be understood as a result of an additional exciton current from the center of the injection area towards the flanks.
In contrast, exciton-exciton \textit{annihilation} leads to a larger reduction of the exciton population with the highest density at the center of the spot.
The change in the shape of the distribution is accompanied by a sub-diffusive behavior of the mean squared displacement that is faster at early times and slows down at later times towards a slope determined by the linear diffusion coefficient. 
Note that in our case subdiffusive behavior appears only at elevated exciton densities therefore we exclude scenarios of non-Markovian exciton dynamics related to population and depopulation of deep traps~\cite{Kurilovich2022}.
This is demonstrated by simulating experimental data using a non-linear drift-diffusion model from Eq.\,\eqref{diffusion:1} and taking into account (a) exciton exciton repulsion only, (b) equal contributions of repulsion and annihilation, and (c) exciton-exciton annihilation only.
The corresponding parameters are:  (a) $U_0 D_0/k_B T = 1\times 10^{-11}$~cm$^{4}$/s, (b) $U_0 D_0/k_B T = 4\times 10^{-12}$~cm$^{4}$/s, $R_A = 5\times 10^{-3}$~cm$^{2}$/s, (c) $R_A = 10\times 10^{-3}$~cm$^{2}$/s.
Importantly, all three scenarios results in the quantitatively similar results.
Disentangling contributions from exciton-exciton annihilation and repulsion thus requires additional analysis of the density- and time-dependent PL. As discussed in Sect.\,\ref{Lifetime}, exciton-exciton annihilation results in the characteristic decrease in PL decay time (linear increase of the recombination rate $r_A=R_A\times n_x$) as a function of density, see \fig{fig:augerrate} a).
This is accompanied by the saturation of the total PL intensity, see \fig{fig:augerrate} b) as extracted from time integrated total luminescence intensities. 
The extracted value for the Auger coefficient from the evaluated Auger rate is $R_A = 5\times10^{-3}$\,cm$^2$/s. Notably the Auger coefficient of  $R_A = 3\times10^{-3}$\,cm$^2$/s as extracted from the PL saturation is close to the one evaluated from the decay rates, though the former method is slightly more accurate \cite{Zipfel2020}.
\\
However, the $R_A$ parameter that is required to reproduce the observed time-dependent mean squared displacement assuming exciton-exciton annihilation as the main origin is twice as large.
It indicates that while the exciton-exciton annihilation contributes substantially to the non-linear effective diffusion, additional mechanisms such as repulsion are required.
\begin{figure}[ht]
	\centering
	\includegraphics[width=0.9\textwidth]{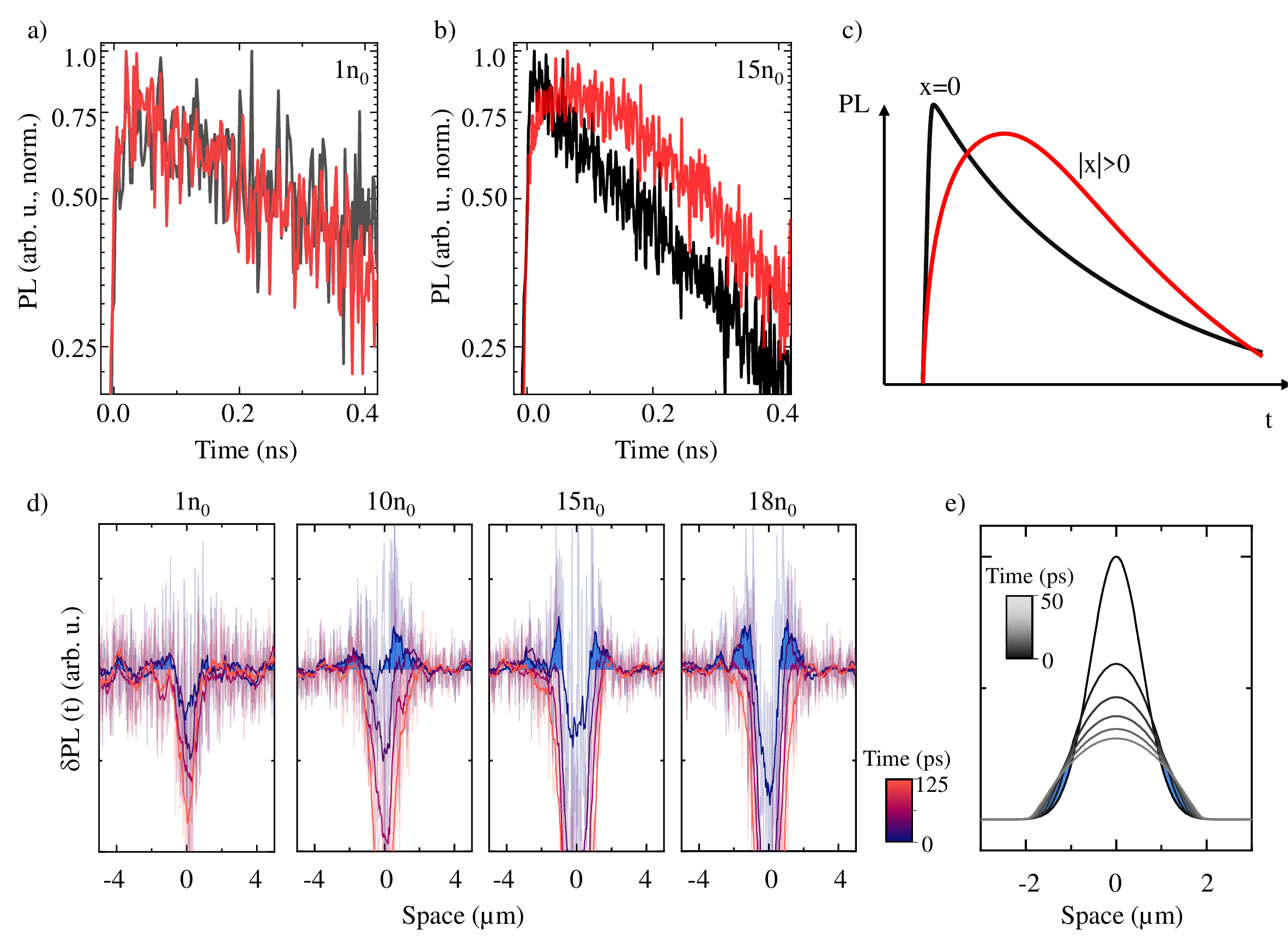}
	\caption{Signatures of exciton-exciton repulsion in differential PL.
		a) PL transients in the middle (black) and on the flank (red) of the spot for the low density regime of $n=$1\,$n_0 = 1.2\times10^{11}$\,cm$^{-2}$ with negligible contributions from repulsion. 
		b) Corresponding data for the high density regime of 15\,$n_0$ with substantial contribution of repulsion.
		c) Schematic illustration of the repulsion-induced delay in the PL transient reaching its maximum.
		d) Differential PL profiles, $\delta PL(x,t) = PL(x,t) - PL(x,0)$, for selected densities.
		Regions with higher relative PL intensity are shaded in light blue. 
		e) Schematic illustration of the impact of exciton-exciton repulsion on the spatial profiles (using a set of simulated profiles evaluated at $t=50$\,ps and for an exaggerated interaction constant of $U_0 D/k_B T = 1\times 10^{-10}$~cm$^{4}$/s.}
	\label{fig:allrep}
\end{figure}
Characteristic signatures of repulsion are identified by analyzing the detected PL response differentially, i.e., in relation to the initial PL signal at $t=0$. 
PL transients in the low and high density regimes are presented in Figs.\,\ref{fig:allrep}\,(a) and (b) for the center ($x=0\,\mu m$) and on the flank ($x=0.5\,\mu m$ and $x=1.5\,\mu m$ relative to the center of the spot), respectively. 
Note that due to the mostly absent drift in the linear regime, PL transients could only be evaluated $x=0.5\,\mu m$ from the spot center with sufficient signal to noise ratio. The increased repulsive action in the nonlinear regime allows for evaluation further away from the spot center, hence the presented flank at $x=1.5\,\mu m$.
Repulsion retains the total number of excitons and leads to their redistribution from the center towards the flanks, as illustrated in Fig.\,\ref{fig:allrep}\,(c). 
At higher densities this is expected to lead to an initial \textit{increase} of the PL intensity in the outside regions, as also observed in the experiment (Fig.\,\ref{fig:allrep}\,(b)).
The outer regions of the spot essentially acquire additional excitons that flow outwards from the center of the spot due to repulsion. 
Moreover, as discussed further below and demonstrated in \fig{fig:delayrep}, the time of the PL transient to reach its maximum increases with both exciton density and distance from the center of the spot.
This contrasts exciton-exciton annihilation, where the main effect is a faster decrease of the exciton population in the center relative to the flanks that does not lead to a spatially-dependent increase of the PL rise time.\\

Alternatively, this behavior is also visualized by evaluating differential PL intensity $\delta PL(x,t) = PL(x,t) - PL(x,0)$ as a function of position $x$ at different times $t$.
This analysis of the experimental data is presented in \fig{fig:allrep}\,(d) for several densities and the first four times ranges of 25\,ps length from 25\,ps up to 125\,ps. 
In the linear regime (with the exciton density $n=1\,n_0$), $\delta PL(x,t)$ is negative for all $x$ and $t$, since the main process determining the absolute density is exciton recombination, i.e., excitons recombine faster than they diffuse from the center of the spot. 
At elevated densities, we observe regions with $\delta PL(t) > 0$ at the flanks of the spot, demonstrating a time-dependent increase in the exciton population and the impact of repulsion, as schematically illustrated in \fig{fig:allrep}\,(e).
These signatures combined, allow us to clearly identify the presence of exciton-exciton repulsion in the studied system. 

\begin{figure}[ht]
	\centering
	\includegraphics[width=\textwidth]{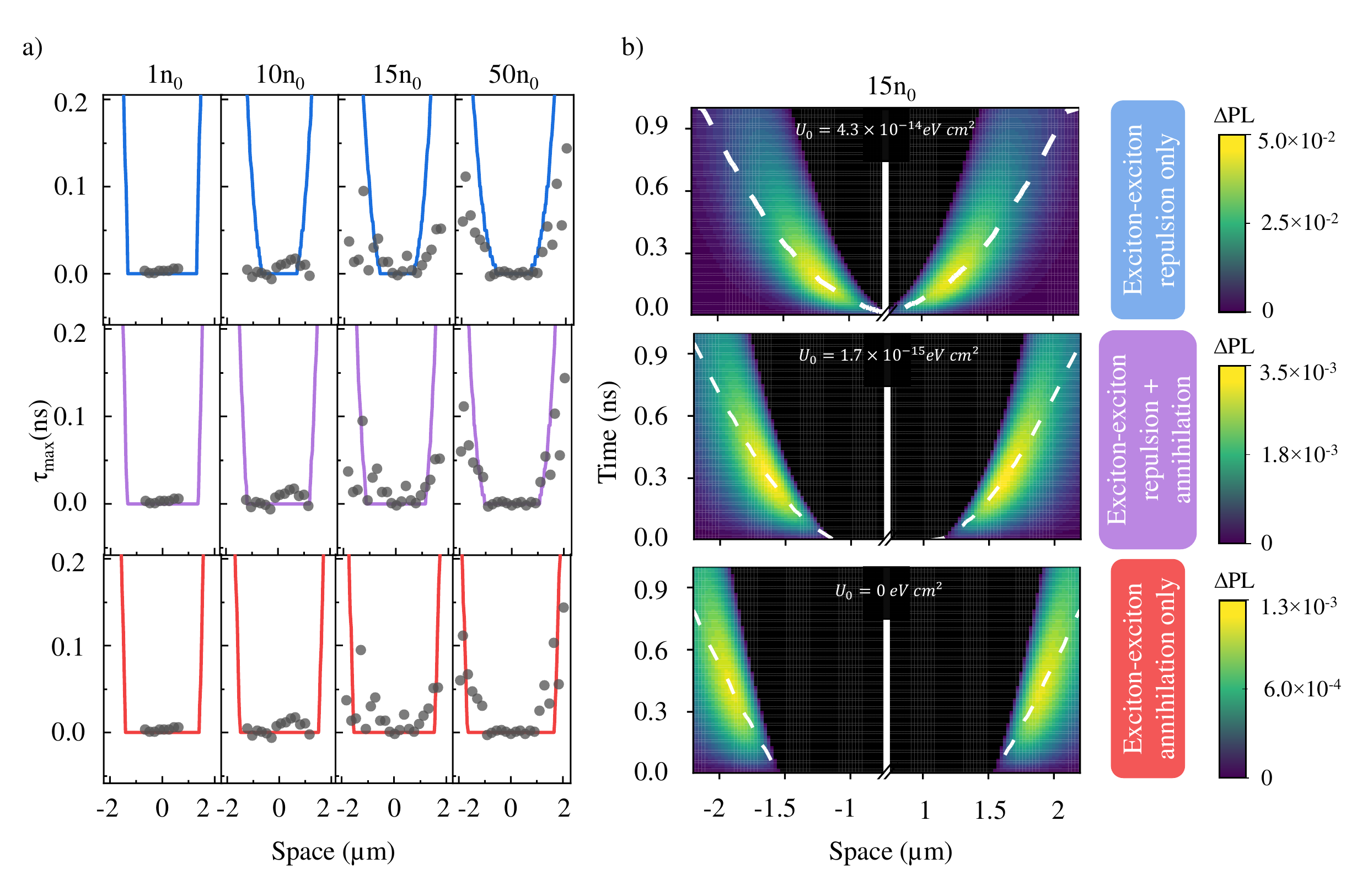}
	\caption{
		a) Extracted time delays for the measured PL transients to reach their respective maxima as function of distance from the center of the spot. 
		The columns correspond to different exciton densities.
		The rows show different simulated curves (solid lines) on top of the same experimental data sets.
		From top to bottom they correspond to simulations taking into account repulsion only (top row, blue), equal contribution of repulsion and annihilation (middle row, purple) and annihilation only (bottom row, red). 
		b) Corresponding 2D plot of $\Delta PL(x,t) = [PL(x,t) - PL(x,0)]/PL_{max}$, to illustrate the position of the time to reach PL maximum, as indicated by the dashed lines.
		The parameters are the same as for the results presented in Fig.\,\ref{fig:SimSigma}.}
	\label{fig:delayrep}
\end{figure}

Importantly, this type of analysis not only distinguishes repulsion from annihilation, but is also used to determine the corresponding repulsion energy constant $U_0$.
To obtain $U_0$, experimentally extracted time delays for the PL to reach its maximum are plotted in \fig{fig:delayrep}\,a) as a function of position for different densities (along the columns).
In each row, results from numerical simulations using Eq.\,\eqref{diffusion:1} are presented including repulsion only (top row), repulsion and annihilation (middle row), and annihilation only (bottom row).
Essentially, they correspond to different values of the repulsion constant $U_0$, as indicated.
Already at one order of magnitude in density above the linear regime, a clear differences in the agreement of the simulated curves with the experimental data becomes visible. 
In particular, for the highest value of the repulsion constant (that would otherwise fit the transient mean squared displacement, see Fig.\,\ref{fig:SimSigma}\,(a)), the simulation overestimates the impact on the spatially-dependent time to reach PL maximum.
In contrast to that, by setting repulsion to zero and including only annihilation and diffusion, the simulated dependence is flat for in a broad region around the center of the spot. 
A good match is then obtained for the repulsion constant $U_0=1.7\times10^{-15}$\,eV\,cm$^2$, that accounts for about half of the sub-diffusive behavior presented in Fig.\,\ref{fig:SimSigma}\,(b) in roughly equal measure with annihilation.
This is further illustrated by the value of the effective diffusivity enhancement $U_0 D_0/(k_B T r_0^2) =6.25\times10^{-4}$\,cm$^2$/s being equal to $R_A/8 =6.25\times10^{-4}$\,cm$^2$/s, see Eq.\,\eqref{D:eff} (with $r_0=0.8\,\mu m$ as used in simulation).

\subsubsection{Discussion in terms of superfluid expansion model}
\label{superfluid}

An interesting prediction for indirect excitons in a TMDC heterostructure is the emergence of a degenerate Bose gas with vanishing viscosity \cite{Fogler2014a}. 
Superfluidity of indirect excitons is expected below the Berezinskii-Kosterlitz-Thouless transition temperature $T_{BKT}$ that can be estimated as
\begin{equation}
	k_B T_{BKT} \approx 1.3 \dfrac{\hbar^2 n_x}{m_x}
\end{equation}
where $n_x$ is the exciton density and $m_x$ is the exciton mass. 
By setting $T_{BKT}$ to be the nominal temperature of the experiment of $T=5$\,K, the superfluidity regime is expected to appear at exciton densities on the order of $10^{11}$\,cm$^{-2}$ or higher, i.e., within the density range in this study.
It is thus useful to consider and test potential interpretation of the observed rapid expansion of the exciton cloud that becomes faster at higher densities in the context of this model.
\begin{figure}[ht]
	\centering
	\includegraphics[width=0.8\textwidth]{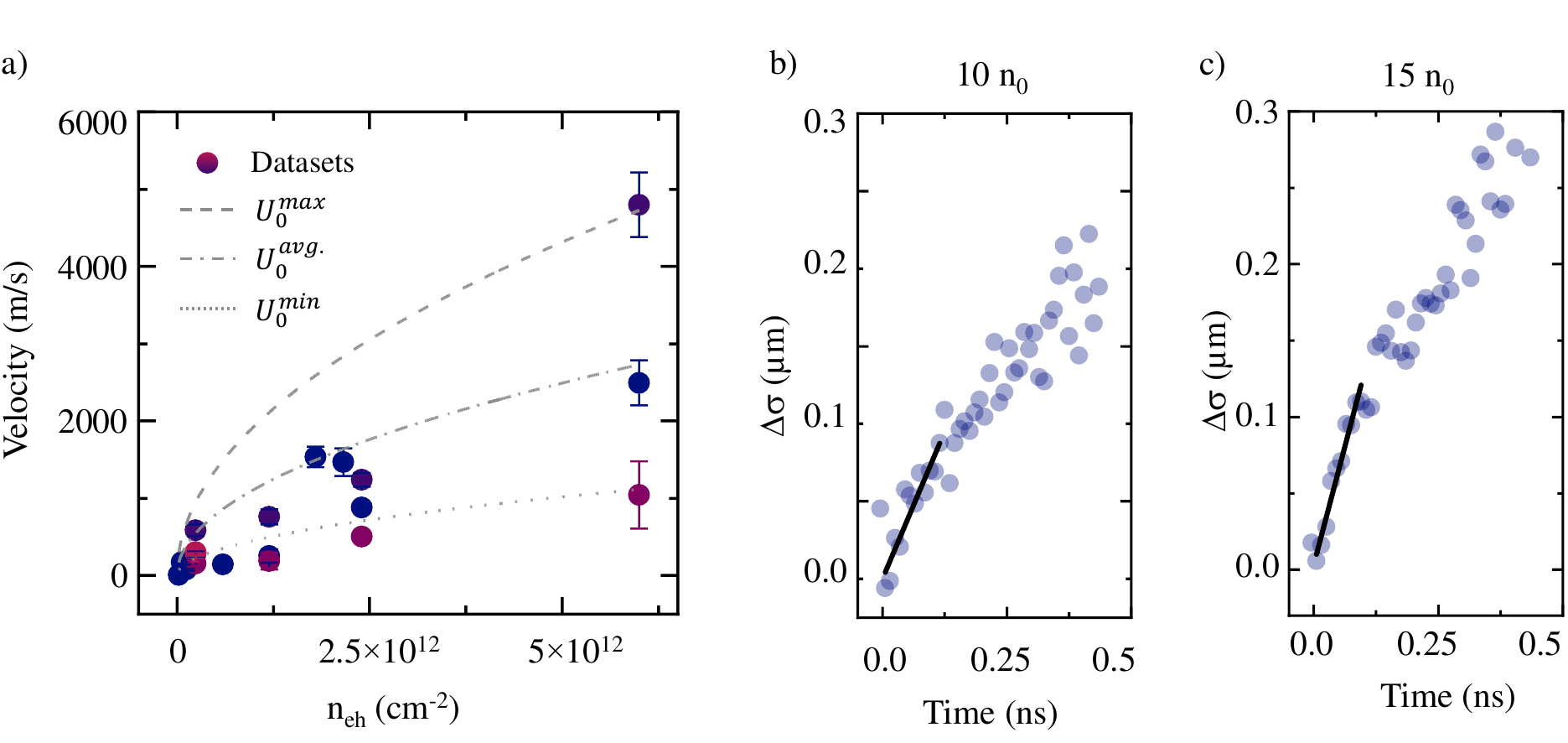}
	\caption{a) Extracted velocity of propagating exciton distribution as a function of injected electron-hole pairs under assumption of ballistic propagation.
		Included are fits according to the superfluid velocity $v_x = (2U_0 n/m)^{1/2}$, see Eq.\,\eqref{SF:velocity}; the three curves correspond to maximum, minimum, and average values of the $U_0$ parameter. 
		b) and c) Broadening of the exciton population as a function of time for a density of 10\,$n_0$ and 15\,$n_0$ respectively. 
		The velocity is obtained from a linear fit at early time below 100\,ps and is shown as a solid line. 
	}
	\label{fig:superfluid}
\end{figure}

We therefore extract the exciton velocity and present it as a function of excitation density in \fig{fig:superfluid} a).
It is obtained from the time-dependent width of the PL profile, $\Delta \sigma (t) = \sigma (t) - \sigma_0$, at early times below 100\,ps, as illustrated in \fig{fig:superfluid} b) and c).
Since $\Delta \sigma (t)$ is smaller than the initial size of the PL spot $\sigma_0$, one cannot unambiguously distinguish diffusive and ballistic propagation.
We also note, that due to the nature of the pulsed experiment and resonant optical excitation into the A-peak of the MoSe$_2$ monolayer, the estimation of injected electron-hole densities from effective absorption, see \fig{fig:Absorption}, should be reasonably accurate up to the $10^{12}$\,cm$^{-2}$ regime.
The extracted velocity increases with density, in close analogy to the analysis using effective diffusion coefficient.
For the studied densities, the absolute values are in the range of a few $10^3$\,m/s. 
They are smaller by several orders of magnitude than those estimated according to $v_x = (2U_0 n/m)^{1/2}$, see Eq.\,\eqref{SF:velocity}, using $U_0$ obtained either from the dipole-dipole interaction model ($434\times10^3$\,m/s), experimental values from energy shifts ($20\times10^3$\,m/s) and repulsion($27\times10^3$\,m/s) or $U_0$ from literature ($90\times10^3$\,m/s \cite{Yuan2020, Sun2022}) (see next section for an overview).

\subsubsection{Interaction constant $U_0$}
\begin{figure}[ht]
	\centering
	\includegraphics[width=\textwidth]{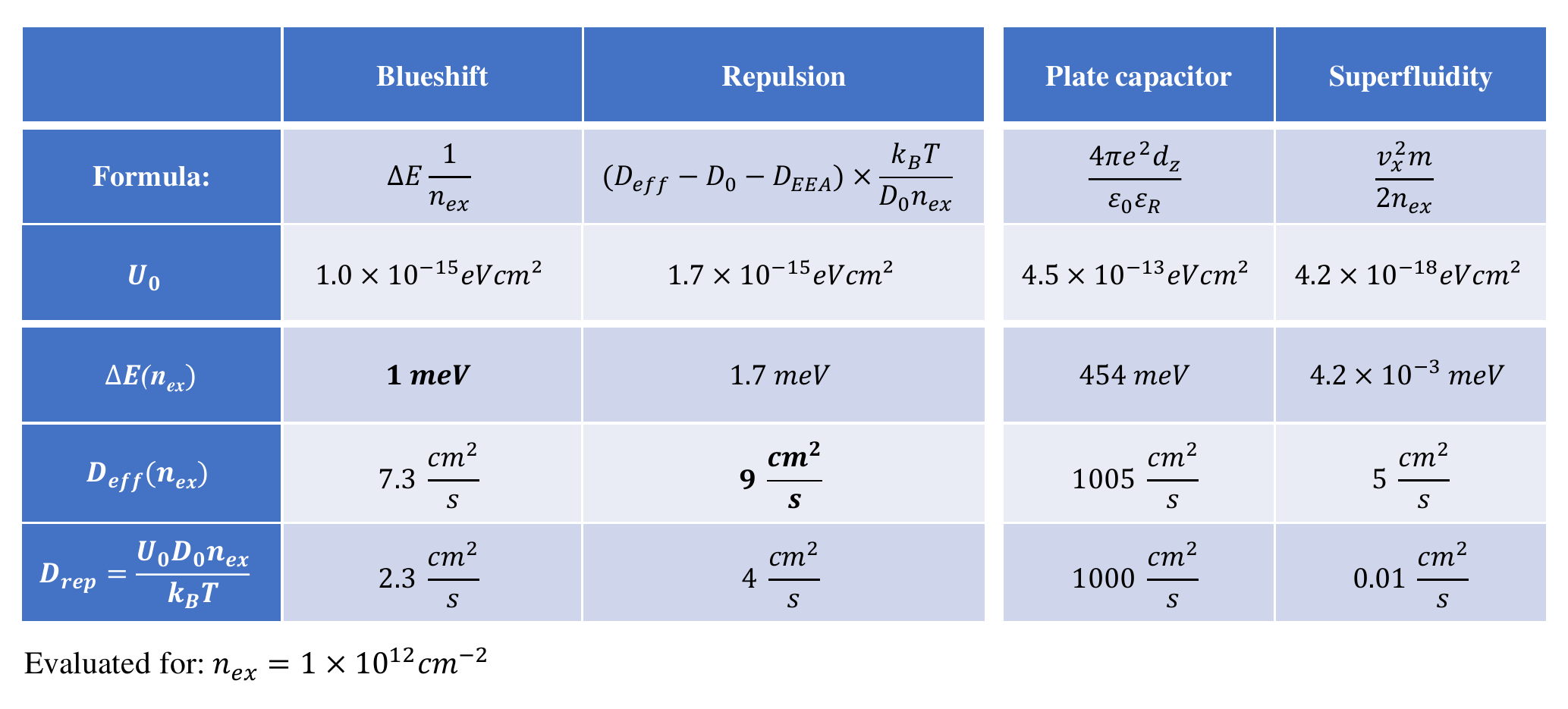}
	\caption{	
		Overview of interaction constants $U_0$, including both measured values and two distinct models.
		The former stem from the analysis of density-dependent exciton energy shift in PL and contribution of repulsion to propagation.
		The latter are extracted either assuming a superfluid scenario and corresponding velocity and in terms of the plate capacitor model for dipole-dipole repulsion (interlayer distance $d_z=1$\,nm and $\varepsilon_R=4.5$ \cite{Kumar2012}). 
		The bottom rows show corresponding, experimentally accessible parameters for an exemplary density of $n_{eh}=10^{12}$\,cm$^{-2}$: energy shift $\Delta E$, effective total diffusion coefficient at short times after the excitation $D_{eff}$ (using annihilation constant of $R_A = 5\times10^{-3}$\,cm$^2$/s), and contribution of the repulsion to the effective diffusion, $D_{rep}$.
		Measured parameters are indicated in bold print.}
	\label{fig:interactiontable}
\end{figure}

The interaction constant $U_0$ governs the contribution of the exciton-exciton repulsion to the non-linear propagation of interlayer excitons.
In the table in \fig{fig:interactiontable} we give an overview of experimentally determined $U_0$ values in our study from spectral shifts and diffusion analysis and place into perspective with estimated values assuming either dipole-dipole repulsion or superfluidity.
The corresponding expressions are given in the first row of the table in \fig{fig:interactiontable}. 
It further includes several experimentally accessible parameters for an exemplary density of $n_{ex}=10^{12}$\,cm$^{-2}$: energy shift $\Delta E$, effective total diffusion coefficient at short times after the excitation $D_{eff}$ (using annihilation constant of $R_A = 5\times10^{-3}$\,cm$^2$/s), and contribution of the repulsion to the effective diffusion, $D_{rep}$.

Importantly, the two values extracted from energy shift and diffusion measurements in the present study agree well with each other, yielding an interaction constant on the order of $10^{-15}$\,eV\,cm$^{2}$. 
This is about one order of magnitude lower that some of the values reported in earlier studies, such as $U_0=1.6\times10^{-14}$\,eV\,cm$^{2}$ \cite{Yuan2020} and  $U_0=2.6\times10^{-14}$\,eV\,cm$^{2}$ \cite{Sun2022}.
We note, that the plate capacitor model was recently shown to present a simplified approximation, since it does not include additional effects such as correlation effects and exchange interaction\,\cite{Erkensten2022}.
Moreover, the extraction of the energy shift from spectra can, in principle, be affected by the interplay of the multi-peak structure of the PL signals and spectral broadening that can both change with increasing density.
For the parameters in our study the capacitor model predicts an interaction constant that is \textit{higher} than experimentally determined ones by more than two orders of magnitude. 
Alternatively, if the scenario of an excitonic superfluid is assumed, the value of $U_0$ would be \textit{smaller} by almost three orders of magnitude, following the discussion in the previous Sect.\,\ref{superfluid}.
Altogether, it means that neither the picture of exclusively dipolar repulsion nor the scenario of superfluidity are appropriate descriptions of the observations.
Here we note that an accurate microscopic calculation of $U_0$ can be demanding and is part of the ongoing work.

\newpage

\section{Extended data}

\subsection{Reproducibility of anomalous diffusion}

\begin{figure}[!htb]
	\centering
	\includegraphics[width=\textwidth]{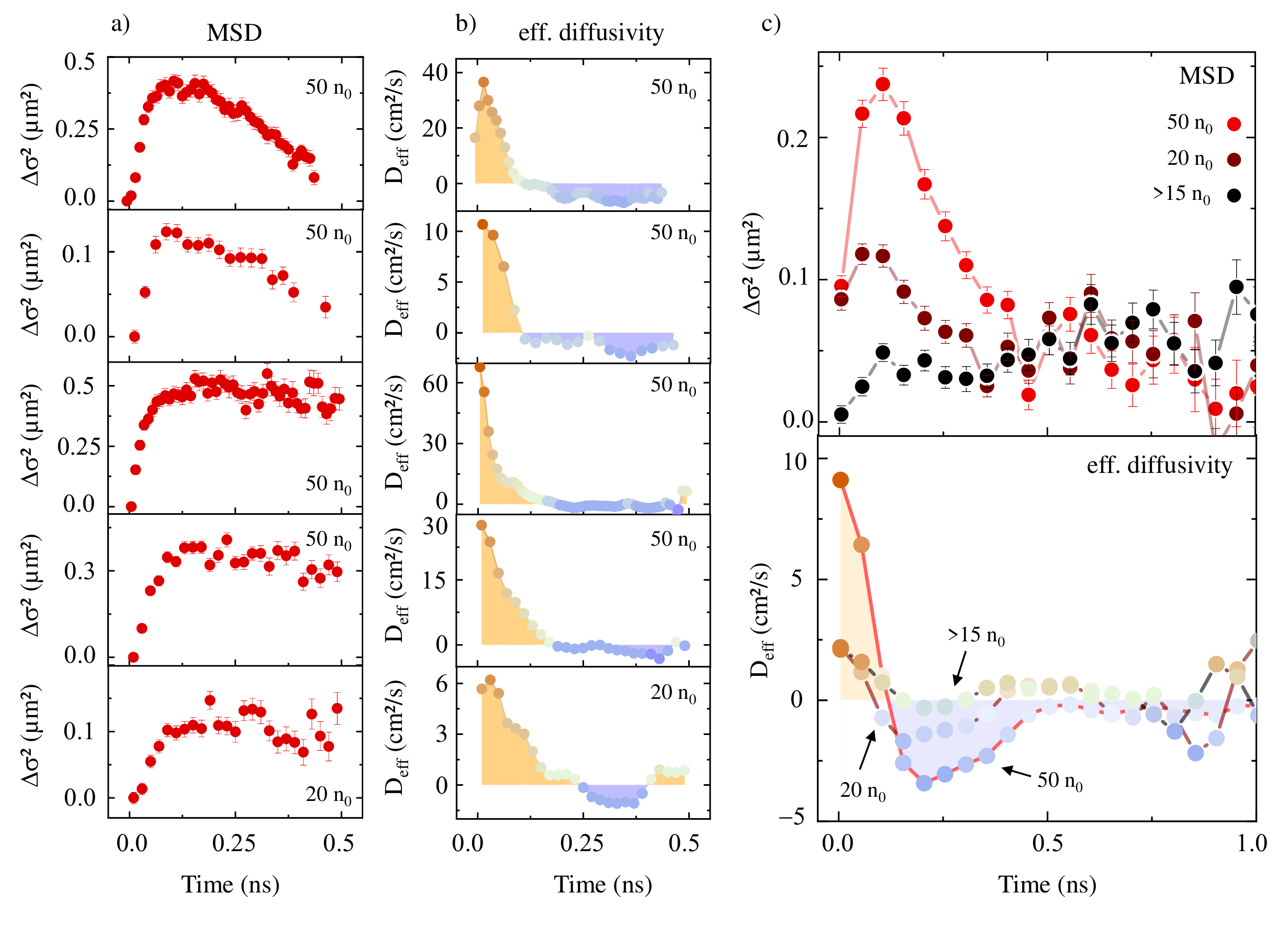}
	\caption{
		a) Mean squared displacement for short time scales of up to 500\,ps 
		b) Corresponding time-dependent, effective diffusion coefficients $D_{\rm{eff}}(t)=\partial\sigma^2(t)/2\partial t$. 	
		The blue shades areas indicate regions of negative effective diffusion coefficients.
		c) Mean squared displacement and d) time-dependent, effective diffusion coefficients for several injection densities, below and above the Mott threshold.}
	\label{fig:NegDiffMSD}
\end{figure}

The anomalous, effectively negative diffusion in the high density regime at the Mott transition is reversible with respect to pump power and a reproducible observation.
A summary of the data collected under these conditions is  presented in \fig{fig:NegDiffMSD} for different measurements and spots on the studied MoSe$_2$/WSe$_2$ sample with $H^h_h$ reconstruction. 
\fig{fig:NegDiffMSD} a) and b) show interlayer exciton mean-squared displacement and effective, time-dependent diffusion coefficients, respectively, for estimated injection densities $n_{ex} \geq 2\times 10^{12}$\,cm$^{-2}$) during the first 500\,ps. The mean squared displacement is given by $\Delta\sigma^2 (t) = \sigma^2(t) - \sigma^2_0$. Typical values for $\sigma_0^2$ are on the order of $0.4\pm0.1$\,$\mu$m$^2$.
Lowest panel correspond to a density just around the Mott threshold with density increasing to the upper panel to $6\times 10^{12}$\,cm$^{-2}$. 
All measurements demonstrate the initially rapid expansion during the first 100\,ps followed by contraction that persists for several 100's of ps. 
The corresponding effective diffusion coefficient initially reaches values of several 10's of cm$^2$/s and then decreases to values below zero down to $-5$\,cm$^2$/s.  \\

The emergence of contracting mean squared displacement and thus effectively negative diffusion coefficients with increasing density is further illustrated in Figs.\,\ref{fig:NegDiffMSD} c) and d). 
It spans the density range of excitation densities being below and above of the estimated threshold for the Mott transition. 
This data shows the sensitivity of the observed behavior to the injection density. 
\begin{figure}[!htb]
	\centering
	\includegraphics[width=0.8\textwidth]{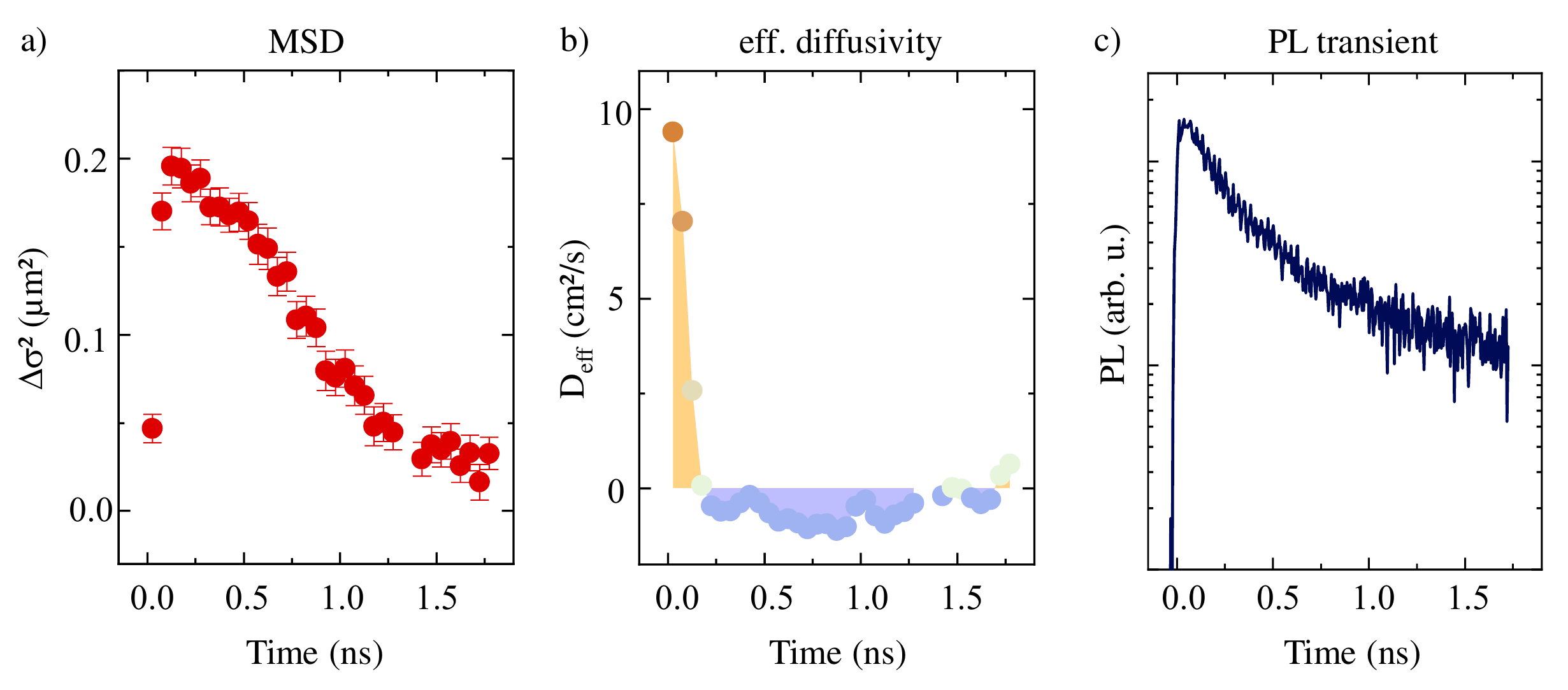}
	\caption{a) Mean squared displacement at Mott transition 50\,$n_0$ up to 1.5\,ns, b) corresponding diffusion coefficient and c) PL transient.}
	\label{fig:NegDifflong}
\end{figure}
Furthermore the contraction is not persistent but fades on timescales of 100\,ps to 700\,ps (depending on the ratio of plasma to excitons in the system.). This is further illustrated in \fig{fig:NegDifflong} a) and b) with a mean-squared displacement and corresponding effective diffusion coefficient respectively for a time window of 1500\,ps. 

\subsection{Density dependent exciton propagation}

\begin{figure}[!htb]
	\centering
	\includegraphics[width=\textwidth]{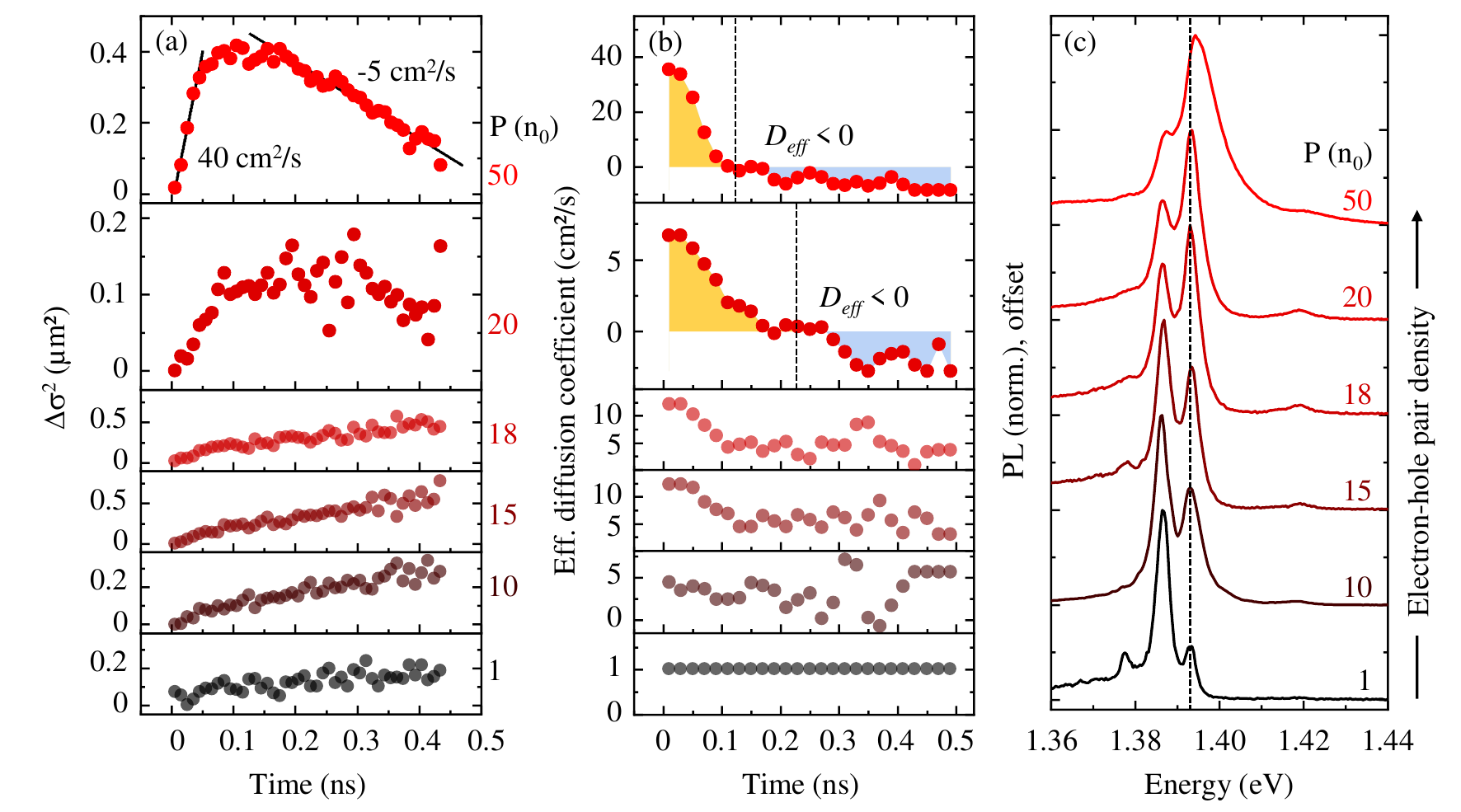}
	\caption{(a) Mean squared displacement of interlayer excitons as function of time for different pump densities in units of $n_0$ = 1.2$\times10^{11}$\,cm$^{-2}$ electron-hole pairs per pulse. 
		The densities range from linear to non-linear regime and up to the Mott transition of exciton ionization. 
		Negative effective diffusion is observed at highest electron-hole pair densities. 
		(b)  Corresponding time-dependent diffusion coefficients and (c) PL spectra, demonstrating only a small shift of the exciton emission and substantial lineshape broadening at the Mott transition.}
	\label{fig:ExtendedPower}
\end{figure}

The extended data from density dependent measurements is presented in \fig{fig:ExtendedPower} to show the gradual changes of a) mean squared displacement, b) time-dependent effective diffusion coefficient, and c) time-integrated PL spectra.
The estimated electron-hole-pair density spans from the linear regime of exciton diffusion up to the Mott transition of exciton ionization. 
Pump densities are given in units of $n_0$ = 1.2$\times10^{11}$\,cm$^{-2}$ electron-hole pairs per pulse and the Mott threshold is expected at densities around 20\,$n_0$. \\

With increasing density, from 1\,$n_0$ up to 18\,$n_0$ the gradual evolution of nonlinear mean squared displacement and effective diffusion can be clearly seen in the \fig{fig:ExtendedPower} a) and b), due to the interplay of exciton repulsion and annihilation.
Upon  reaching the Mott threshold the behavior changes drastically as negative effective diffusion emerges, accompanied by a substantial lineshape broadening. 
After crossing the Mott threshold the effective negative diffusion becomes more substantial and reaches an earlier onset. 
\newpage

\section{Theoretical descriptions of the high density regime}

\subsection{Estimation of Mott transition density from bandgap renormalization}

In this section we describe the theoretical approach to account for many-body effects at finite electron-hole densities in a TMDC heterobilayer and estimate the transition from bound excitons to dense electron-hole plasma.
At given temperature and density, a gas of photoexcited electron-hole pairs establishes a certain balance between excitons and unbound carriers. 
This balance can be approximately understood in the so-called chemical picture \cite{semkat_ionization_2009, steinhoff_exciton_2017} as an adaption of the chemical potentials of electrons, holes and excitons.
The fraction of electron-hole pairs bound as excitons is essentially determined by the ratio of the exciton binding energy, defined as the difference between quasi-particle band gap and exciton energy, to the thermal energy. 
With increasing excitation density, both the quasi-particle gap and the exciton energy are renormalized due to carrier-carrier interaction.
In particular, this leads to an effective reduction of the exciton binding energy at increasing electron-hole pair densities. 
Around the density where the binding energy vanishes, a transition from coexisting bound and unbound electron-hole pairs to a fully ionized plasma takes place. 
Hence, this so-called excitonic Mott transition can be approximately identified with the quasi-particle band gap energy being equal to the absolute energy of the exciton resonance. 
We therefore estimate the exciton Mott density by considering a plasma of unbound electrons and holes and computing the density-dependent bandgap to compare it to the exciton energy. To this end, we combine material-realistic band structures and Coulomb interaction matrix elements from first principles with a many-body theory for photoexcited carriers based on nonequilibrium Green functions.
\\
\\\textbf{Density functional theory calculations, spin-orbit coupling and Coulomb matrix elements}

\begin{figure}
	\centering
	\includegraphics[width=.8\columnwidth]{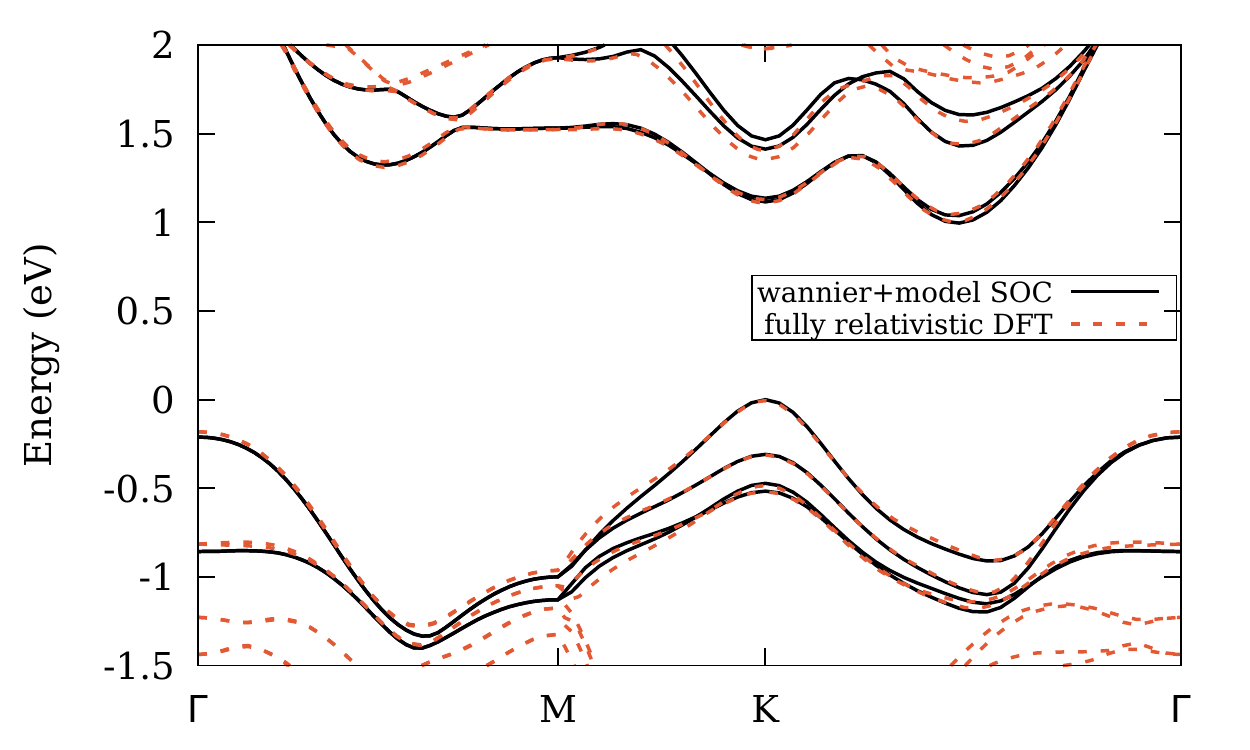}
	\caption{Band structure of freestanding MoSe$_2$/WSe$_2$ heterobilayer as obtained from a fully relativistic DFT calculation (dashed red lines) compared to a non-relativistic Wannier construction augmented by a $\boldsymbol{L\cdot S}$-Hamiltonian (solid black lines). 
		In the case of the Wannier construction, we show only the bands around the fundamental gap used for the many-body description of the photoexcited carriers.}
	\label{fig:DFT_bands}
\end{figure}

\begin{figure}
	\centering
	\includegraphics[width=.8\columnwidth]{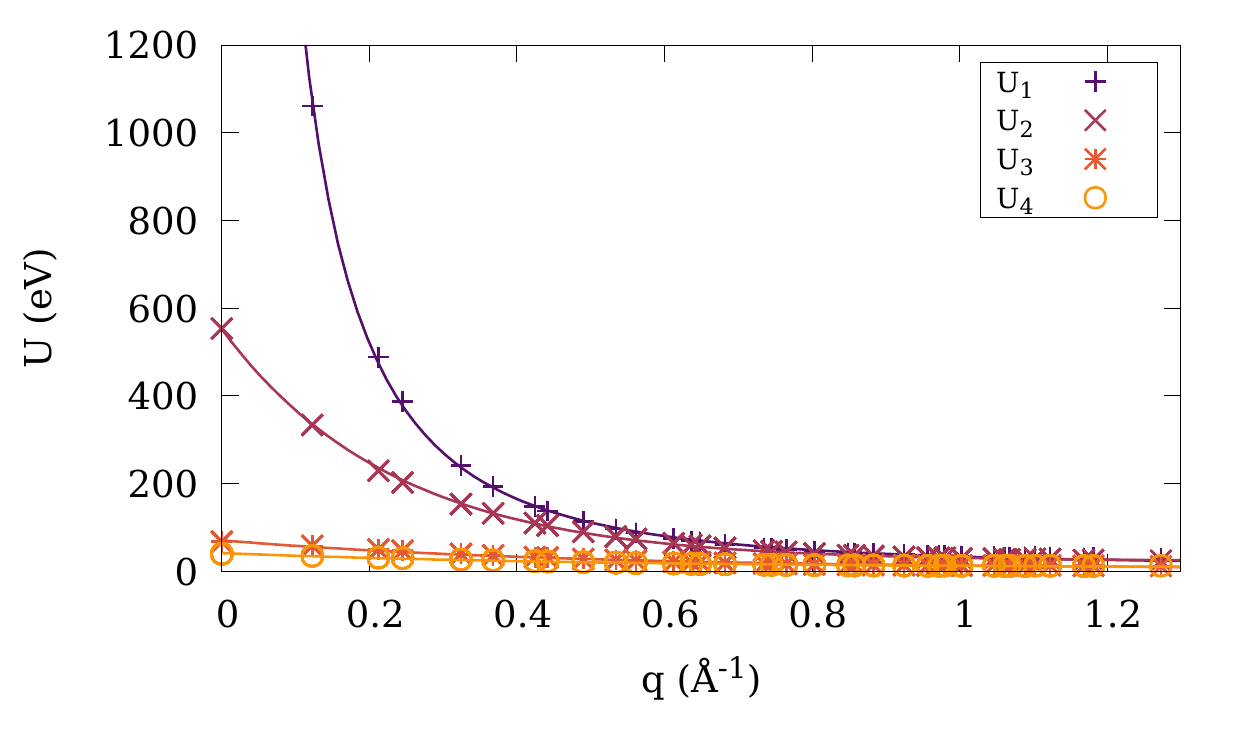}
	\caption{Leading four eigenvalues of the bare Coulomb matrix for a freestanding MoSe$_2$/WSe$_2$ heterobilayer (symbols) and analytic fit functions (solid lines) as discussed in the text.}
	\label{fig:U_fit}
\end{figure}

\begin{figure}
	\centering
	\includegraphics[width=.8\columnwidth]{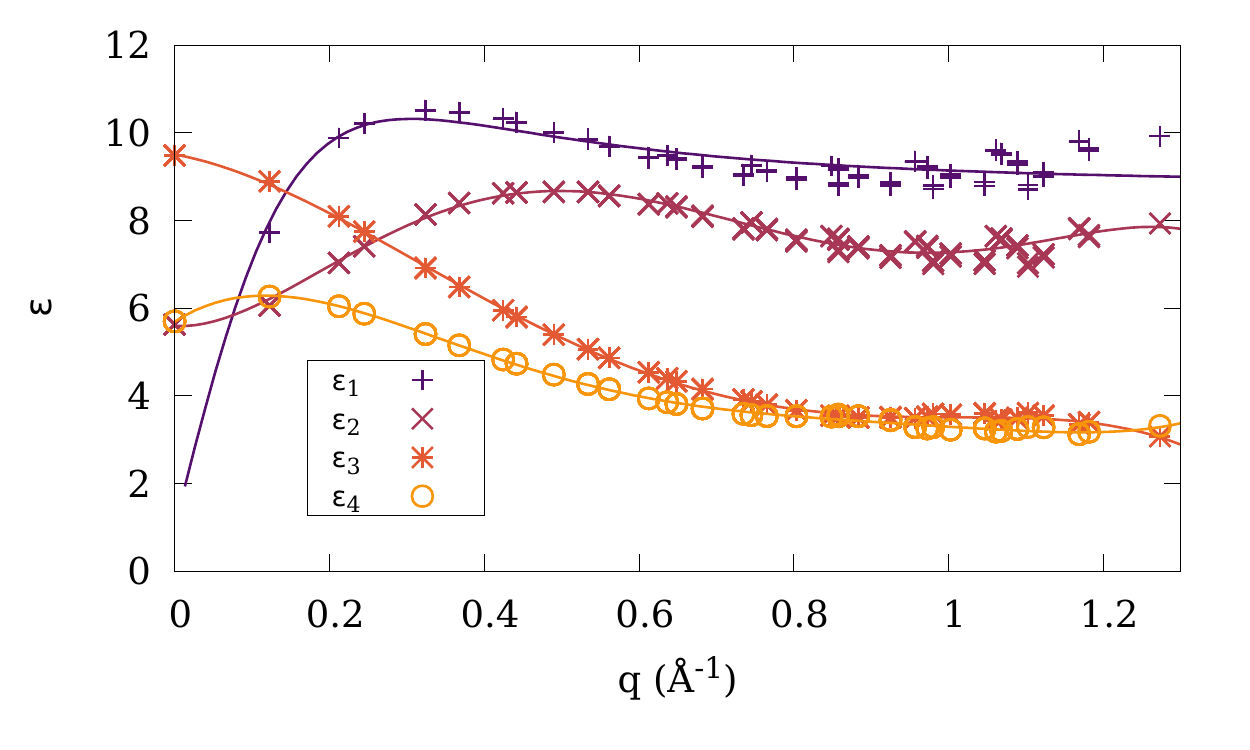}
	\caption{Leading four eigenvalues of the dielectric matrix for a freestanding MoSe$_2$/WSe$_2$ heterobilayer (symbols) and analytic fit functions (solid lines) as discussed in the text.}
	\label{fig:eps_fit}
\end{figure}

\begin{figure}
	\centering
	\includegraphics[width=.8\columnwidth]{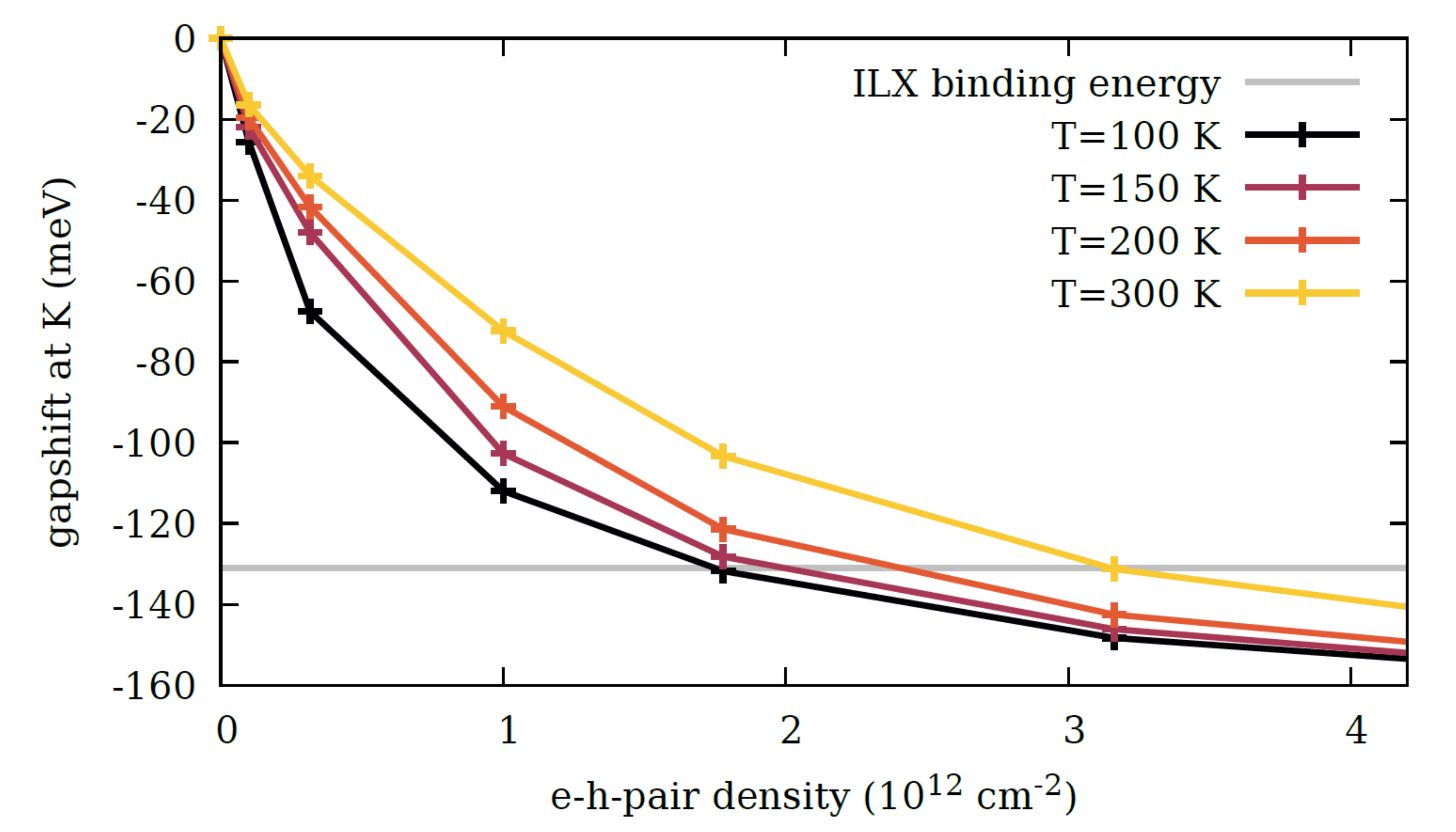}
	\caption{Temperature and excitation-density dependence of the bandgap energy shift in an hBN-encapsulated, $H^h_h$ reconstructed MoSe$_2$/WSe$_2$ heterostructure. 
		For comparison, the absolute energy of the bright inter-layer exciton (ILX) is shown, based on the low-density value and considering negligible energy shifts on the order of a few meV at the highest studied density (c.f.Fig.\,\ref{fig:ExtendedPower}\,(c).}
	\label{fig:gapshift_results}
\end{figure}

Density functional theory (DFT) calculations for a freestanding H$^h_h$ MoSe$_2$/WSe$_2$ heterobilayer are carried out using QUANTUM ESPRESSO V.6.6 \cite{giannozzi_quantum_2009, giannozzi_advanced_2017}. 
We apply the generalized gradient approximation (GGA) by Perdew, Burke, and Ernzerhof (PBE) \cite{perdew_generalized_1997} and use optimized norm-conserving Vanderbilt pseudopotential~\cite{van_setten_pseudodojo:_2018} at a plane-wave cutoff of $80$~Ry. 
Uniform meshes (including the $\Gamma$-point) with $18\times18\times1$ k-points are combined with a Fermi-Dirac smearing of $5$~mRy. 
Using a fixed lattice constant of $a=3.29$~\AA\, \cite{gillen_interlayer_2018} and a fixed cell height of $35$~\AA, forces are minimized below $5\cdot 10^{-3}$~eV/\AA. The D3 Grimme method~\cite{grimme_consistent_2010} is used to include van-der-Waals corrections.
We use RESPACK \cite{nakamura_respack:_2021} to construct a lattice Hamiltonian $H_0(\bk)$ in a 22-dimensional localized basis of Wannier orbitals (d$_{z^2}$, d$_{xz}$, d$_{yz}$, d$_{x^2-y^2}$ and d$_{xy}$ for Mo and W, respectively, p$_x$, p$_y$ and p$_z$ for Se) from the DFT results. 
We also calculate the dielectric function as well as bare and screened Coulomb matrix elements in the localized basis. 
For the polarization function, a cutoff energy of $8$~Ry, $192$~bands as well as $70$~frequency points on a logarithmic grid are taken into account.
The values of the bare and screened Coulomb interaction are extrapolated from vacuum heights of $35$~\AA\,to $55$~\AA\,.
Spin-orbit interaction is included using an on-site $\boldsymbol{L\cdot S}$-coupling Hamiltonian, which is added to the non-relativistic Wannier Hamiltonian:
\begin{equation}
	\begin{split}
		H(\bk)=I_2 \otimes H_0(\bk)+H_{\textrm{SOC}}\,.
	\end{split}
	\label{eq:H_tot}
\end{equation}
Here, $I_2$ is the $2\times2$ identity matrix in the Hilbert space spanned by eigenstates $\ket{\uparrow}$ and $\ket{\downarrow}$ of the spin z component (perpendicular to the bilayer). 
We assume that the Coulomb matrix in Wannier representation is spin-independent and that spin-up and spin-down states are not mixed.
Diagonalization of $H(\bk)$ yields the band structure $\varepsilon_{\bk}^{\lambda}$ and the Bloch states $\ket{\psi_{\bk}^{\lambda}}=\sum_{\alpha} c^{\lambda}_{\alpha,\bk}\ket{\bk,\alpha} $, where the coefficients $c^{\lambda}_{\alpha,\bk}$ describe the momentum-dependent contribution of the orbital $\alpha$ to the Bloch band $\lambda$.
The Bloch sums $\ket{\bk,\alpha}$ are connected to the localized basis via $\ket{\bk,\alpha}=\frac{1}{\sqrt{N}}\sum_{\bR}e^{i\bk\cdot\bR}\ket{\bR,\alpha}$ with the number of unit cells $N$ and lattice vectors $\bR$.
The SOC-Hamiltonian is given by 
\begin{equation}
	\begin{split}
		H_{\textrm{SOC}}=\frac{1}{\hbar^2}\boldsymbol{\tilde{L}\cdot S}=\frac{1}{2\hbar}\boldsymbol{\tilde{L}\cdot \sigma}
	\end{split}
	\label{eq:H_SOC}
\end{equation}
with the Pauli matrices $\boldsymbol{\sigma}=(\sigma_x,\sigma_y,\sigma_z)$ and the angular momentum operator
\begin{equation}
	\boldsymbol{\tilde{L}}=
	\begin{pmatrix}
		\lambda_{\textrm{Mo}}\boldsymbol{L}_{l=2} & 0   &  0 &  0 &  0 &  0 \\
		0 & \lambda_{\textrm{Se}}^{(1)}\boldsymbol{L}_{l=1}   & 0 &  0 &  0 &  0 \\
		0 & 0 &  \lambda_{\textrm{Se}}^{(1)}\boldsymbol{L}_{l=1} &  0 &  0 &  0 \\
		0 & 0 & 0 & \lambda_{\textrm{W}}\boldsymbol{L}_{l=2} & 0   &  0 \\
		0 & 0 & 0 & 0 & \lambda_{\textrm{Se}}^{(2)}\boldsymbol{L}_{l=1}   & 0 \\
		0 & 0 & 0 & 0 & 0 &  \lambda_{\textrm{Se}}^{(2)}\boldsymbol{L}_{l=1}
	\end{pmatrix}\,.
	\label{eq:L_op}
\end{equation}
The latter contains intra-atomic coupling parameters $\lambda_{\textrm{Mo}}$ and $\lambda_{\textrm{W}}$ for the $(l=2)$-subspace (d$_{z^2}$, d$_{xz}$, d$_{yz}$, d$_{x^2-y^2}$ and d$_{xy}$) and $\lambda_{\textrm{Se}}^{(i)} $ for the $(l=1)$-subspace (p$_x$, p$_y$ and p$_z$).
We assume that the effective coupling parameters for the Se-atoms are different for the Mo- ($i=1$) and the W-($i=2$)based layers. 
The angular momentum operators in the localized basis are given in Ref.~\cite{steinhoff_microscopic_2021}.
The coupling constants are chosen as $\lambda_{\textrm{Mo}}=102$ meV, $\lambda_{\textrm{W}}=318$ meV, $\lambda_{\textrm{Se}}^{(1)}=85$ meV and $\lambda_{\textrm{Se}}^{(2)}=-247$ meV to reproduce the spin-orbit splittings at the K-point, as obtained from DFT calculations including spin-orbit coupling using fully relativistic pseudopotentials.
Fig.~\ref{fig:DFT_bands} shows the good agreement between bands around the fundamental gap obtained directly from fully relativistic DFT calculations and those from
diagonalization of the spin-augmented Wannier Hamiltonian (\ref{eq:H_tot}).
\\Starting from the density-density-like bare Coulomb interaction matrix elements in the Wannier basis,
\begin{equation}
	\begin{split}
		U_{\alpha\beta}(\bq)=\sum_{\bR}e^{i\bq\cdot\bR}U_{\alpha\beta\beta\alpha}(\bR)
		=\sum_{\bR}e^{i\bq\cdot\bR}\bra{\boldsymbol{0},\alpha}\bra{\bR,\beta} U(\br,\br') \ket{\bR,\beta}\ket{\boldsymbol{0},\alpha}\,,
	\end{split}
	\label{eq:U}
\end{equation}
and the corresponding (statically) screened matrix elements $V_{\alpha\beta}(\bq)$, we obtain an analytic description of Coulomb interaction in freestanding TMDC heterobilayers that can be augmented by screening from a dielectric environment\cite{schonhoff_interplay_2016,steinhoff_exciton_2017,steinke_coulomb-engineered_2020}. 
To this end, we diagonalize the bare Coulomb matrix $\boldsymbol{U}(\bq)$ to obtain eigenvalues $U_i(\bq)$ and eigenvectors $\boldsymbol{e}_i(\bq)$. 
The TMDC bilayer is embedded in a three-dimensional unit cell with an effective vacuum distance $h_{\textrm{vac}}$ separating adjacent bilayers. 
The Coulomb interaction between two slabs decreases inversely proportional to $h_{\textrm{vac}}$. 
Therefore, we extrapolate the bare Coulomb eigenvalues to infinite vacuum distance using three different distances of 35 \AA\,, 45 \AA\, and 55 \AA\, and assuming a linear dependence on $h_{\textrm{vac}}^{-1}$.
The leading four eigenvalues are shown in Fig.~\ref{fig:U_fit}.
All further values are of similar size as $U_4(\bq)$. For the analytic description of the leading eigenvalue, we use
\begin{equation}
	\begin{split}
		U_1(q)=\frac{n_{\textrm{orb}} e^2}{2\varepsilon_0 A_{\textrm{UC}}}\frac{1}{q(1+\gamma q + \delta q^2 + \eta q^3)}\,,
	\end{split}
	\label{eq:U_1}
\end{equation}
where the area of the hexagonal unit cell $A_{\textrm{UC}}=\frac{\sqrt{3}}{2}a^2$ and the number of orbitals $n_{\textrm{orb}}=22$ ensure the proper normalization of the Coulomb matrix elements. 
The eigenvalues $U_2(q)$ through $U_{22}(q)$ are fitted by exponential functions
\begin{equation}
	\begin{split}
		U_i(q)=a_i\,\textrm{exp}(-b_i q)+c_i\,.
	\end{split}
	\label{eq:U_n}
\end{equation}
The matrix elements of the screened interaction $\boldsymbol{V}(q)$ in the eigenbasis of the bare interaction are then obtained via
\begin{equation}
	\begin{split}
		V_i(q)=\varepsilon^{-1}_i(q)\,U_i(q)\,,
	\end{split}
	\label{eq:V_from_U}
\end{equation}
where the dielectric matrix $\boldsymbol{\varepsilon}(q)$ accounts for both the material-specific internal polarizability and the screening by the environment. 
As for the bare Coulomb matrix, we extrapolate the dielectric matrix eigenvalues to infinite vacuum distance.
First, we introduce an analytic description for the freestanding bilayer dielectric function, i.e. in the absence of external screening.
While the eigenvalues $\varepsilon_2(q)$ through $\varepsilon_{22}(q)$ are well described by fifth-order polynomials, the leading eigenvalue is expressed by
\begin{equation}
	\begin{split}
		\varepsilon_1(q)=\varepsilon_{\infty}(q)\frac{1-\beta^2 e^{-2qd}}{1+2\beta e^{-qd}+\beta^2 e^{-2qd}}\,,
	\end{split}
	\label{eq:eps_1}
\end{equation}
with $\beta= \frac{\varepsilon_{\infty}(q)-1}{\varepsilon_{\infty}(q)+1}$ \cite{rosner_wannier_2015} and the bulk dielectric constant given by a modified Resta model \cite{resta_thomas-fermi_1977}
\begin{equation}
	\begin{split}
		\varepsilon_{\infty}(q)=\frac{a+q^2}{\frac{a\, \textrm{sin}(qc)}{qbc}+q^2}+e \,.
	\end{split}
	\label{eq:eps_inf}
\end{equation}
As layer thickness, we use the distance $d=0.99$ nm between the outermost Se atoms in our unit cell (corresponding distance between the Mo and W atoms within the unit cell is 0.65\,nm).
The leading four eigenvalues are shown in Fig.~\ref{fig:eps_fit}.
After all fitting parameters are obtained, environmental screening can be taken into account according to the Wannier function continuum electrostatics approach \cite{rosner_wannier_2015}.
It combines a macroscopic electrostatic model for the screening by the dielectric environment with a localized description of Coulomb interaction.
The leading eigenvalue $\varepsilon_1(q)$, which is most sensitive to macroscopic screening, is modified by replacing the dielectric function of a freestanding bilayer 
with that of an arbitrary vertical heterostructure. 
The latter is obtained by solving Poisson's equation for a test charge in a slab with thickness $d$ and dielectric  function $\varepsilon_{\infty}(q)$ embedded in a z-dependent dielectric profile \cite{florian_dielectric_2018}.
After calculating the eigenvalues $V_i(q)$, the Coulomb matrix in the Wannier basis is obtained by using the eigenvectors $\boldsymbol{e}_i(\bq)$. Due to the complex-valuedness of the eigenvectors at finite momenta, the matrix elements acquire an imaginary part.
However, we find that the imaginary part is significantly smaller than the real part, in particular at long wavelength (small momenta). 
To simplify the many-body description discussed in the following section, we discard the imaginary contributions to retain real-valued matrix elements.
\\We finally compute screened Coulomb matrix elements in the Bloch-state representation by a unitary transformation using the coefficients $c^{\lambda}_{\alpha,\bk}$:
\begin{equation}
	\begin{split}
		V^{\lambda,\nu,\nu',\lambda'}_{\bk_1, \bk_2, \bk_3, \bk_4} = \frac{1}{N}\sum_{\alpha, \beta} \big( c_{\alpha, \bk_1}^{\lambda} \big)^{*} \big( c_{\beta, \bk_2}^{\nu} \big)^{*} V^{\alpha\beta}_{\bk_1-\bk_4}c_{\beta, \bk_3}^{\nu'} c_{\alpha, \bk_4}^{\lambda'}  \, ,
	\end{split}
	\label{eq:Coul_ME}
\end{equation}
where $\bk_4=\bk_1+\bk_2-\bk_3+\bG$ accounts for momentum conservation and the factor $1/N$ stems from the normalization of Wannier functions. 
Note that $V^{\alpha\beta}_{\bq}$ is periodic with respect to reciprocal lattice vectors $\bG$.
In the following, we pull $1/(N A_{\textrm{UC}})=1/\mathcal{A}$ out of the matrix elements. 
\\Following the discussion in \cite{gong_magnetoelectric_2013}, we assume that Bloch states are approximately spin-diagonal. 
We assign a definite spin to each band according to the dominant contribution given by the coefficients $c^{\lambda}_{\alpha,\bk}$. 
Furthermore, we make use of the fact that 
Coulomb interaction is spin-conserving, so that we can set Coulomb matrix elements $V^{\lambda,\nu,\nu',\lambda'}_{\bk_1 \bk_2 \bk_3 \bk_4}$ to zero if $\lambda$ and $\lambda'$ or $\nu$ and $\nu'$ belong to different spins.
\\
\\\textbf{Bandgap renormalization}
\\
\\To compute changes of the quasiparticle bandgap energy on a GW level we consider the Schwinger-Keldysh self-energy\,\cite{kremp_quantum_2005}:
\begin{equation}
	\begin{split} 
		\Sigma(1,1')=\Sigma^{\textrm{H}}(1,1')+\Sigma^{\textrm{GW}}(1,1')
		=-i\hbar\int d2\,V(1,2)G(2,2^+)\delta(1,1')+i\hbar G(1,1')W(1',1)\,,
		\label{eq:selfenergy}
	\end{split}
\end{equation}
where $1$ is a combined index for the position $\boldsymbol{r}_1$, the spin $\sigma_1$ and the time $t_1$ on the Keldysh contour.
To evaluate excitation-induced renormalization effects in a two-dimensional crystal in quasi-equilibrium, the retarded component of the self-energy is transformed to the Bloch representation and into the frequency domain.
Assuming that the photoexcited electrons and holes can be described as quasi-particles with an interaction-induced finite lifetime, renormalized quasi-particle energies are then given by the self-consistency relation:
\begin{equation}
	\begin{split} 
		E^{\lambda}_{\bk}&=\varepsilon_{\bk}^{\lambda}+\Sigma_{\bk}^{\textrm{H},\lambda}+\textrm{Re}\,\Sigma_{\bk}^{\textrm{GW},\textrm{ret},\lambda}(E^\lambda_{\bk})\\
		&=\varepsilon_{\bk}^{\lambda}+\Sigma_{\bk}^{\textrm{H},\lambda}+\Sigma_{\bk}^{\textrm{F},\lambda}+\textrm{Re}\,\Sigma_{\bk}^{\textrm{MW},\textrm{ret},\lambda}(E^{\lambda}_{\bk})\,.
		\label{eq:GW_energy}
	\end{split}
\end{equation}
The corresponding quasi-particle damping energies are 
$\Gamma^{\lambda}_{\bq}=-\textrm{Im}\,\Sigma_{\bk}^{\textrm{MW},\textrm{ret},\lambda}(E^{\lambda}_{\bk})$.
We have split the GW self-energy into an instantaneous Fock term and the so-called Montroll-Ward term according to the decomposition of the 
retarded screened Coulomb interaction matrix \cite{kremp_quantum_2005}:
\begin{equation}
	\begin{split} 
		W^{\textrm{ret}}_{\alpha\beta,\bq}(t)=W^{\delta}_{\alpha\beta,\bq}\delta(t)+\theta(t)\Big[W^{>}_{\alpha\beta,\bq}(t) - W^{<}_{\alpha\beta,\bq}(t) \Big]\,.
		\label{eq:W_ret}
	\end{split}
\end{equation}
Note that the Coulomb interaction $W$ contains both, background screening from the semiconductor and its dielectric environment in the ground state and screening from photoexcited carriers \cite{erben_excitation-induced_2018}. 
The Montroll-Ward self-energy is explicitly given by:
\begin{equation}
	\begin{split} 
		\Sigma_{\bk}^{\textrm{MW},\textrm{ret},\lambda}(\omega) = i\hbar\int_{-\infty}^{\infty}\frac{d\omega'}{2\pi}
		\frac{1}{\mathcal{A}}\sum_{\bk'\lambda'}\frac{(1-f^{\lambda'}_{\bk'})W^{>,\lambda'\lambda\lambda'\lambda}_{\bk'\bk\bk'\bk}(\omega')+ 
			f^{\lambda'}_{\bk'}W^{<,\lambda'\lambda\lambda'\lambda}_{\bk'\bk\bk'\bk}(\omega')}{\hbar\omega-E^{\lambda'}_{\bk'}+i\Gamma^{\lambda'}_{\bk'}-\hbar\omega'}
		\,,
		\label{eq:MW}
	\end{split}
\end{equation}
with the Fermi distribution functions for electrons and holes $f^{\lambda}_{\bk}$. The band sum is limited such that electron-hole exchange is not taken into account.
The plasmon propagators in Bloch representation are connected to the Coulomb matrix by:
\begin{equation}
	\begin{split} 
		W^{\gtrless,\lambda_1\lambda_2\lambda_3\lambda_4}_{\bk_1\bk_2\bk_3\bk_4}(\omega)=\sum_{\alpha,\beta} 
		(c^{\lambda_1}_{\alpha,\bk_1})^*(c^{\lambda_2}_{\beta,\bk_2})^*c^{\lambda_3}_{\beta,\bk_3}c^{\lambda_4}_{\alpha,\bk_4}
		W^{\gtrless}_{\alpha\beta,\bk_3-\bk_2}(\omega)
		\,.
		\label{eq:W_prop_Bloch}
	\end{split}
\end{equation}
In quasi-equilibrium, the propagators fulfill the Kubo-Martin-Schwinger relation \cite{kremp_quantum_2005}
\begin{equation}
	\begin{split} 
		W^{>}_{\alpha\beta,\bq}(\omega)= e^{\frac{\hbar\omega}{k_{\textrm{B}} T}} W^{<}_{\alpha\beta,\bq}(\omega)\,.
		\,
		\label{eq:W_prop_KMS}
	\end{split}
\end{equation}
Combining this with the Kramers-Kronig relations for the inverse dielectric function and Eq.~(\ref{eq:W_ret}), the propagators can be expressed in terms of the retarded Coulomb interaction:
\begin{equation}
	\begin{split} 
		W^{>}_{\alpha\beta,\bq}(\omega)&=(1+n_{\textrm{B}}(\omega))\,2i\,\textrm{Im}\,W^{\textrm{ret}}_{\alpha\beta,\bq}(\omega), \\
		W^{<}_{\alpha\beta,\bq}(\omega)&=   n_{\textrm{B}}(\omega) \,2i\,\textrm{Im}\,W^{\textrm{ret}}_{\alpha\beta,\bq}(\omega)
		\,
		\label{eq:W_prop_ret}
	\end{split}
\end{equation}
with the Bose distribution function $n_{\textrm{B}}(\omega)$. 
The retarded Coulomb matrix is obtained using the inverse dielectric matrix for photoexcited carriers as:
\begin{equation}
	\begin{split} 
		W^{\textrm{ret}}_{\alpha\beta,\bq}(\omega)&=\sum_{\gamma} \varepsilon^{-1,\textrm{ret},\alpha\gamma}_{\textrm{exc},\bq}(\omega)V^{\gamma\beta}_{\bq}
		\,.
		\label{eq:W_ret_eps}
	\end{split}
\end{equation}
The dielectric matrix itself is given by
\begin{equation}
	\begin{split} 
		\varepsilon^{\textrm{ret},\alpha\beta}_{\textrm{exc},\bq}(\omega)=\delta_{\alpha\beta}-\sum_{\gamma}V^{\alpha\gamma}_{\bq} P^{\gamma\beta}_{\textrm{exc},\bq}(\omega)
		\label{eq:eps_ab}
	\end{split}
\end{equation}
where we describe the polarization matrix for photoexcited carriers in the Lindhard (random-phase) approximation
\begin{equation} 
	\begin{split}  
		P^{\alpha\beta}_{\textrm{exc},\bq}(\omega)=\frac{1}{\mathcal{A}}\sum_{\lambda,\lambda',\bk} c_{\alpha, \mathbf{k}}^{\lambda} c_{\beta, \mathbf{k}-\mathbf{q}}^{\lambda'}  \Big(c_{\beta, \mathbf{k}}^{\lambda} \Big)^{*} \Big(c_{\alpha, \mathbf{k}-\mathbf{q}}^{\lambda'} \Big)^{*}
		\frac{f_{\bk-\bq}^{\lambda'}-f_{\bk}^{\lambda}}{\varepsilon_{\bk-\bq}^{\lambda'}-\varepsilon_{\bk}^{\lambda}+\hbar\omega+i\gamma}~.
		\label{eq:lindhard}
	\end{split}
\end{equation}
We use a phenomenological damping $\gamma =$ min($10$ meV, $\hbar\omega$) to ensure the correct analytic behavior in the static limit $\omega\rightarrow 0$.
\\In summary, the Montroll-Ward self-energy can be written as
\begin{equation}
	\begin{split} 
		\Sigma_{\bk}^{\textrm{MW},\textrm{ret},\lambda}(\omega) = i\hbar\int_{-\infty}^{\infty}\frac{d\omega'}{2\pi}
		\sum_{\bk'\lambda'}&\sum_{\alpha\beta}
            (c^{\lambda'}_{\alpha,\bk'})^*(c^{\lambda}_{\beta,\bk})^*c^{\lambda'}_{\beta,\bk'}c^{\lambda}_{\alpha,\bk}
		2i\,\textrm{Im}\,\Big\{ W^{\textrm{ret}}_{\alpha\beta,\bk-\bk'}(\omega')\Big\}
		\frac{1-f^{\lambda'}_{\bk'}+n_{\textrm{B}}(\omega')}{\hbar\omega-E^{\lambda'}_{\bk'}+i\Gamma^{\lambda'}_{\bk'}-\hbar\omega'}
		\,.
		\label{eq:MW_final}
	\end{split}
\end{equation}
The Fock self-energy is given by
\begin{equation}
	\begin{split} 
		\Sigma_{\bk}^{\textrm{F},\lambda} = -\sum_{\bk'\lambda'} V^{\lambda\lambda'\lambda\lambda'}_{\bk\bk'\bk\bk'} f^{\lambda'}_{\bk'}
		+\sum_{\bk'\bar{\lambda}'} U^{\lambda\bar{\lambda}'\lambda\bar{\lambda}'}_{\bk\bk'\bk\bk'} f^{\bar{\lambda}'}_{\bk'} \,,
		\label{eq:Fock}
	\end{split}
\end{equation}
where the index $\lambda'$ runs over bands within the same carrier species as $\lambda$, while $\bar{\lambda}'$ sums over carriers with opposite charge.
As discussed in \cite{erben_excitation-induced_2018}, electron-hole exchange is described by unscreened Coulomb matrix elements.
Using the same notation, the Hartree self-energy can be written as
\begin{equation}
	\begin{split} 
		\Sigma_{\bk}^{\textrm{H},\lambda} = \sum_{\bk'\lambda'} V^{\lambda\lambda'\lambda'\lambda}_{\bk\bk'\bk'\bk} f^{\lambda'}_{\bk'}
		- \sum_{\bk'\bar{\lambda}'} V^{\lambda\bar{\lambda}'\bar{\lambda}'\lambda}_{\bk\bk'\bk'\bk}f^{\bar{\lambda}'}_{\bk'} \,.
		\label{eq:Hartree}
	\end{split}
\end{equation}
Due to the Coulomb singularity at long wavelength ($\bq=0$), the Hartree interaction requires a separate treatment in the Wannier representation. 
For a charge-neutral system the macroscopic (leading) term drops out and only microscopic contributions to Hartree interaction remain. Following the procedure in \cite{schobert_ab_2023} we calculate Hartree-type matrix elements by setting the macroscopic eigenvalue to zero before transforming the matrix element to the Wannier representation.
\\
\\\textbf{Exciton binding energy}
\\
\\The exciton eigenfunctions $\phi^{he}_{\alpha,\bk}(\bq)$ and eigenenergies $E_{\alpha,\bq}$ in the absence of photoexcited carriers are solutions of the Bethe-Salpeter equation (BSE)
\begin{equation}
	\begin{split}
		(\varepsilon^{\textrm{e}}_{\bk-\bq}+\varepsilon^{\textrm{h}}_{\bk}-E_{\alpha,\bq})\phi^{he}_{\alpha,\bk}(\bq)
		-\frac{1}{\mathcal{A}}\sum_{\bk',h',e'}
		V^{e,h',h,e'}_{\bk-\bq,\bk',\bk,\bk'-\bq}\phi^{h'e'}_{\alpha,\bk'}(\bq)=0\,.
	\end{split}
	\label{eq:wannier}
\end{equation}
By diagonalizing Eq.~(\ref{eq:wannier}) for zero center-of-mass momentum $\bq$ and comparing the lowest eigenenergy for electrons and holes with the same spin to the corresponding bandgap at the K point, we obtain the binding energy of bright inter-layer excitons.
\\
\\\textbf{Results}
\\
\\To guide the interpretation of diffusion experiments with respect to different density regimes, we compute the density-dependent bandgap renormalization in an hBN-encapsulated heterobilayer for different  temperatures.
To this end, we solve the self-consistency equation (\ref{eq:GW_energy}) using the Montroll-Ward (\ref{eq:MW_final}), Fock (\ref{eq:Fock}) and Hartree (\ref{eq:Hartree}) self-energies. 
We assume that environmental screening from the hBN encapsulation layers is described by a dielectric constant of $\varepsilon_{\textrm{hBN}} = \sqrt{4.95 \cdot 2.86}$ \cite{artus_natural_2018}. 
In addition, a narrow gap of $0.3$ nm between the bilayer and the surrounding hBN layers has been taken into account \cite{florian_dielectric_2018}. 
To resolve the Fermi surface of electrons and holes at all temperatures, a uniform $90\times90\times1$ Monkhorst-Pack grid is used for Brillouin zone sampling. 
The bandgap shift is compared to the exciton binding energy resulting from the BSE (\ref{eq:wannier}), for which we obtain $E_{\textrm{B}}=131$ meV. 
The results are collected in Fig.~\ref{fig:gapshift_results}.

\subsection{Two-component diffusion model}

The model is proposed to illustrate a scenario, where one can observe effective contraction of the spatially-resolved PL due to the presence of more than one component in the propagation.
It is reasonable to consider presence of unbound electron-hole plasma that extends rapidly and a small residual fraction of excitons with slower propagation and longer lifetime.
As emphasized in the main text, this presents a phenomenological description of a scenario that could occur at these conditions.

To describe the exciton and electron-hole plasma diffusion we build on the two-component drift-diffusion model formulated in Ref.~\cite{Zipfel2020}:
	\begin{subequations}
		\label{X:eh:diffusion}
		\begin{align}
			\frac{\partial n}{\partial t} + \frac{n}{\tau} + R_A n^2 = {D_0} \Delta n + \frac{U_0 {D_0} }{k_BT} \bm \nabla\cdot(n\bm \nabla n) + \gamma_x n_{eh}^2,\label{diffusion:X}\\
			\frac{\partial n_{eh}}{\partial t} + \frac{n_{eh}}{\tau_{eh}} + R_{eh} n^2 = D_{eh} \Delta n_{eh} + \frac{U_{eh} D_{eh}}{k_BT} \bm \nabla\cdot(n_{eh}\bm \nabla n_{eh}) - \gamma_x n_{eh}^2.\label{diffusion:eh}
		\end{align}
	\end{subequations}
	Equation~\eqref{diffusion:X} describes the diffusion of bound excitons [cf. Eq.~\eqref{diffusion:1}] with $n$, as before, being the exciton density. Equation~\eqref{diffusion:eh} describes the diffusion of the electron-hole plasma, subscripts $eh$ denote quantities corresponding to the plasma: $n_{eh}$ is the density of electrons or holes in plasma (electric neutrality condition implies electron and hole densities being equal), $D_{eh}$ is the plasma diffusion coefficient, $\tau_{eh}$ describes geminate recombination rate in plasma, $R_{eh}$ describes bimolecular recombination rate, $U_{eh}$ takes into account repulsive interactions in plasma, and $\gamma_x$ describes electron and hole binding in excitons. In our model~Eq.~\eqref{X:eh:diffusion}, for simplicity, we neglect mutual drag of excitons and plasma and assume plasma to be non-degenerate, but the latter assumption affects only the value of $U_{eh}$. We take the following initial conditions, cf. Eq.~\eqref{initial},
	\begin{equation}
		\label{eh:X:init}
		n(\bm r,0) = \xi \frac{N_0}{\pi r_0^2} e^{-r^2/r_0^2}, \quad n_{eh}(\bm r,0) = (1-\xi) \frac{N_0}{\pi r_0^2} e^{-r^2/r_0^2},
	\end{equation}
	where the factor $0 \leqslant \xi \leqslant 1$ gives the fraction of excitons. The emission intensity is presented in the form
	\begin{equation}
		\label{eh:X:intensity}
		I(\bm r, t) = \alpha_X n(\bm r,t) + \alpha_{eh} n^2_{eh}(\bm r,0),
	\end{equation}
	with $\alpha_X$ and $\alpha_{eh}$ being the coefficients describing respective contributions.\\

\begin{figure}[ht]
	\centering
	\includegraphics[width=\textwidth]{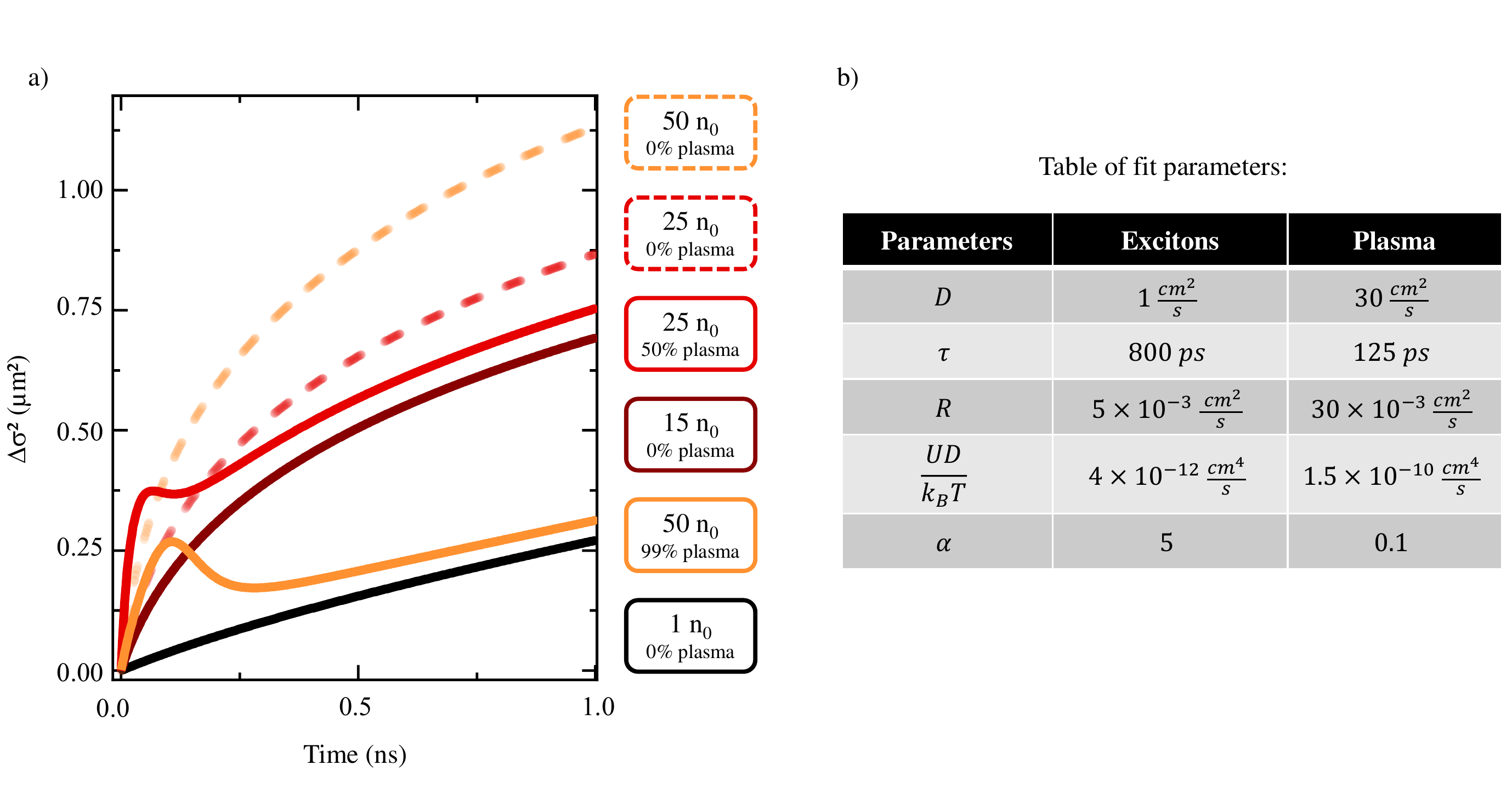}
	\caption{a) Simulated mean-squared displacement curves. Solid lines representing simulations with parameters corresponding to experiment, e.g. included plasma ratio for densities above Mott transition. Dashed lines represent behavior excluding presence of plasma. b) Corresponding parameters used in simulation for both exciton and plasma components.}
	\label{fig:Twocomponents}
\end{figure}

A selection of pump density dependent mean-squared displacement curves from this simulation is presented in \fig{fig:Twocomponents} a), corresponding parameters for exciton and plasma respectively are listed in table of  \fig{fig:Twocomponents} b). For densities above Mott transition ($n_{ex}\geq25n_0$) plasma is included (excluded) in the solid (dashed) lines.

Naturally, the two-component, excitons and plasma, diffusion is not the only one mechanism leading to observations of effectively negative diffusion. 
For example, an interplay of trapping and detrapping of excitons can result in the negative diffusivity\,\cite{Kurilovich2022}.
This can be, however, excluded, since these processes should also occur and be even more pronounced at low excitation densities.
Another possible scenario can be related to the instability of the exciton-plasma system towards the formation of droplets where particles can diffuse towards the droplet due to gain in the interaction energy\,\cite{Karpov1995,Argyrakis2009}.
Similarly, negative diffusivity of the electron-hole plasma may indicate uncompensated long-range forces, negative compressibility, and exciton formation\,\cite{Efros2008}.
Thus, we emphasize that the suggested two-component model is one, arguably simplest, possibility to consider as an explanation for our observations.

\newpage



%